\documentclass[12pt,a4paper]{article}

\usepackage[latin1]{inputenc}
\usepackage{graphicx} 
\usepackage{amsmath}
\usepackage{amssymb}
\usepackage{cite}
\usepackage[colorlinks=true, pdftex, linktocpage, citecolor=red, urlcolor=blue]{hyperref}
\usepackage{multirow}

{\end{list}}
\newcounter{enumct}

\newlength{\abstwidth}
\renewcommand{\b}{{\mathrm b}}
\renewcommand{\d}{\mathrm{d}}
\newcommand{\g}{{\mathrm g}}
\newcommand{\q}{{\mathrm q}}
\newcommand{\qbar}{\overline{\mathrm q}}
\renewcommand{\t}{{\mathrm t}}
\newcommand{\W}{{\mathrm W}}
\newcommand{\Z}{{\mathrm Z}}
\newcommand{\gammaZ}{\gamma^* / {\mathrm Z^0}}

\newcommand{\as}{\alpha_{\mathrm{s}}}
\newcommand{\aem}{\alpha_{\mathrm{em}}}
\newcommand{\pT}{p_{\perp}}
\newcommand{\pTs}{p^2_{\perp}}
\newcommand{\pTe}{p_{\perp\mathrm{evol}}}
\newcommand{\pTse}{p^2_{\perp\mathrm{evol}}}
\newcommand{\pTmin}{p_{\perp\mathrm{min}}}

\newcommand{\pTcut}{p_{\perp\mathrm{cutoff}}}
\newcommand{\pThat}{\hat{p}_{\perp}}
\newcommand{\zmin}{z_{\mathrm{min}}}
\newcommand{\zmax}{z_{\mathrm{max}}}
\newcommand{\mdip}{m_{\mathrm{dip}}}
\newcommand{\mred}{m_{\mathrm{red}}}

\begin{document}
\sloppy

% No page number on title page.
\pagestyle{empty}
\hypersetup{pageanchor=false}
 
\begin{flushright}
LU TP 17-28\\
MCnet-17-14\\
%arXiv:17??.????? [hep-ph]\\
September 2017
\end{flushright}

\vspace{\fill}

\begin{center}
{\LARGE\bf Some Dipole Shower Studies}\\[10mm]
{\Large Baptiste Cabouat\footnote{Now at School of Physics and Astronomy, 
Schuster Building, Oxford Road, University of Manchester, 
Manchester M13 9PL, United Kingdom}
 and Torbj\"orn Sj\"ostrand}\\[3mm]
{\it Theoretical Particle Physics,
Department of Astronomy and Theoretical Physics,}\\[1mm]
{\it Lund University, SE-223 62 Lund, Sweden}
\end{center}

\vspace{\fill}

\begin{center}
\begin{minipage}{\abstwidth}
{\bf Abstract}\\[2mm]
Parton showers have become a standard component in the description of
high-energy collisions. Nowadays most final-state ones are of the dipole 
character, wherein a pair of partons branches into three, with energy
and momentum preserved inside this subsystem. For initial-state
showers a dipole picture is also possible and commonly used, but the
older global-recoil strategy remains a valid alternative, wherein
larger groups of partons share the energy--momentum preservation task.
In this article we introduce and implement a dipole picture also for 
initial-state radiation in \textsc{Pythia}, and compare with the existing 
global-recoil one, and with data. For the case of Deeply Inelastic 
Scattering we can directly compare with matrix element expressions 
and show that the dipole picture gives a very good description over 
the whole phase space, at least for the first branching.
\end{minipage}
\end{center}

\vspace{\fill}

\phantom{dummy}

\clearpage

\pagestyle{plain}
\hypersetup{pageanchor=true}
\setcounter{page}{1}

\section{Introduction}

In the current description of high-energy collisions, such as those at the 
LHC, parton showers play a key role \cite{Buckley:2011ms,Olive:2016xmw}. 
The natural starting point for a description of the perturbative stage 
of the collisions is to use matrix-element (ME) calculations, but with
increasing parton multiplicity these rapidly become quite time-consuming.
A practical limit lies around eight final-state partons for leading-order
(LO) calculations and four for next-to-leading-order (NLO) ones.
By contrast, a high-$\pT$ LHC collision could contain a hundred partons
above a 1~GeV lower cutoff scale. It is therefore natural to combine the 
ME calculations for a few energetic and well separated partons 
with the parton-shower ones, that in an approximate manner can add 
further soft and collinear emissions. 
 
The concept of parton showers is implicit already in the DGLAP evolution 
equations \cite{Gribov:1972ri,Altarelli:1977zs,Dokshitzer:1977sg}, 
and over the years 
many shower algorithms have been written. In its simplest incarnation, 
a shower implements a set of successive partonic branchings $a \to b + c$,
where the two daughters $b$ and $c$ can branch further in their turn. 
Showers may differ in a number of respects, such as how emissions are 
ordered by an evolution variable, how energy and momentum is shared 
between the daughters of a branching, and how overall energy and momentum 
conservation is ensured. It is also necessary to distinguish between 
initial-state radiation (ISR) and final-state radiation (FSR), where 
the former involves a succession of spacelike partons stretching from
the original incoming protons to the hard interaction, while the latter
describes a cascade of timelike partons occurring afterwards. The naive 
choice of evolution variable, to order possible emissions, is the spacelike 
or timelike virtuality $Q$ of a parton, since by Heisenberg's uncertainty 
relation the proper lifetime of it should be of order $1 / Q$ (for 
$\hbar = 1$), such that lower $Q$'s should correspond to earlier times
for ISR and later for FSR. 

The virtuality choice does not take into account the possibility of 
destructive interference in the soft-gluon radiation pattern 
surrounding a pair of colour-correlated hard partons, however.
This can be solved by instead evolving in terms of a gradually
decreasing emission angle \cite{Marchesini:1983bm,Marchesini:1987cf}.
With modest updates \cite{Gieseke:2003rz} this algorithm remains the
default in the \textsc{Herwig} event generator
\cite{Bahr:2008pv,Bellm:2015jjp}, and has been successful over the years. 
To note is that the algorithm is not completely Lorentz-frame-independent 
and that overall energy--momentum conservation is only ensured at the 
very end by some nontrivial transformations.

\begin{figure}[t!] 
\centering
\includegraphics[width=0.5\textwidth]{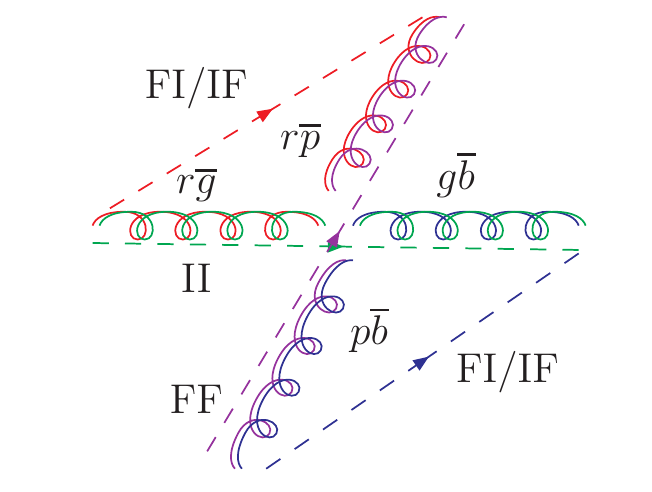}
\caption{Colour flow for the process 
$\g(r\overline{g}) + \g(g\overline{b}) \to \g(r\overline{p}) 
+ \g(p\overline{b})$. Here, the $N_{\mathrm{C}} \to \infty$ limit is
used so that $p$ stands for the new colour purple. The dashed lines
represent the colour lines stretching between the dipole ends. The
type of dipole is indicated.}
\label{FigExDip}
\end{figure}

An alternative is the dipole approach \cite{Gustafson:1986db}, 
first implemented in the \textsc{Ariadne} algorithm 
\cite{Gustafson:1987rq,Lonnblad:1992tz}, which also achieves a correct
handling of soft-gluon interference aspects. In it the $1 \to 2$
branching paradigm is replaced by a $2 \to 3$ one, where the original
dipole is defined by a pair of matching colour--anticolour partons, 
as defined in the $N_{\mathrm{C}} \to \infty$ limit \cite{tHooft:1973alw},
where each colour label is unique. Often it is convenient to split the 
full radiation pattern into two dipole-end contributions, where one of 
the two partons acts as radiator and the other as recoiler, with 
four-momentum preserved inside the dipole. The terminology then is 
to refer to FF, II, FI and IF emissions, depending on whether the
radiator and recoiler are in the final (F) or initial (I) state.
The FI and IF cases occur when a colour line flows from the initial 
to the final state. An example of every type of dipoles is given in 
Fig.~\ref{FigExDip}. The dipole approach is, in many variants, standard 
in generators such as \textsc{Sherpa}\cite{Schumann:2007mg,Gleisberg:2008ta}, 
\textsc{Vincia} \cite{Giele:2007di,Fischer:2016vfv} and \textsc{Dire} 
\cite{Hoche:2015sya}, and is an option in \textsc{Herwig} 
\cite{Platzer:2009jq}. For the extensions to ISR, often the 
Catani--Seymour dipole kinematics is used \cite{Catani:1996vz}.

\begin{figure}[t!] 
\centering
\includegraphics[width=0.3\textwidth]{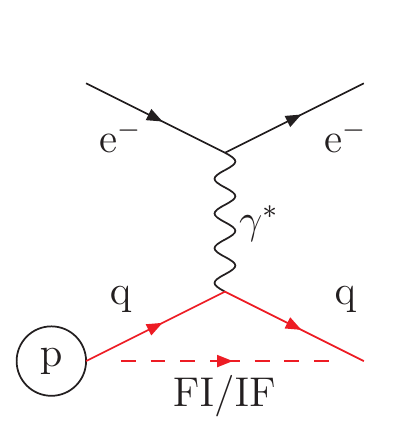}
\caption{Deeply inelastic scattering: an incoming electron scatters
one of the quark within the incoming proton. The dashed line
represents the colour line stretching between the two dipole ends.}
\label{FigDIS}
\end{figure}

The \textsc{Pythia} generator
\cite{Sjostrand:2004ef,Sjostrand:2006za,Sjostrand:2014zea}, is also
dipole-based for FSR, both FF and FI topologies, but ISR is
implemented in the so-called global recoil scheme that is implicit in
an II dipole setup, wherein all final-state particles share the recoil
of an ISR emission. This is a perfectly valid approach for a process
like $\gammaZ$ production at hadron colliders, insofar as it attaches
well with a ME-based view of the production process. A consistent
FI/IF dipole handling is essential for a description of showers in
Deeply Inelastic Scattering (DIS), however
\cite{Carli:2010cg,Hoche:2015sya}. For this case, it can be seen in
Fig.~\ref{FigDIS} that a FI/IF dipole naturally stretches between
the incoming quark and the final scattered quark. In the current
article, therefore, we develop and implement a description of the IF
emission topology, and combine it with the FI contribution. As it
turns out, it is possible to set up kinematics such that the IF
contribution matches the DIS gluon-emission ME, thereby providing an
economical description. The new framework also allows a comparison of
dipole vs. global recoil e.g.\ for $\gammaZ$ production at hadron
colliders. 

Do note that the showers described in this article are formally
accurate only to leading log (LL), although many aspects of
next-to-leading-log (NLL) physics are implemented, such as the use of
$\pTs$ as $\as$ scale. By contrast, while early attempts to develop
NLL parton showers petered out (NLLJET \cite{Kato:1990as}), the main
thrust of current shower development is in that direction, with first
implementations in \textsc{Vincia} \cite{Li:2016yez} and \textsc{Dire}
\cite{Hoche:2017iem,Hoche:2017hno}. An important aspect of this game
is also to correctly include subleading colour corrections
\cite{Platzer:2012np}. 

The plan of the article is as follows. In Section~2 we describe the
current default framework for ISR and FSR in \textsc{Pythia}, to set
the stage. Section~3 introduces the new alternative framework for the
FI + IF dipole handling, with special emphasis on the comparison with
the DIS MEs. A first comparison with data is presented in Section~4,
both for DIS and for $\gammaZ$ and jets production at hadron colliders.
Finally Section~5 provides a summary and outlook.

 \section{The existing framework}

Given a hard process as starting point, \textsc{Pythia} will create a
parton-level event by interleaving ISR, FSR and MPI (multiparton 
interaction) activity in a combined downwards evolution in transverse 
momentum 
\begin{eqnarray} 
\frac{\d \mathcal{P}}{\d \pT} & = & \left( 
\frac{\vphantom{\left(\right)} \d\mathcal{P}_{\mathrm{MPI}}}{\d \pT} 
+ \sum \frac{\vphantom{\left(\right)} \d\mathcal{P}_{\mathrm{ISR}}}{\d \pT} 
+ \sum \frac{\vphantom{\left(\right)} \d\mathcal{P}_{\mathrm{FSR}}}{\d \pT} 
\right) \nonumber \\ 
& \times & \exp \left( - \int_{\pT}^{p_{\perp\mathrm{max}}} \left(
\frac{\vphantom{\left(\right)} \d\mathcal{P}_{\mathrm{MPI}}}{\d \pT'}
+ \sum \frac{\vphantom{\left(\right)} \d\mathcal{P}_{\mathrm{ISR}}}{\d \pT'} 
+ \sum \frac{\vphantom{\left(\right)} \d\mathcal{P}_{\mathrm{FSR}}}{\d \pT'}
\right) \d \pT' \right) ~,
\label{eq:combinedevol}
\end{eqnarray} 
that probabilistically determines what the next step
will be \cite{Corke:2010yf}. Here the ISR sum runs over all incoming
partons, two for each already produced MPI, including the hard
interaction itself, the FSR sum runs over all outgoing partons, and
$p_{\perp\mathrm{max}}$ is the $\pT$ of the previous step. The
Sudakov-style \cite{Sudakov:1954sw} exponential ensures that
probabilities are bounded by unity. While FSR is described by
evolution from the hard process forwards, ISR is described by
evolution from it backwards to the shower initiators
\cite{Sjostrand:1985xi}. The decreasing $\pT$ scale therefore is not
a simple time variable, but can instead be viewed as an evolution
towards increasing resolution power. That is, given that the event
has a particular structure when activity above some $\pT$ scale is
resolved, how might that picture change when the resolution cutoff is
reduced by some infinitesimal $\d \pT$? 

The ISR and FSR branching probabilities in eq.~(\ref{eq:combinedevol})
are provided by standard DGLAP evolution equations, where the
evolution variable is a modified $\pT$ scale
\begin{eqnarray} 
\pTse & = & z (1 - z) Q^2 ~~ \mathrm{for~FSR} ~, \\
\pTse & = & \phantom{z} (1 - z) Q^2 ~~ \mathrm{for~ISR} ~. 
\end{eqnarray}  
Here $Q^2$ is the timelike or spacelike virtuality of
the off-shell parton for FSR and ISR, respectively
\cite{Sjostrand:2004ef}. (For simplicity we only show the formulae in
the massless case.) The $\pTe$ would agree with the conventional
$\pT$ of the daughters in a branching if $z$ had been defined as the
fraction of the lightcone momentum $E + p_{\mathrm{longitudinal}}$. Now 
it is not, as we shall see, which leads to modest mismatches between 
$\pTe$ and $\pT$. In eq.~(\ref{eq:combinedevol}) ISR and FSR is actually 
to be written in terms of its respective $\pTe$, while the MPI $\pT$ 
remains the normal one.

\begin{figure}[t!] \centering
\includegraphics[width=0.7\textwidth]{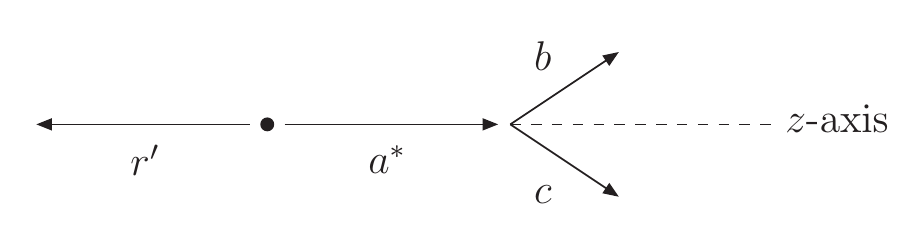}
\caption{Sketch of the branching $a\to b + c$ with a recoiler $r$. The
$z$-axis is chosen to be along the direction of the momentum of parton
$a$.}
\label{FigFF}
\end{figure}

FSR on its own is handled by dipole showering. Each coloured parton
$a$ is assigned a recoiler $r$ that carries the corresponding
anticolour in an $N_C \to \infty$ representation of the colour
flow. (Exceptions exist, such as in the decay $\t \to \b \W^+$, where
the $\W$ is the recoil partner of the $\b$, so as to preserve the
$\t$ mass.) In a branching $a \to b + c$ the dipole invariant mass is
preserved by the recoiler energy being scaled down, while its
direction is maintained. Kinematically, the branching can be split in
two steps: $a + r \to a^* + r' \to b + c + r'$, where $a^*$ is the
intermediate off-shell parton of virtuality $Q^2$ (see Fig.~\ref{FigFF}). 
In the first step four-vectors are modified according to
\begin{eqnarray} 
p_{a^*} & = & p_a + \frac{Q^2}{m_{ar}^2} p_r ~, \label{eq:eafsrshift} \\ 
p_{r'} & = & \left( 1 - \frac{Q^2}{m_{ar}^2} \right) p_r ~.
\label{eq:erfsrshift} 
\end{eqnarray} 

The $z$ definition in the second step is most easily related to the
kinematics in the dipole rest frame by $E_b = z E_{a^*}, E_c = (1 -
z) E_{a^*}$. These $\pTse$ and $z$ definitions have the advantage
that they exactly match on to the singularity structure of MEs, such
as the textbook $\gammaZ \to \q(1) + \qbar(2) + \g(3)$ one, when
radiation from the two dipole ends is combined
\cite{Bengtsson:1986hr,Corke:2010yf},
\begin{equation}
\frac{\d p_{\perp\mathrm{evol},\q}^2}{p_{\perp\mathrm{evol},\q}^2} \, 
\frac{\d z_{\q}}{1 - z_{\q}} 
+ \frac{\d p_{\perp\mathrm{evol},\qbar}^2}{p_{\perp\mathrm{evol},\qbar}^2} \,
\frac{\d z_{\qbar}}{1 - z_{\qbar}} 
= \frac{\d x_1 \, \d x_2}{(1 - x_2) x_3} 
+ \frac{\d x_1 \, \d x_2}{(1 - x_1) x_3} 
= \frac{\d x_1 \, \d x_2}{(1 - x_1) (1 - x_2)} ~,
\label{eq:split}
\end{equation} 
with $x_i = 2 E_i/E_{\mathrm{tot}}$. Matrix element
corrections therefore are easily implemented (also when generalized
to massive kinematics \cite{Norrbin:2000uu,Sjostrand:2004ef}). This
would not be the case if the true $\pT$ had been used instead of
$\pTe$ \cite{Cabouat:2017xx}, at least with this $z$ definition.
 
ISR on its own is handled with backwards evolution and a global
recoil. That is, consider a collision $b + r \to F$, where $F$ may
represent a multibody final state. If $b$ comes from a previous
branching $a \to b + c$, by backwards evolution, the full process
reads $a + r \to b^* + c + r \to F' + c$. Note that $r$ remains
unchanged by the branching in this case. Here $z = m^2_{br} / m^2_{ar}$, 
which gives a good match to relevant (Mandelstam) ME
variables. Considering e.g.\ emission in a $\q + \qbar \to \Z^0$
process, giving $\q + \qbar \to \Z^0 + \g$, 
$\hat{s} = m^2_{\mathrm{Z}} / z$ and 
\begin{equation} 
\frac{\d\pTse}{\pTse} = \frac{\d Q^2}{Q^2} = \frac{\d \hat{t}}{\hat{t}} 
~~\mathrm{or}~~ \frac{\d \hat{u}}{\hat{u}}~,
\end{equation} 
simplifying ME reweighting also here
\cite{Miu:1998ju}. The $F'$ system is a boosted and rotated copy of
$F$, i.e. the internal topology is unchanged. As the backwards
evolution continues, the new $F$ system also contains the $c$ parton
of the previous branching. 

The ISR and FSR descriptions can be separated so long as colour does
not flow between the initial and the final state. Notably, if $F$ is
a colour singlet state, the ISR approach above is a valid $a + r$ II
dipole-language description of the radiation. At hadron colliders
this is seldom the case, however, and therefore an FI and IF handling
need to be introduced, one way or another, for the colour dipoles
stretched between the initial and the final state.

The kinematics of an FI branching gives some differences relative to
an FF one. In the dipole rest frame a fraction $Q^2 / m_{ar}^2$ of the
recoiler energy is given from the recoiler to the emitter, exactly as
in eq.~(\ref{eq:eafsrshift}). But the recoiler is not a final-state
particle, so the increase of $a$ momentum is not compensated anywhere
in the final state. Instead the incoming parton that the recoiler
represents must have its momentum increased, not decreased, by the
same amount as the emitter. That is, its momentum fraction $x$ needs 
to be scaled up as
\begin{equation} 
x_{r'} = \left( 1 + \frac{Q^2}{m_{ar}^2} \right) x_r ~.
\label{eq:fraction}
\end{equation} 
Note that the direction along the incoming beam axis is
not affected by this rescaling, and that the kinematics construction
therefore inevitably comes to resemble that of Catani--Seymour
dipoles \cite{Catani:1996vz}. The dipole mass $m_{ar}$ and the
squared subcollision mass $\hat{s}$ are increased in the process, the
latter by the same factor as $x_r$. As with ISR, the increased $x$
value leads to an extra PDF weight
\begin{equation} 
\frac{x_{r'} f_r(x_{r'},\pT^2)}{x_r f_r(x_r,\pT^2)}
\label{eq:pdfratio}
\end{equation} 
in the emission probability and Sudakov form
factor. This ensures a proper damping of radiation in the $x_{r'} \to
1$ limit.

So far \textsc{Pythia} has had no implementation of IF dipole ends;
all ISR is handled by the II approach. To first approximation this is
no problem for the total emission rate, so long as each incoming
parton is allowed to radiate according to its full colour charge. In
more detail, however, one must beware of a double- or undercounting of
the full radiation pattern when it is combined with the FI
contribution. Note that this pattern should depend on the scattering
angle of the colour flow in a hard process: if colour flows from an
incoming parton $i$ to a final parton $f$ then 
$m^2_{if} = E_i E_f (1 - \cos\theta_{if})$ 
sets the phase space available for emission. In
\cite{Corke:2010yf} an approximate prescription is introduced to
dampen FI radiation that otherwise could be doublecounted, but no
corresponding procedure is implemented on the ISR side. What is done
with ISR, on the other hand, is to implement azimuthal asymmetries in
the radiation pattern from colour coherence
considerations\cite{Webber:1986mc}, that lines up radiation off the
$i$ parton with the azimuthal angle of the $f$, in the same spirit as
a dipole would, but presumably not as accurately.

\begin{figure}[t!] \centering
\includegraphics[width=0.45\textwidth]{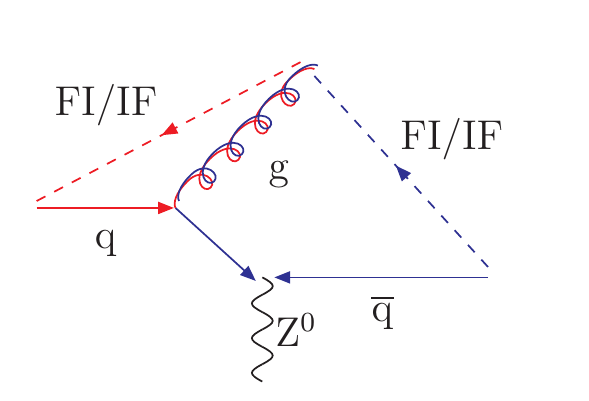}
\caption{Colour flow for the process $\q + \qbar \to \gammaZ +
\g$. The dashed lines represent the colour lines stretching between
the dipole ends.}
\label{FigZg}
\end{figure}

While it thus would seem that the dipole IF + FI approach is superior
to the global-recoil one, the issue is not always as one-sided. The
prime example is $\q + \qbar \to \gammaZ$ production. Once a gluon
has been emitted from the original $\q + \qbar$ II dipole, any further
emission will be related to the resulting $\q + \g$ and $\g + \qbar$
dipoles represented in Fig.~\ref{FigZg}. Therefore the $\gammaZ$
only receives a recoil in the first step for the dipole approach. With
Feynman diagrams, on the other hand, the $\gammaZ$ takes a recoil
that is modified as further gluon emissions are considered. In this
respect the global-recoil shower strategy is analogous with how
resummation techniques \cite{Dokshitzer:1978hw} are used to sum up
the effects of infinitely many gluon emissions on the $\pT$ spectrum
of the $\gammaZ$. This clear defect of the dipole picture has been a
main reason to maintain the older global-recoil strategy, with modest
improvements. 

Nowadays showers are not used on their own when high precision is
required, however, but are matched/merged with higher-order MEs
\cite{Buckley:2011ms}. With the kinematics of the hardest four or so
emissions based on MEs, and only subsequent ones described by showers,
it is reasonable to assume that the $\gammaZ$ $\pT$ spectrum is not
impaired by the lack of further recoils. On a philosophical level, it
still reminds us that the dipole picture also is an approximation,
and that different approaches should be developed as a means to assess
uncertainties.

Finally, it should be mentioned that \textsc{Pythia} also contains a
global-recoil option for FSR, not only for ISR. That is, when one
final parton radiates, all other final partons are boosted, as a unit,
so as to preserve total four-momentum. This option is mainly intended
to simplify matching/merging with NLO results, the way they are
calculated with the \textsc{MadGraph5\_aMC@NLO} program
\cite{Alwall:2014hca}. Typically global recoil is therefore only used
in the first one or two branchings, whereafter one switches to the
dipole picture. A similar strategy could be envisioned for ISR, even
if it has not been studied here. 

\section{The new approach} 

\subsection{Kinematics for IF emissions}
\label{subsec:kinematics}

\begin{figure}[t!] \centering
\includegraphics[width=0.7\textwidth]{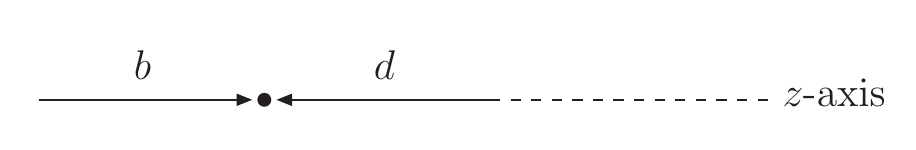}\\
(a)\\
\includegraphics[width=0.7\textwidth]{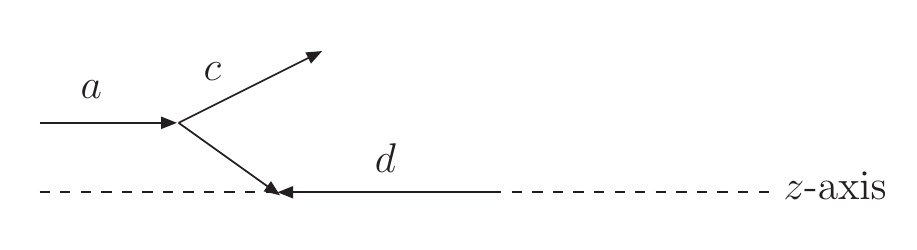}\\
(b)
\caption{ISR kinematics. (a) Before branching: partons $b$ and $d$ 
incoming in the $\pm z$ direction. (b) After the branching $a \to b + c$, 
now with $a$ and $d$ are along the $z$-axis.}
\label{FigIF}
\end{figure}

Let us consider a collision in the event frame between two incoming
partons $b$ and $d$ with four-momenta
$p_{b,d} = x_{b,d}(\sqrt{s}/2)(1;0,0,\pm 1)$, where $\sqrt{s}$ is the
total centre-of-mass energy and $x_{b,d}$ are the four-momentum
fractions. The two partons are taken as massless. A sketch of the
process is given in Fig.~\ref{FigIF}(a).

When evolving backwards in time, the parton $b$ is seen as coming from 
the branching $a \to b + c$. The parton $b$ hence obtains a spacelike 
virtuality $Q^2 > 0$, and the previously established kinematics has to 
be modified. It is now the parton $a$ which should be the incoming
one, with four-momentum $p'_a = x_a(\sqrt{s}/2)(1;0,0,1)$ in the
event frame, Fig.~\ref{FigIF}(b) (primed labels denote momenta after 
the branching has been considered). The parton $d$ keeps its original 
four-momentum, so $p'_d = p_d$. The variable $z$ is defined as 
$z = x_b/x_a$ or, in terms of invariant masses, as 
$z = m_{bd}^2/m_{ad}^2$. This holds since $a$, $b$ and $d$ are always 
taken as massless, so that $m_{bd}^2 = (p_b + p_d)^2 = x_b\,x_d\,s$ and
$m_{ad}^2 = (p'_a + p'_d)^2 = x_a\,x_d\,s$ \cite{Sjostrand:2004ef}. 
Therefore also $p'_a = p_b / z$. 

In the default global-recoil approach, the whole final state created by 
$\{b + d\}$ obtains changed momenta. In the new scheme, the recoil is 
instead taken by the single final parton $f$, which is the one 
colour-connected to parton $b$. In the following, this parton $f$ is 
referred to as the colour partner.

Before the branching, four-momentum conservation gives
\begin{equation} 
p_b + p_d = p_f + p_F,
\end{equation}
where $F$ represents the system of all final partons except for the 
colour partner. After the branching $a \to b + c$ instead
\begin{equation} 
p'_a + p'_d = p'_f + p'_F + p'_c.
\end{equation}
The local recoil ansatz implies that $p'_d = p_d$ and $p'_F = p_F$,
while $p'_f \neq p_f$. The difference between the above two equations 
gives
\begin{equation}
p'_a - p_b = p'_f - p_f + p'_c,
\label{IFcons} 
\end{equation}
where $p_b$, $p'_a$ and $p_f$ are known. Together $p'_f$ and $p'_c$ 
contain eight unknowns. Equation (\ref{IFcons}) gives four constraints,
and three others are 
\begin{equation} 
p_c'^2 = m_c^2, \hspace{75pt} p_f'^2 = p_f^2 = m_f^2,
\label{mcol}
\end{equation}
\begin{equation} 
p_b'^2 = (p'_a - p'_c)^2 = -Q^2.
\end{equation}
The remaining degree of freedom is the azimuthal angle $\varphi$ of the 
emitted parton $c$, which can be generated isotropically in the dipole
rest frame. This is one of the advantages of the new approach: azimuthal 
asymmetries due to colour coherence effects are automatically generated 
when the system is boosted and rotated back to the event rest frame.

Given the $Q^2$ and $z$ variables, the unknown four-momenta can be
expressed in the $\{b + f\}$ rest frame, here denoted as $\hat{p}$. 
Before the branching
\begin{eqnarray}
\hat{p}_b & = & \left(\frac{\mdip^2 - m_f^2}{2\mdip};0,0,
\frac{\mdip^2 - m_f^2}{2\mdip}\right), \\
\hat{p}_f & = & \left(\frac{\mdip^2 + m_f^2}{2\mdip};0,0,
 - \frac{\mdip^2 - m_f^2}{2\mdip}\right),
\end{eqnarray}
with $\mdip^2 = (p_b + p_f)^2$. Note that the dipole mass is not
conserved during the branching. After the branching
\begin{eqnarray}
\hat{p}'_c & = & \left( \frac{1}{z} - 1 \right) \hat{p}_b 
+ \hat{p}_{\mathrm{shift}}, \\
\hat{p}'_f & = & \hat{p}_f - \hat{p}_{\mathrm{shift}},
\end{eqnarray}
with
\begin{equation}
\hat{p}_{\mathrm{shift}} = \left( 
\frac{(2z - 1)Q^2}{2\mdip} + z\frac{m_c^2}{\mdip};
\hat{p}_{\perp} \cos\varphi, \hat{p}_{\perp} \sin\varphi, 
- \frac{Q^2}{2\mdip} - z\frac{m_f^2}{\mdip} 
\frac{Q^2 + m_c^2}{\mdip^2 - m_f^2} \right),
\label{eq:shift}
\end{equation}
where $\hat{p}_{\perp}$ is the transverse momentum of parton $c$ 
with respect to the dipole axis
\begin{equation}
\hat{p}_{\perp}^2 = \left((1 - z)(Q^2 + m_c^2) - m_c^2\right)
\left(1 - z\frac{Q^2 + m_c^2}{\mdip^2 - m_f^2}\right)
- m_f^2\left(z\frac{Q^2 + m_c^2}{\mdip^2 - m_f^2}\right)^2.
\label{eq:pTs}
\end{equation}
The same set of rotations and boosts as used to get to the $\{b + f\}$
rest frame can then be inverted to bring $\hat{p}'_c$ and $\hat{p}'_f$ 
back to $p'_c$ and $p'_f$ in the event rest frame.

\subsection{Gluon emission in DIS}

Now that the kinematics has been set up, the emission pattern 
of IF systems can be analyzed, as described by
$\d \mathcal{P}_{\mathrm{ISR}}$ in eq.~(\ref{eq:combinedevol}),
using standard DGLAP splitting kernels and backwards evolution as for 
II dipoles \cite{Sjostrand:2004ef}. For simplicity the PDF corrections, 
cf.\ eq.~(\ref{eq:pdfratio}), are omitted in the following discussion. 

\begin{figure}[t!]
 \centering
\includegraphics[width=0.7\textwidth]{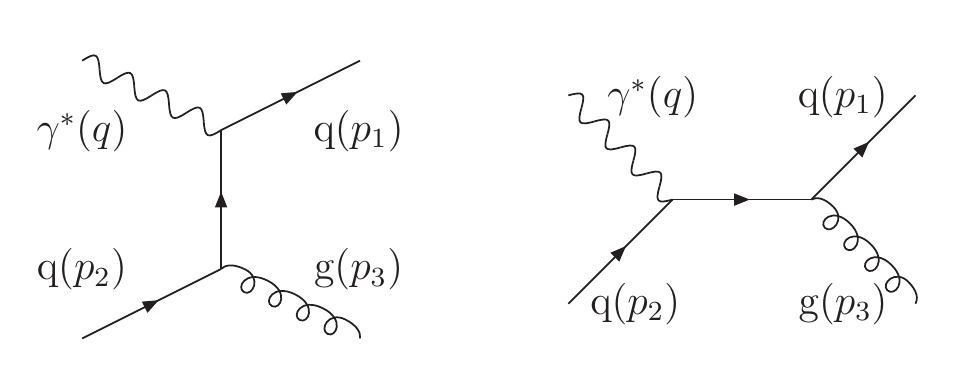}
\caption{The two Feynman graphs contributing to the process
$\gamma^* + \q \to \q + \g$ at $\mathcal{O}(\aem\as)$. The assigned
four-momenta of the particles are given in brackets.}
\label{FigDISFeyn}
\end{figure}

As already explained, the prime example is gluon emission in DIS,
where a single FI/IF dipole naturally appears. At $\mathcal{O}(\aem\as)$
two Feynman graphs lead to this process, Fig.~\ref{FigDISFeyn}.

From a parton-shower point of view, the final state $\q + \g$ can be
generated either via gluon emission off the IF system or off the FI
one. The standard FSR machinery implemented in the existing
\textsc{Pythia} already includes FI systems. The aim now is to
calculate the contribution of an IF emission in this case and combine
it with the contribution from the FI emission in order to compare with
the full matrix element, which also includes interference effects. This 
is done by using the IF kinematics previously established.

To study this kind of processes, the $(x,z_1)$ variables used by Catani
and Seymour \cite{Catani:1996vz} are convenient:
\begin{equation} x = \frac{|q^2|}{2\,p_2\cdot q}, \hspace{75pt}
z_1 = \frac{p_{1}\cdot p_2}{p_2\cdot q}.
\end{equation}
The usual $Q^2$ and $z$ variables can be expressed in terms of $(x,z_1)$, 
noting that $\hat{p}'_a = p_2$, $\hat{p}'_f = p_{1}$, $\hat{p}'_c = p_{3}$ and 
$q = p_{1} + p_{3} - p_2$. Setting $m_f = m_c = 0$ for simplicity, 
one finds that
\begin{equation} 
z = x, \hspace{50pt} Q^2 = \mdip^2\frac{1 - z_1}{x}, \hspace{50pt}
\d Q^2 \, \d z = \frac{\mdip^2}{x} \, \d x \, \d z_1. 
\end{equation}
Therefore the probability for an IF-type branching $\q \to \q + \g$ is 
\begin{equation} 
\d \mathcal{P}_{\q \to \q \g}^{\mathrm{IF}} 
= \frac{4}{3} \, \frac{\as}{2\pi} \, \frac{\d Q^2}{Q^2} \, 
\frac{1 + z^2}{1 - z} \, \d z = \frac{4}{3}\,\frac{\as}{2\pi}\,
\frac{(1 + x^2)\,\d x\,\d z_1}{(1 - x)(1 - z_1)}.
\label{eq:q2qgDIS}
\end{equation}
The soft ($z = 1$) and collinear ($Q^2 = 0$) singularities are
mapped onto the singularities $x = 1$ and $z_1 = 1$, respectively. 
This is a striking result, since the full matrix-element expression for
$\gamma^* + \q \to \q + \g$  has exactly the same two singularities.
Overall, only the numerator is slightly different, $z_1^2 + x^2$ 
\cite{Catani:1996vz} instead of our $1 + x^2$ . (There are also finite 
terms that we leave aside here.) That is, the IF-type branching 
$\q \to \q + \g$ generates both singularities of the full cross-section 
on its own. 

The recoil of an FF dipole emission is not uniquely specified 
\cite{Gustafson:1987rq}. In \textsc{Pythia} the FF dipole is 
artificially split into two dipole ends according to 
eq.~(\ref{eq:split}), which allows to have two different 
phase-space mappings, where the recoiler in each case does not change 
its direction of motion. For the FI/IF dipole there is no such freedom: 
the two incoming partons must always be parallel with the beam axis,
whether an emission is viewed as an FI or an IF one 
\cite{Sjostrand:2004ef, Cabouat:2017xx}. Also, the momentum fraction 
of the initial-state dipole end has to be increased after the branching, 
cf.\ eq.~(\ref{eq:fraction}), in order to absorb the virtuality.
Therefore there will only be one phase-space mapping. The full emission 
rate could still be viewed as a sum of one IF and one FI contribution, 
by splitting the expression in eq.~(\ref{eq:q2qgDIS}) in the spirit of 
eq.~(\ref{eq:split}). A corresponding reweighting of the IF rate would be 
easily achieved. Unfortunately the $(Q^2,z)$ variables of an FI 
dipole-end emission are not trivially related to the $(x,z_1)$ ones.
This could be overcome by a reweighting with the appropriate Jacobian,
but would be more cumbersome and not bring any benefits relative to using 
only $\d \mathcal{P}_{\q \to \q \g}^{\mathrm{IF}}$, which on its own 
reproduces the full answer so well. In the end, working only with IF 
emissions then seems reasonable. 

\subsection{Generalisation}

The previous example was for the branching $\q \to \q + \g$. It is now
important to verify whether these features, which appear for this
specific branching, are also present for the other kinds of
branchings. Therefore, the emission probabilities for IF systems and
FI systems will be compared. The objective is to check whether the
emission pattern of the FI type can be described by the IF type
only, at least as far as the singularity structure goes. Invariant 
masses will be used as variables to make the comparison easier between 
ISR and FSR. For an IF branching $a \to b + c$  they are 
$m_{ac}^2 = (p'_a + p'_c)^2$ and $m_{fc}^2 = (p'_f + p'_c)^2$. 
In the FI case we have $m_{bc}^2 = (p'_b + p'_c)^2$ and 
$m_{rc}^2 = (p'_{r} + p'_c)^2$, where $r$ is the recoiling colour partner 
in the initial state (recall eq.~(\ref{eq:fraction}) and 
Fig.~\ref{FigFF}). For massless partons this gives
\begin{equation}
z = \frac{\mdip^2}{\mdip^2 + m_{fc}^2},
\hspace{75pt} Q^2 = m_{ac}^2
\label{IFkin}
\end{equation}
for IF, and
\begin{equation}
z = \frac{\mdip^2(\mdip^2 + m_{bc}^2)
- m_{rc}^2(\mdip^2 - m_{bc}^2)}{(\mdip^2 + m_{bc}^2)^2},
\hspace{50pt} Q^2 = m_{bc}^2
\label{FIkin}
\end{equation} 
for FI. The limits $m_{ac}^2\to 0$ and $m_{rc}^2\to 0$ can be associated
with IF emissions, and the limits $m_{bc}^2\to 0$ and $m_{fc}^2\to 0$ 
with FI ones. Table~\ref{table:sing} summarizes the singularity structure 
of the branching probabilities for IF and FI.

\begin{table}[t] 
\centering
\begin{tabular}{|c|c|c|c|} 
\hline \multirow{3}{*}{Branching $a\to bc$} 
& \multirow{3}{*}{\begin{tabular}{c} Singularities \\ 
of $P_{a\to bc}(z)$ \end{tabular}} &
\multirow{3}{*}{\begin{tabular}{c} Singularities \\ 
of $\d \mathcal{P}_{a\to bc}^{\mathrm{IF}}$ \end{tabular}}
& \multirow{3}{*}{\begin{tabular}{c} Singularities \\ 
of $\d \mathcal{P}_{a\to bc}^{\mathrm{FI}}$ \end{tabular}}
\\ & & & \\ & & & \\ \hline 
\multirow{3}{*}{$\q \to \q \g$} &
\multirow{3}{*}{$\displaystyle \frac{1}{1 - z}$} &
\multirow{3}{*}{$\displaystyle
\frac{\d m_{ac}^2\,\d m_{fc}^2}{m_{ac}^2\,m_{fc}^2}$}
& \multirow{3}{*}{$\displaystyle
\frac{\d m_{bc}^2\,\d m_{rc}^2}{m_{bc}^2}$} 
\\ & & & \\ & & & \\ \hline 
\multirow{3}{*}{$\q \to \g \q$} &
\multirow{3}{*}{$\displaystyle \frac{1}{z}$} &
\multirow{3}{*}{$\displaystyle
\frac{\d m_{ac}^2\,\d m_{fc}^2}{m_{ac}^2}$} &
\multirow{3}{*}{$\displaystyle
\frac{\d m_{bc}^2\,\d m_{rc}^2}{m_{bc}^2}$} 
\\ & & & \\ & & & \\ \hline 
\multirow{3}{*}{$\g \to \g\g$} &
\multirow{3}{*}{$\displaystyle \frac{1}{z(1 - z)}$} &
\multirow{3}{*}{$\displaystyle
\frac{\d m_{ac}^2\,\d m_{fc}^2}{m_{ac}^2\,m_{fc}^2}$}
& \multirow{3}{*}{$\displaystyle
\frac{\d m_{bc}^2\,\d m_{rc}^2}{m_{bc}^2}$} 
\\ & & & \\ & & & \\ \hline 
\multirow{3}{*}{$\g \to \q\qbar$} & 
\multirow{3}{*}{1} &
\multirow{3}{*}{$\displaystyle
\frac{\d m_{ac}^2\,\d m_{fc}^2}{m_{ac}^2}$} &
\multirow{3}{*}{$\displaystyle
\frac{\d m_{bc}^2\,\d m_{rc}^2}{m_{bc}^2}$} 
\\ & & & \\ & & & \\ \hline
\end{tabular}
\caption{Singularity structure of the probability of emission for IF
and FI. For IF, the parton $b$ in the branching $a\to b + c$ is the
one which was incoming before the backward evolution. The mapping
between IF and FI labels is the following: $a\leftrightarrow r$,
$f\leftrightarrow b$ and $c\leftrightarrow c$.}
\label{table:sing}
\end{table}

For the branchings $\q \to \q \g$ and $\g \to \g \g$,
$\d \mathcal{P}_{a\to bc}^{\mathrm{IF}}$ contains both
the singularities $m_{ac}^2 = 0$ and $m_{fc}^2 = 0$. The first one is
expected since it is an IF system, but the singularity $m_{fc}^2 = 0$
is actually the same as the singularity $m_{bc}^2 = 0$ which shows up
in $\d \mathcal{P}_{a\to bc}^{\mathrm{FI}}$. Therefore, by analogy with 
the DIS case, the probability $\d \mathcal{P}_{a\to bc}^{\mathrm{IF}}$ 
seems sufficient to describe the emission pattern of both IF and FI 
systems in those cases. For the branchings $\q \to \g \q$ and 
$\g \to \q \qbar$, on the other hand, 
$\d \mathcal{P}_{a\to bc}^{\mathrm{IF}}$ does not obtain any additional 
singularity that could be associated with FI emissions. Here the 
flavour configurations are also separate, see further below, so IF 
and FI anyway have to be considered separately. All possible 
flavour configurations have been studied, see Table~\ref{table:conf}.

\begin{table}[t!] 
\centering
\begin{tabular}{|c|c|c|} 
\hline \multirow{3}{*}{\begin{tabular}{c}
Dipole configuration: \\ initial end $-$ final end \end{tabular}} &
\multirow{3}{*}{Branching $a\to bc$} &
\multirow{3}{*}{\begin{tabular}{c} Emission pattern \\ 
described with: \end{tabular}} 
\\ & & \\ & & \\ \hline
\multirow{6}{*}{$\q_{\mathrm{i}} - \q_{\mathrm{f}}$} & 
\multirow2{*}{$\q_{\mathrm{f}}\to \q \g$} &
\multirow2{*}{$\d \mathcal{P}_{\q \to \q_{\mathrm{i}}\g}^{\mathrm{IF}}$} 
\\ & & \\
\cline{2-3} & \multirow2{*}{$\q \to\q_{\mathrm{i}} \g$} & 
\multirow2{*}{$\d \mathcal{P}_{\q \to  \q_{\mathrm{i}}\g}^{\mathrm{IF}}$} 
\\ & & \\
\cline{2-3} & \multirow2{*}{$\g \to \q_{\mathrm{i}} \qbar$} &
\multirow2{*}{$\d \mathcal{P}_{\g\to \q_{\mathrm{i}}\qbar}^{\mathrm{IF}}$} 
\\ & & \\ \hline 
\multirow{8}{*}{$\g_{\mathrm{i}} - \g_{\mathrm{f}}$} &
\multirow2{*}{$\g_{\mathrm{f}} \to \g \g$} &
\multirow2{*}{$\d \mathcal{P}_{\g\to \g_{\mathrm{i}} \g}^{\mathrm{IF}}$} 
\\ & & \\
\cline{2-3} & \multirow2{*}{$\g \to \g_{\mathrm{i}}\g$} &
\multirow2{*}{$\d \mathcal{P}_{\g \to \g_{\mathrm{i}}\g}^{\mathrm{IF}}$} 
\\ & & \\
\cline{2-3} & \multirow2{*}{$\q \to \g_{\mathrm{i}} \q$}
& \multirow2{*}{$\d \mathcal{P}_{\q \to \g_{\mathrm{i}} \q}^{\mathrm{IF}}$} 
\\ & & \\
\cline{2-3} & \multirow2{*}{$\g_{\mathrm{f}}\to\q \qbar$} &
\multirow2{*}{$\d \mathcal{P}_{\g_{\mathrm{f}} \to \q \qbar}^{\mathrm{FI}}$} 
\\ & & \\ \hline
\multirow{8}{*}{$\q_{\mathrm{i}} - \g_{\mathrm{f}}$} & 
\multirow2{*}{$\g_{\mathrm{f}} \to \g \g$} &
\multirow2{*}{$\d \mathcal{P}_{\q \to \q_{\mathrm{i}}\g}^{\mathrm{IF}}$} 
\\ & & \\
\cline{2-3} & \multirow2{*}{$\q \to\q_{\mathrm{i}} \g$} & 
\multirow2{*}{$\d \mathcal{P}_{\q \to \q_{\mathrm{i}}\g}^{\mathrm{IF}}$} 
\\ & & \\
\cline{2-3} & \multirow2{*}{$\g \to \q_{\mathrm{i}} \qbar$} &
\multirow2{*}{$\d \mathcal{P}_{\g \to \q_{\mathrm{i}}\qbar}^{\mathrm{IF}}$} 
\\ & &
\\ \cline{2-3} & \multirow2{*}{$\g_{\mathrm{f}}\to\q \qbar$} &
\multirow2{*}{$\d \mathcal{P}_{\g_{\mathrm{f}} \to \q \qbar}^{\mathrm{FI}}$} 
\\ & & \\ \hline
\multirow{6}{*}{$\g_{\mathrm{i}} - \q_{\mathrm{f}}$} & 
\multirow2{*}{$\q_{\mathrm{f}} \to \q \g$} &
\multirow2{*}{$\d \mathcal{P}_{\g\to \g_{\mathrm{i}}\g}^{\mathrm{IF}}$} 
\\ & & \\
\cline{2-3} & \multirow2{*}{$\g\to\g_{\mathrm{i}}\g$} &
\multirow2{*}{$\d \mathcal{P}_{\g\to \g_{\mathrm{i}}\g}^{\mathrm{IF}}$} 
\\ & & \\
\cline{2-3} & \multirow2{*}{$\q \to\g_{\mathrm{i}} \q$} & 
\multirow2{*}{$\d \mathcal{P}_{\q \to \g_{\mathrm{i}}\q}^{\mathrm{IF}}$} 
\\ & & \\ \hline
\end{tabular}
\caption{The four configurations of an original FI/IF dipole, with 
all the branchings that can occur for it. The probability of emission 
which is used to describe the branching has been specified.}
\label{table:conf}
\end{table}

The general strategy is to use as much as possible the branching
probabilities of the IF type. Take the example of 
$\q_{\mathrm{i}} - \g_{\mathrm{f}}$.  A gluon emission might either come 
from the ISR $\q \to\q_{\mathrm{i}} \g$ or from the FSR 
$\g_{\mathrm{f}}\to \g \g$. The same final configuration is obtained
in both cases. As for the DIS case, the double-singularity structure 
of $\d \mathcal{P}_{\q \to \q_{\mathrm{i}}\g}^{\mathrm{IF}}$ can be used 
to describe both the ISR and the FSR, with a smooth transition between 
the two. The only problem is a slight mismatch in colour factors
between $\q \to\q_{\mathrm{i}} \g$ and $\g_{\mathrm{f}}\to \g\g$, 
which will be addressed in the next section.

Now instead consider the ISR branching $\g \to\q_{\mathrm{i}}\qbar$ off 
the same original $\q_{\mathrm{i}} - \g_{\mathrm{f}}$ dipole. This leads 
to a final flavour configuration that cannot be obtained by FSR off 
$\g_{\mathrm{f}}$. The emission pattern is then described with 
$\d \mathcal{P}_{\g\to \q_{\mathrm{i}}\qbar}^{\mathrm{IF}}$, which
has only one singularity, as wanted. The converse applies for the FSR
$\g_{\mathrm{f}}\to\q \qbar$, which can only be described by
$\d \mathcal{P}_{\g_{\mathrm{f}}\to \q \qbar}^{\mathrm{FI}}$ since there is 
no ISR which would give an equivalent final configuration. 

In summary we see that the dipole picture works elegantly for the
emission of gluons, but is less elegant when the quark flavour 
content is changed, a well-known observation since long
\cite{Andersson:1989ki}.

\subsection{Some technical aspects}

Some technical issues are addressed in this section. They relate to 
the way the basic ideas are implemented in \textsc{Pythia}. These
aspects are important, but not essential to understand the main
ideas of this article.

\subsubsection{Phase-space cuts}

The kinematics for an IF emission has been derived in 
Section~\ref{subsec:kinematics}. Also the allowed $(\pTse, z)$ 
phase-space region has to be known. Firstly, a lower cutoff
$\pTse > \pTcut^2$ is imposed, where $\pTcut \approx 1$~GeV 
represents a scale where perturbation theory breaks down and 
confinement takes over. (Actually, a smooth damping of perturbative
emissions is used rather than a sharp cutoff.) The range 
$[\zmin,\zmax]$ of allowed $z$ values is obtained from the 
physical condition $\hat{p}_{\perp}^2 > 0$ \cite{Cabouat:2017xx}. To this end 
eq.~(\ref{eq:pTs}) is rewritten in terms of the evolution variable
$\pTse$. For a massless emitted parton ($m_c = 0$) the evolution 
variable is $\pTse = (1 - z) Q^2$ and
\begin{equation}
\hat{p}_{\perp}^2 = 
\pTse \left( 1 - \frac{z}{1 - z} \, \frac{\pTse}{\mred^2} \right) 
- m_f^2 \left( \frac{z}{1 - z} \, \frac{\pTse}{\mred^2} \right)^2.
\end{equation}
with
$\mred^2 = \mdip^2 - m_f^2$. The constraint $\hat{p}_{\perp}^2 > 0$ 
then gives
\begin{equation}
\zmax(\pTse) = \frac{2 + \left(\pTse - \pTe\sqrt{\pTse + 4m_f^2}\right)
/ \mred^2}{2\left(1 + \frac{\pTse}{\mred^2} - \frac{m_f^2}{\mred^2}
\frac{\pTse}{\mred^2}\right)}.
\end{equation}
An overestimate independent of $\pTse$ is required in the veto algorithm 
used for the downwards evolution in $\pTse$ \cite{Sjostrand:2006za}. 
Since $\zmax$ is strictly decreasing with $\pTse$ the value at 
$\pTcut^2$ can be used to this end. The lower limit comes from
$x_a = x_b / z \leq 1$, which implies $z \geq x_b$.

The range previously found is valid for a massless emitted parton. 
The case where $m_c\neq 0$ occurs e.g.\ for 
$\g \to \mathrm{Q}\overline{\mathrm{Q}}$, with 
$\mathrm{Q} = \mathrm{c},\b$. The procedure is the same as above, 
but now $\pTse = (1 - z)(Q^2 + m_c^2)$ and 
\begin{equation}
\hat{p}_{\perp}^2 
= (\pTse - m_c^2) \left(1 - \frac{z}{1 - z}\, \frac{\pTse}{\mred^2} \right) 
- m_f^2 \left(\frac{z}{1 - z}\,\frac{\pTse}{\mred^2}\right)^2,
\end{equation} 
which gives 
\begin{equation}
\zmax(\pTse) = \frac{2(\pTse - m_c^2) + \frac{\pTse}{\mred^2}
\left(\pTse - m_c^2 - \sqrt{(\pTse - m_c^2)^2 + 4m_f^2(\pTse - m_c^2)}
\right)}{2\left(\pTse - m_c^2 + \frac{\pTse(\pTse - m_c^2)}{\mred^2} 
- m_f^2 \frac{\pTe^4}{\mred^4}\right)}.
\end{equation}
This expression is rather cumbersome. If the colour partner 
is a gluon or a light quark, $m_f = 0$, however, it simplifies to
\begin{equation}
\zmax(\pTse) = \frac{2(\pTse - m_c^2)}{2\left(\pTse - m_c^2 
+ \frac{\pTse(\pTse - m_c^2)}{\mred^2}\right)}
 = \frac{\mred^2}{\mred^2 + \pTse}.
\end{equation}
The $\zmax$ function is strictly decreasing in that specific case,  
and can be overestimated by $\tilde{z}_{\mathrm{max}} = \zmax(m_c^2)$, 
since the evolution is such that $\pTse\geq m_c^2$ in the massive case.

For $m_f\neq 0$ the function $\zmax$ is not strictly decreasing anymore. 
It is bounded from above by the function for the $m_f = 0$ case, however.
Therefore the overestimate
$\tilde{z}_{\mathrm{max}} = \mred^2/(\mred^2 + m_c^2)$
can be used also for $m_f > 0$. The lower limit remains
$\tilde{z}_{\mathrm{min}} = x_b$.

\subsubsection{Colour factors}

When a dipole is stretched between a quark and a gluon the two radiate
with different colour factors, $C_F = 4/3$ for the former and 
$C_A/ 2 = 3/2$ for the latter, where the $1/2$ for the gluon comes from
its radiation being split between two dipoles. More precisely, one can 
write the $\g \to \g\g$ splitting kernel as 
\cite{Gustafson:1987rq}:
\begin{equation}
P_{\g \to \g\g}(z)  = C_A\,\frac{(1-z(1-z))^2}{z(1-z)}
 = \frac{C_A}{2}\left( \frac{1+z^3}{1-z} + \frac{1+(1-z)^3}{z} \right)
= C_A\,\frac{1+z^3}{1-z} ,
\end{equation}
where the last equality is by relabelling symmetry of the two gluons.
In the dipole approach, the differences between a $\q \to \q\g$ and a
$\g \to \g\g$ branching thus are the colour factors, $C_F$ vs.\ $C_A/2$, 
and the numerators of the splitting kernels, \mbox{$1 + z^2$} vs.\ $1 + z^3$. 
In the soft-gluon limit, $z \to 1$, only the former difference survives.

For a $\q_{\mathrm{i}} - \g_{\mathrm{f}}$ dipole, a description purely in terms
of IF radiation therefore will underestimate the $\g_{\mathrm{f}}\to \g\g$
rate by a factor $2 C_F / C_A = 8 / 9$. The idea is to find a compensating
smooth weight, which is unity for a gluon emission off $\q_{\mathrm{i}}$ and
$C_A/(2C_F) = 9/8$ for one off $\g_{\mathrm{f}}$. Using $1/m^2$ as a 
measure of proximity, we have chosen the weight
\begin{equation} 
w_{\q_{\mathrm{i}} - \g_{\mathrm{f}}} = \frac{m_{fc}^2
+ \frac{C_A}{2C_F}\,m_{ac}^2}{m_{fc}^2 + m_{ac}^2}.
\end{equation}
With this choice, $w_{\q_{\mathrm{i}} - \g_{\mathrm{f}}}\to 1$ 
for $m_{ac}^2\to 0$ (emission from $\q_{\mathrm{i}}$) and
$w_{\q_{\mathrm{i}} - \g_{\mathrm{f}}}\to C_A/(2C_F)$ 
for $m_{fc}^2\to 0$ (emission from $\g_{\mathrm{f}}$). 

Let us now see in more detail how the IF branching probability $w_{\q_{\mathrm{i}} - \g_{\mathrm{f}}}\,\d \mathcal{P}_{\q\to \q_{\mathrm{i}}\g}^{\mathrm{IF}}$ leads to the gluon-radiation pattern of the full $\q_{\mathrm{i}} - \g_{\mathrm{f}}$ dipole on its own. The IF kinematics (\ref{IFkin}) leads to the following Jacobian
\begin{equation}
\d Q^2\,\d z = \frac{\mdip^2}{(\mdip^2+m_{fc}^2)^2}\d m_{fc}^2\,\d m_{ac}^2.
\end{equation}
\noindent Therefore, the branching probability can be written as
\begin{equation}
\begin{split}
w_{\q_{\mathrm{i}} - \g_{\mathrm{f}}}\,\d \mathcal{P}_{\q \to \q_\mathrm{i} \g}^{\mathrm{IF}}&=w_{\q_{\mathrm{i}} - \g_{\mathrm{f}}}\,C_F \, \frac{\as}{2\pi} \, \frac{\d Q^2}{Q^2} \,\frac{1 + z^2}{1 - z} \, \d z \\
&= \frac{m_{fc}^2 + \frac{C_A}{2C_F}\,m_{ac}^2}{m_{fc}^2 + m_{ac}^2}\,C_F \, \frac{\as}{2\pi}\,\frac{\d m_{ac}^2}{m_{ac}^2}\,\frac{\d m_{fc}^2}{m_{fc}^2}
\left(1+\left(\frac{\mdip^2}{\mdip^2+m_{fc}^2}\right)^2\right)\,\frac{\mdip^2}{\mdip^2+m_{fc}^2}.
\end{split} 
\label{wdP}
\end{equation}
\noindent When $m_{ac}^2 \to 0$, the emission can be associated to $\q_\mathrm{i}$ and one gets
\begin{equation}
w_{\q_{\mathrm{i}} - \g_{\mathrm{f}}}\,\d \mathcal{P}_{\q \to \q_\mathrm{i} \g}^{\mathrm{IF}} \sim C_F \, \frac{\as}{2\pi}\,\frac{\d m_{ac}^2}{m_{ac}^2}\,\frac{\d m_{fc}^2}{m_{fc}^2}
\left(1+\left(\frac{\mdip^2}{\mdip^2+m_{fc}^2}\right)^2\right)\,\frac{\mdip^2}{\mdip^2+m_{fc}^2},
\end{equation}
\noindent which leads to the right colour factor $C_F$ with the right singularity structure. For $m_{fc}^2 \to 0$, the emission is seen as coming from $\g_\mathrm{f}$ and equation (\ref{wdP}) gives
\begin{equation}
w_{\q_{\mathrm{i}} - \g_{\mathrm{f}}}\,\d \mathcal{P}_{\q \to \q_\mathrm{i} \g}^{\mathrm{IF}}\sim\frac{C_A}{2}\,\frac{\as}{2\pi}\,\frac{\d m_{ac}^2}{m_{ac}^2}\,\frac{\d m_{fc}^2}{m_{fc}^2}\times 2.
\label{LimIF}
\end{equation}
Let us now compare with the expected behaviour of the probability in this region of phase-space i.e. $\d \mathcal{P}_{\g_\mathrm{f} \to \g \g}^{\mathrm{FI}}$, defined by
\begin{equation}
\d \mathcal{P}_{\g_\mathrm{f} \to \g \g}^{\mathrm{FI}}= \frac{C_A}{2}\,\frac{\as}{2\pi} \, \frac{\d Q^2}{Q^2} \,\frac{1 + z^3}{1 - z} \, \d z.
\end{equation}
The FI kinematics (\ref{FIkin}) leads to
\begin{equation}
\d Q^2\,\d z = \frac{\mdip^2-m_{bc}^2}{(\mdip^2+m_{bc}^2)^2}\d m_{bc}^2\,\d m_{rc}^2,
\end{equation}
and
\begin{equation}
\d \mathcal{P}_{\g_\mathrm{f} \to \g \g}^{\mathrm{FI}}=\frac{C_A}{2}\,\frac{\as}{2\pi}\,\frac{\d m_{bc}^2\,\d m_{rc}^2}{m_{bc}^2}\,
\frac{(1 + z^3)(\mdip^2-m_{bc}^2)}{m_{bc}^2(\mdip^2+m_{bc}^2)+m_{rc}^2(\mdip^2-m_{bc}^2)}
\end{equation}
\noindent The limit $m_{fc}^2 \to 0$ in the IF case corresponds to $m_{bc}^2 \to 0$ in the FI case. In this limit,
\begin{equation}
\d \mathcal{P}_{\g_\mathrm{f} \to \g \g}^{\mathrm{FI}}\sim \frac{C_A}{2}\,\frac{\as}{2\pi}\,\frac{\d m_{bc}^2}{m_{bc}^2}\,\frac{\d m_{rc}^2}{m_{rc}^2}\,\left(1+\left(1-\frac{m_{rc}^2}{\mdip^2}\right)^3\right).
\label{LimFI}
\end{equation}
It clearly appears that the singularity structure of equation (\ref{LimFI}) is reproduced by the singularities present in equation (\ref{LimIF}), as desired. Moreover, the weight defined previously ensures that the probability defined in equation (\ref{LimIF}) comes with the right colour factor $C_A/2$. The extra non singular term $\left(1+\left(1-m_{rc}^2/\mdip^2\right)^3\right)$ in equation (\ref{LimFI}) actually approaches the value 2 when $m_{rc}^2 \to 0$, as in equation (\ref{LimIF}). This shows that gluon emissions off a $\q_{\mathrm{i}} - \g_{\mathrm{f}}$ dipole can be fully described by the probability $w_{\q_{\mathrm{i}} - \g_{\mathrm{f}}}\,\d \mathcal{P}_{\q \to \q_\mathrm{i} \g}^{\mathrm{IF}}$ only, without any double counting.

In terms of the usual variables, $\pTse = (1 - z)Q^2$ and $z$, one obtains
\begin{equation} 
m_{fc}^2 = \mdip^2\frac{1 - z}{z}, 
\hspace{50pt} m_{ac}^2 = Q^2 = \frac{\pTse}{1 - z},
\end{equation}
which leads to the weight
\begin{equation} 
w_{\q_{\mathrm{i}} - \g_{\mathrm{f}}} 
= \frac{\mdip^2(1 - z)^2 + \frac{C_A}{2C_F}\,z\,\pTse}
{\mdip^2(1 - z)^2 + z\,\pTse}.
\label{weightqi}
\end{equation} 
The same procedure can be applied for the configuration
$\g_{\mathrm{i}}$ - $\q_{\mathrm{f}}$ with the ISR $\g\to\g_{\mathrm{i}} \g$ 
and the FSR $\q_{\mathrm{f}}\to \q\g$. The weight here is
\begin{equation} 
w_{\g_{\mathrm{i}} - \q_{\mathrm{f}}} = \frac{\mdip^2(1 - z)^2
+ \frac{2C_F}{C_A}\,z\,\pTse}{\mdip^2(1 - z)^2 + z\,\pTse}.
\end{equation}
The two other dipole configurations ($\q_{\mathrm{i}} - \q_{\mathrm{f}}$ and $\g_{\mathrm{i}} - \g_{\mathrm{f}}$) do not need any correction since the two dipole ends there have the same flavour. Indeed, it has been shown that for a $\q_{\mathrm{i}} - \q_{\mathrm{f}}$ dipole, the first-order matrix element is explicitly reproduced by $\d \mathcal{P}_{\q \to \q \g}^{\mathrm{IF}}$ in the case of DIS. Therefore, the collinear limit and soft-gluon limit are reproduced without any double counting. Since this behaviour is universal, this feature can be easily generalized to other processes than DIS. The case of the $\g_{\mathrm{i}} - \g_{\mathrm{f}}$ dipole is completely similar to the $\q_{\mathrm{i}} - \q_{\mathrm{f}}$ one, only with a different colour factor.

\subsubsection{Gluon polarization}

\begin{figure}[t!] 
\centering
\includegraphics[width=0.3\textwidth,angle=-90]{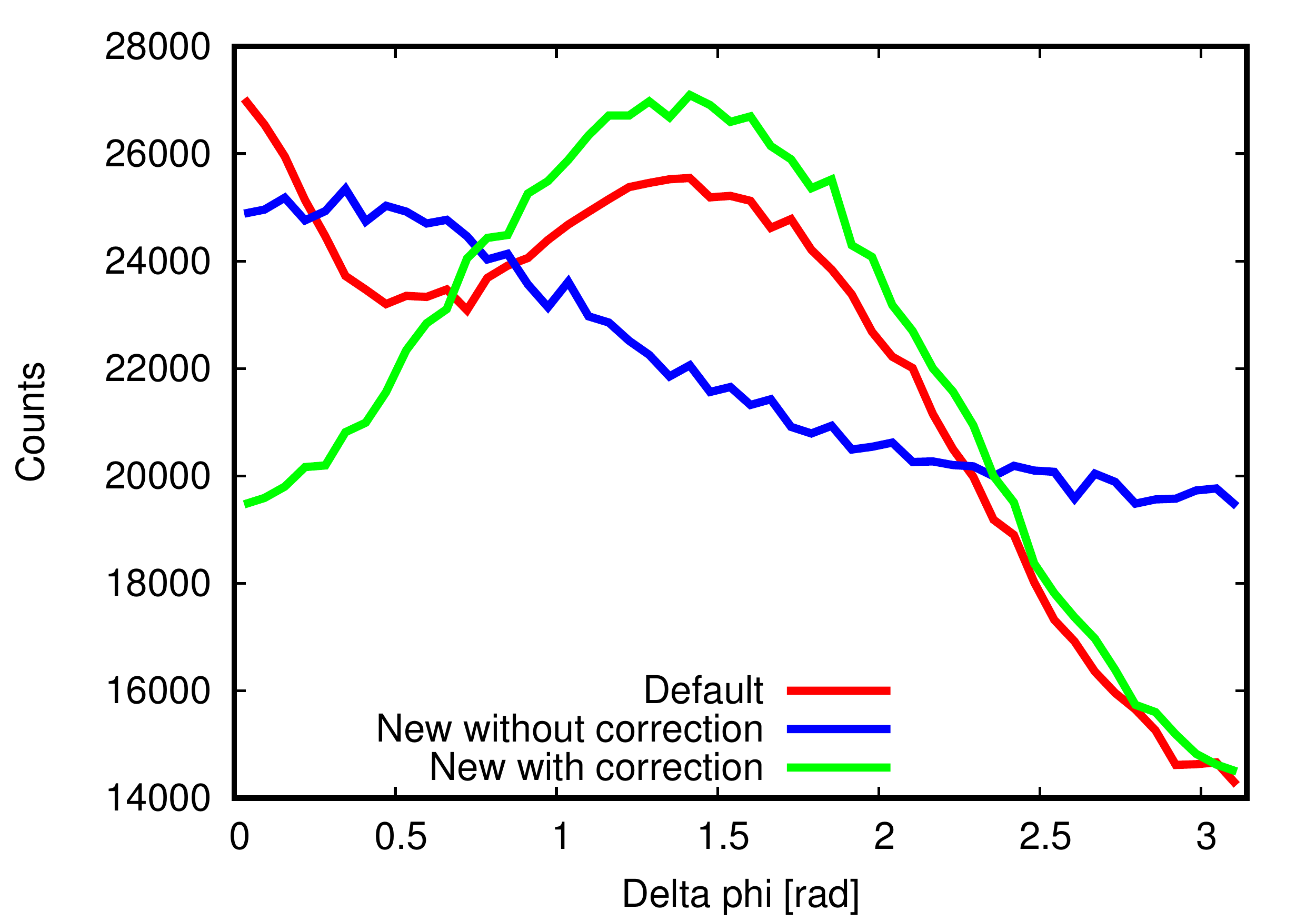}
\includegraphics[width=0.3\textwidth,angle=-90]{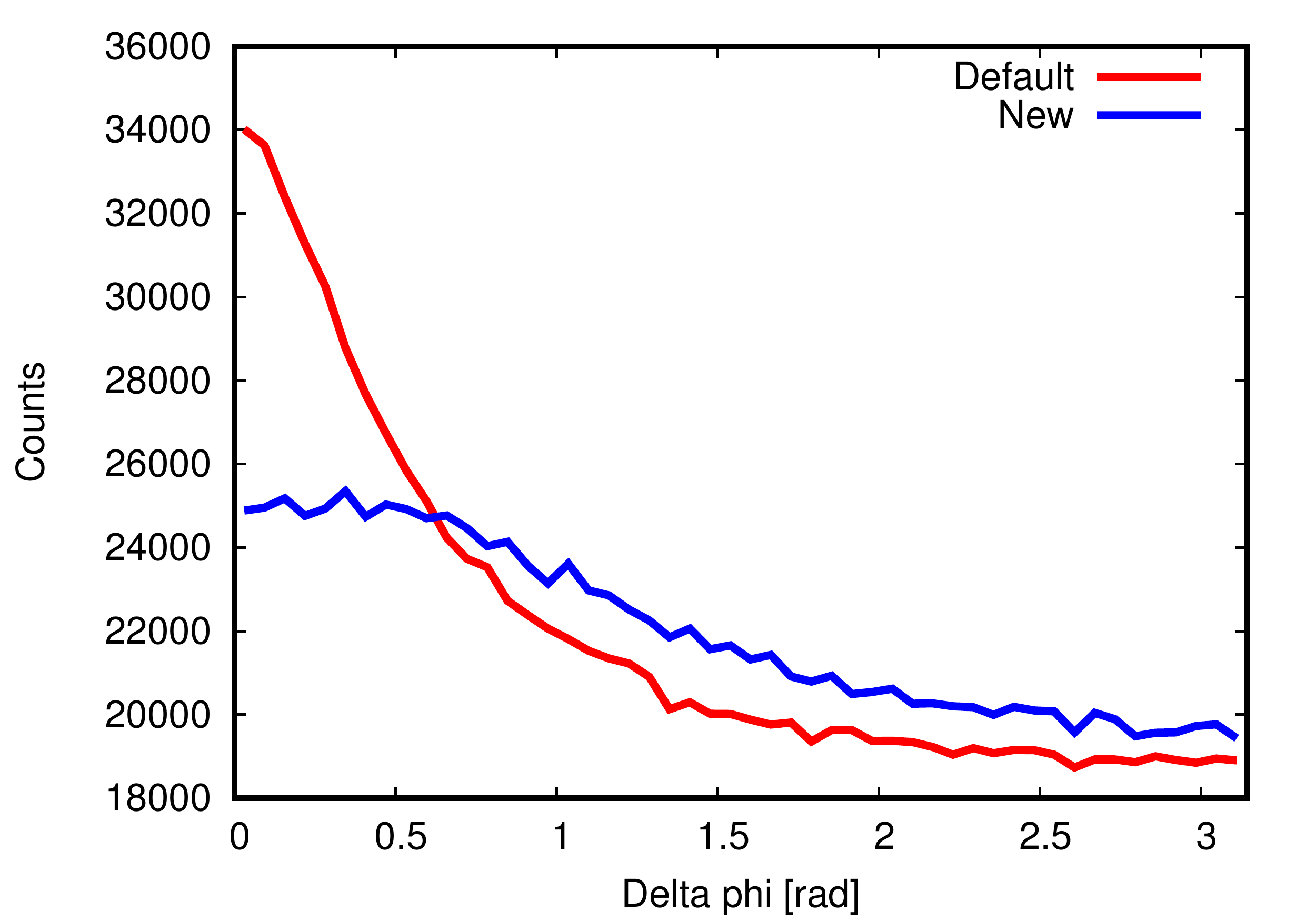}\\
\vspace{5pt} 
(a) \hspace{200pt} (b)
\caption{Histograms of the $\Delta\varphi$ variable, as defined in 
the text, for $\gammaZ$ production. In (a), the red curve is for the old 
global-recoil scheme whereas blue and green are for the new dipole
scheme with or without gluon polarization effects included. In (b), 
the gluon polarization effects are removed so only the colour-coherence 
azimuthal asymmetries remain. Note the suppressed zero on the vertical
axis.}
\label{Fig:Dphi}
\end{figure}

The global-recoil shower implements two sources of azimuthal 
asymmetries: colour coherence and gluon plane polarization.
The former is automatically included in the dipole formulation.
That is, radiation off a $\{b + f\}$ dipole is assumed isotropic
in azimuth, but after a boost to the event rest frame the radiation
is biased in the azimuthal direction of $f$, even the one that 
would be thought of as ISR off the $b$.

The gluon polarization has to be considered separately, however.
It has the effect of correlating the production and decay planes
of a gluon. To be more specific, assume that parton $b$ is a gluon,
produced by $a \to b + c$, and branching by $b \to g + h$.
In a frame where $b$ is aligned along the $z$ axis the angle
$\Delta\varphi = \varphi_c - \varphi_g$ should follow a distribution
\cite{Webber:1986mc}
\begin{equation}
\frac{\d \mathcal{P}_{\varphi}}{\d \varphi} \propto
1 + c_{\mathrm{pol}}\cos(2\Delta\varphi),
\end{equation}
where $c_{\mathrm{pol}}$ depends on flavours and kinematics at 
the production and decay vertices of the gluon. (Note that
$\varphi_a  = \varphi_c$ and that $\varphi_h = \varphi_g + \pi$  
gives the same $\cos(2\Delta\varphi)$ as $\varphi_g$.)  

There is some ambiguity which frame to use when $b$ is set along the 
$z$ axis. The natural choice, and the one we have used, is the 
$\{b + d\}$ rest frame, where $b$ and the other-side incoming 
parton $d$ are along the $\pm z$ axis. The disadvantage is that it 
may partly counteract the colour-coherence azimuthal asymmetry,
induced by the boost from the $\{b + f\}$ rest frame. 
This problem would have been solved had the latter frame been used, 
where only gluon polarization gives azimuthal anisotropies. 
That frame does not have any obvious relation with the $b \to g + h$ 
decay, on the other hand, so would also be imperfect.

\begin{figure}[t!] \centering
\includegraphics[width=0.45\textwidth]{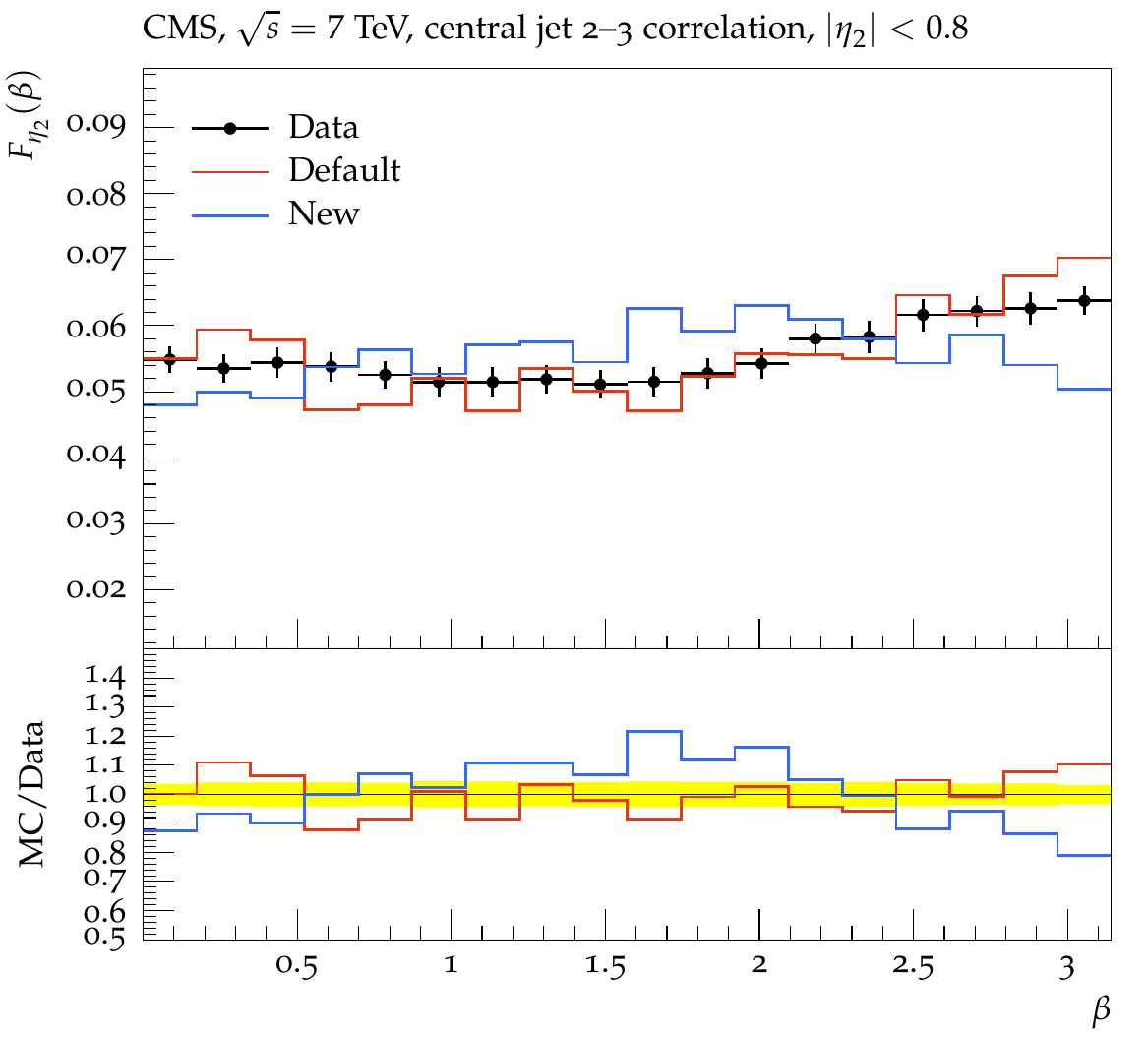}
\includegraphics[width=0.45\textwidth]{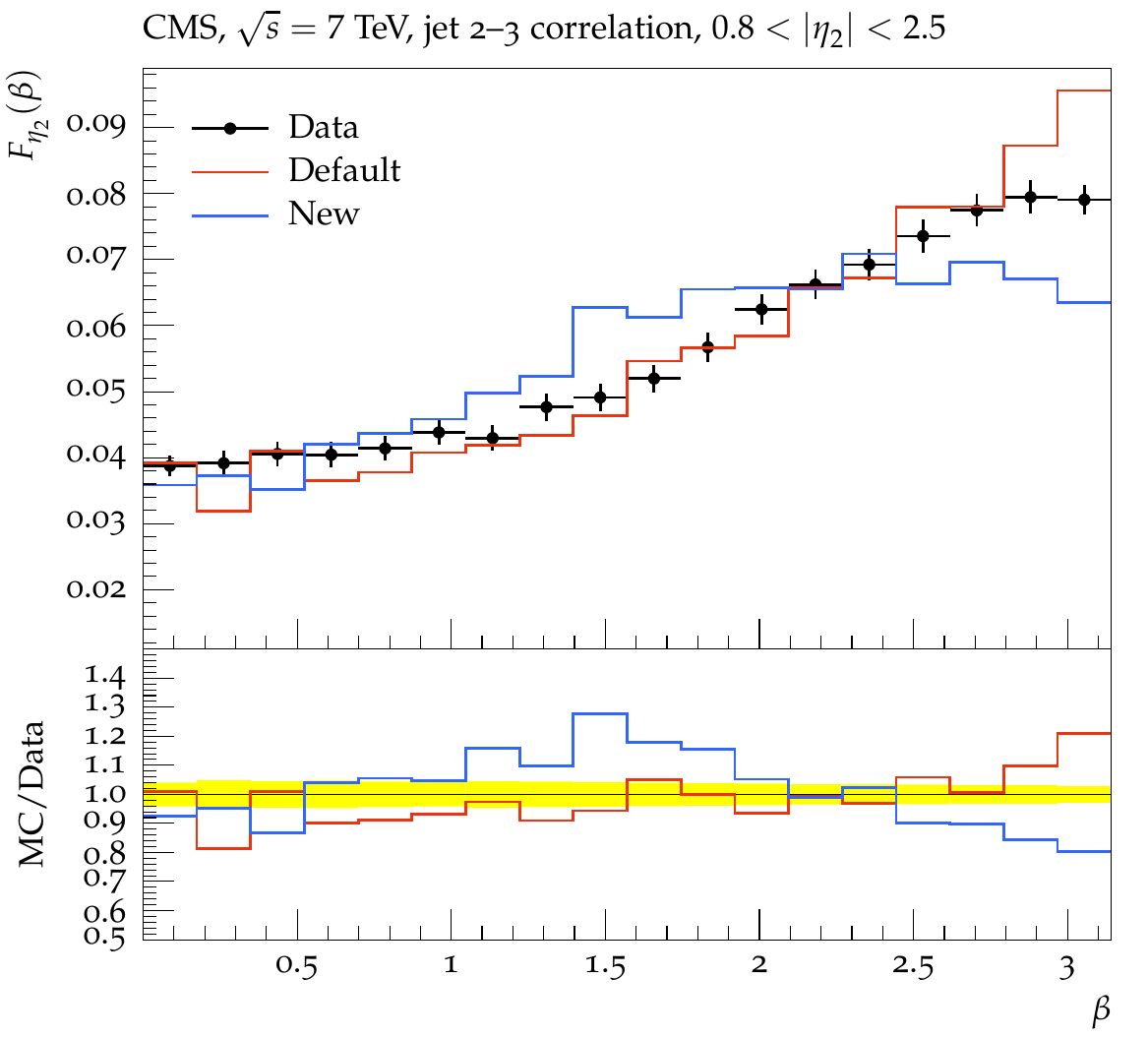}\\
\vspace{5pt} 
(a) \hspace{200pt} (b)
\caption{Angular correlation between the second and third jets 
measured by CMS for pp collisions at 7 TeV 
\cite{Buckley:2010ar,Chatrchyan:2013fha}. The new and default 
procedures are compared, with all azimuthal asymmetries included. 
(Results are for QCD $2 \to 2$ events, generated with $\pThat > 60$~GeV, 
that survive the experimental selection.)}
\label{Fig:ColPhi}
\end{figure}

Results are given in Fig.~\ref{Fig:Dphi}(a) for the hard process  
$\q + \qbar \to \gammaZ$ at 7 TeV. FSR, MPI and hadronization are
turned off. The new procedure without polarization correction 
moderately favours small $\Delta\varphi$, whereas the polarization 
effects favour $\Delta\varphi \sim \pi/2$. Overall the latter curve
is closer to the old default one, which is known to describe azimuthal
asymmetries decently \cite{Buckley:2011ms}. A significant difference 
arises for $\Delta\varphi = 0$, however. This is due to the other
source of azimuthal asymmetries, colour coherence, as shown in 
Fig.~\ref{Fig:Dphi}(b), where gluon polarization effects have been 
switched off. The colour-coherence azimuthal distribution implemented 
in the default scheme clearly gives a stronger contribution for 
$\Delta\varphi = 0$ than the ones automatically generated by the 
new scheme by the boost to the $\{b + d\}$ rest frame. 

In Fig.~\ref{Fig:ColPhi}, the difference between the two schemes
is clearly visible in comparisons with data. The $\beta$ angle 
measures the distribution of the third jet around the second in
three-jet QCD events, and is very sensitive to colour coherence
effects \cite{Chatrchyan:2013fha}. The default procedure here gives
a reasonably good description, while the new approach fares visibly
worse. This is a surprising and unfortunate result, since the dipole 
approach ought to give the best description of colour-coherence 
azimuthal asymmetries. We have already seen that the gluon polarization 
effects can counteract the colour coherence ones, but for these 
distributions such effects appear to be small and do not offer an 
explanation. Therefore further studies will be necessary to understand 
this issue.

\section{Comparisons with data and with other approaches}

\subsection{Gauge boson production}

\begin{figure}[t!] \centering
\includegraphics[width=0.3\textwidth,angle=-90]{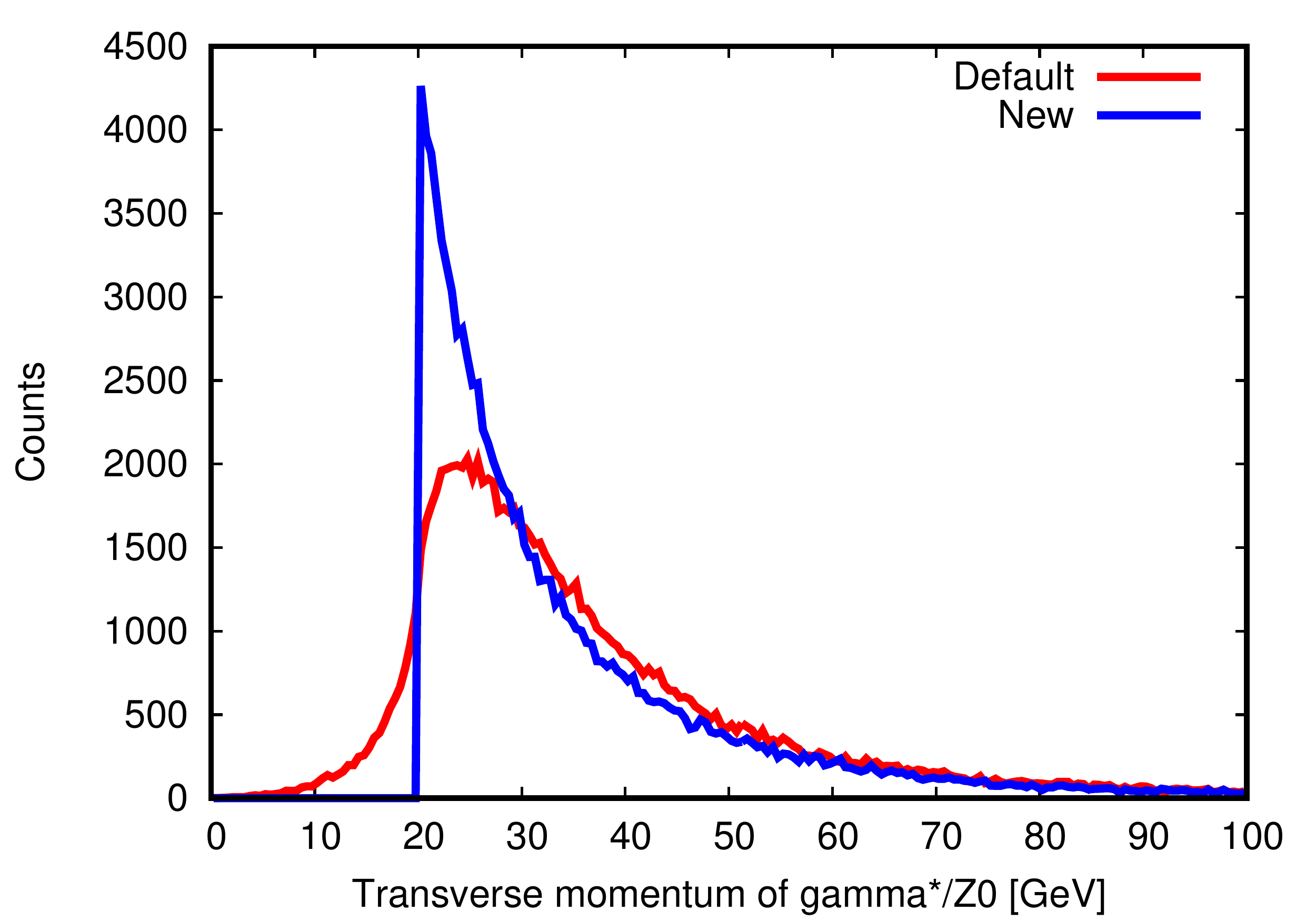}
\includegraphics[width=0.3\textwidth,angle=-90]{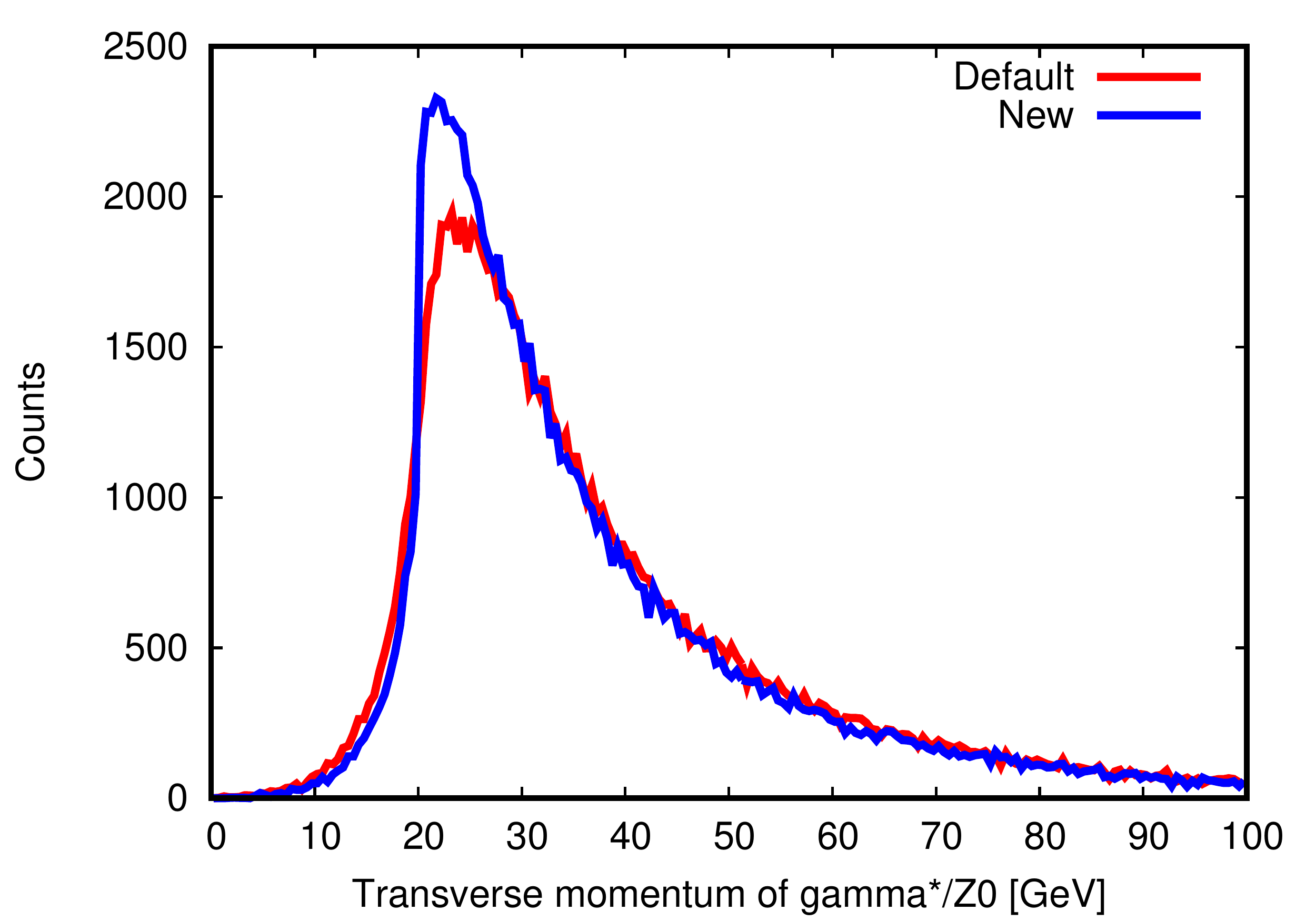}\\
\vspace{5pt}
(a) \hspace{200pt} (b)
\caption{Transverse momentum of $\gammaZ$ at 7~TeV LHC, with 
$\pThat > 20$~GeV in the $2 \to 2$ process: 
(a) for $\q + \qbar \to \gammaZ + \g$, 
(b) for $\q + \g \to \gammaZ + \q$. The new dipole approach is compared
with the old default one. $\pT$ shifts due to primordial $k_{\perp}$ are 
not included here for simplicity.}
\label{Fig:pTZ}
\end{figure}

\begin{figure}[t!] \centering
\includegraphics[width=0.45\textwidth]{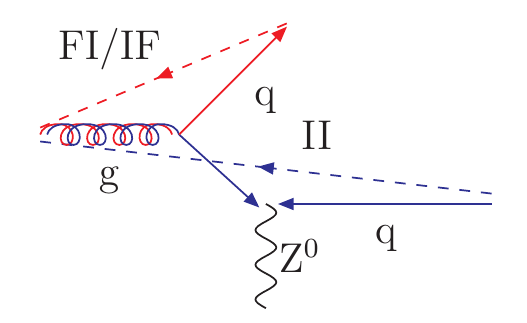}
\caption{Colour flow for the process $\q + \g \to \gammaZ + \q$. 
The dashed lines represent the colour lines stretched between
the dipole ends.}
\label{Fig:Zq}
\end{figure}

The process $\q + \qbar \to \gammaZ + \g$ allows a clean comparison
between the dipole approach and the default global-recoil
procedure. Indeed, the emission of a gluon off the $\q + \qbar$ dipole
leads to the formation of two FI/IF dipoles, as shown in
Fig.~\ref{FigZg}. Therefore, with the new scheme, the $\pT$ of a 
$\gammaZ$ is fixed by the hard $2 \to 2$ process and is not altered 
by further emissions. The lower $\pT$ limit is then set by the choice 
of phase-space cuts. In contrast, with the global-recoil procedure, 
the $\gammaZ$ $\pT$ can be increased, but also reduced, in consecutive 
branchings. Some typical results are given in Fig.~\ref{Fig:pTZ}(a). 

In Fig.~\ref{Fig:pTZ}(b), the $\gammaZ$ $\pT$ spectrum is also
given for the process $\q + \g \to \gammaZ + \q$. This process is
interesting because it leads to the formation of one FI/IF dipole
$\mathrm{g}_{\mathrm{i}}-\mathrm{q}_{\mathrm{f}}$ and one II dipole
$\mathrm{g}_{\mathrm{i}}-\mathrm{q}_{\mathrm{i}}$, as illustrated in
Fig.~\ref{Fig:Zq}. Therefore, an emission off
$\mathrm{g}_{\mathrm{i}}$ can be described either in the IF framework
or in the II picture involving global recoils. In the first case,
the $\gammaZ$ will not get any recoil when the new scheme is used, 
but in the second it will. This is illustrated in Fig.~\ref{Fig:pTZ}(b),
where the new scheme now closer agrees with the older one.

\begin{figure}[t!] \centering
\includegraphics[width=0.45\textwidth]{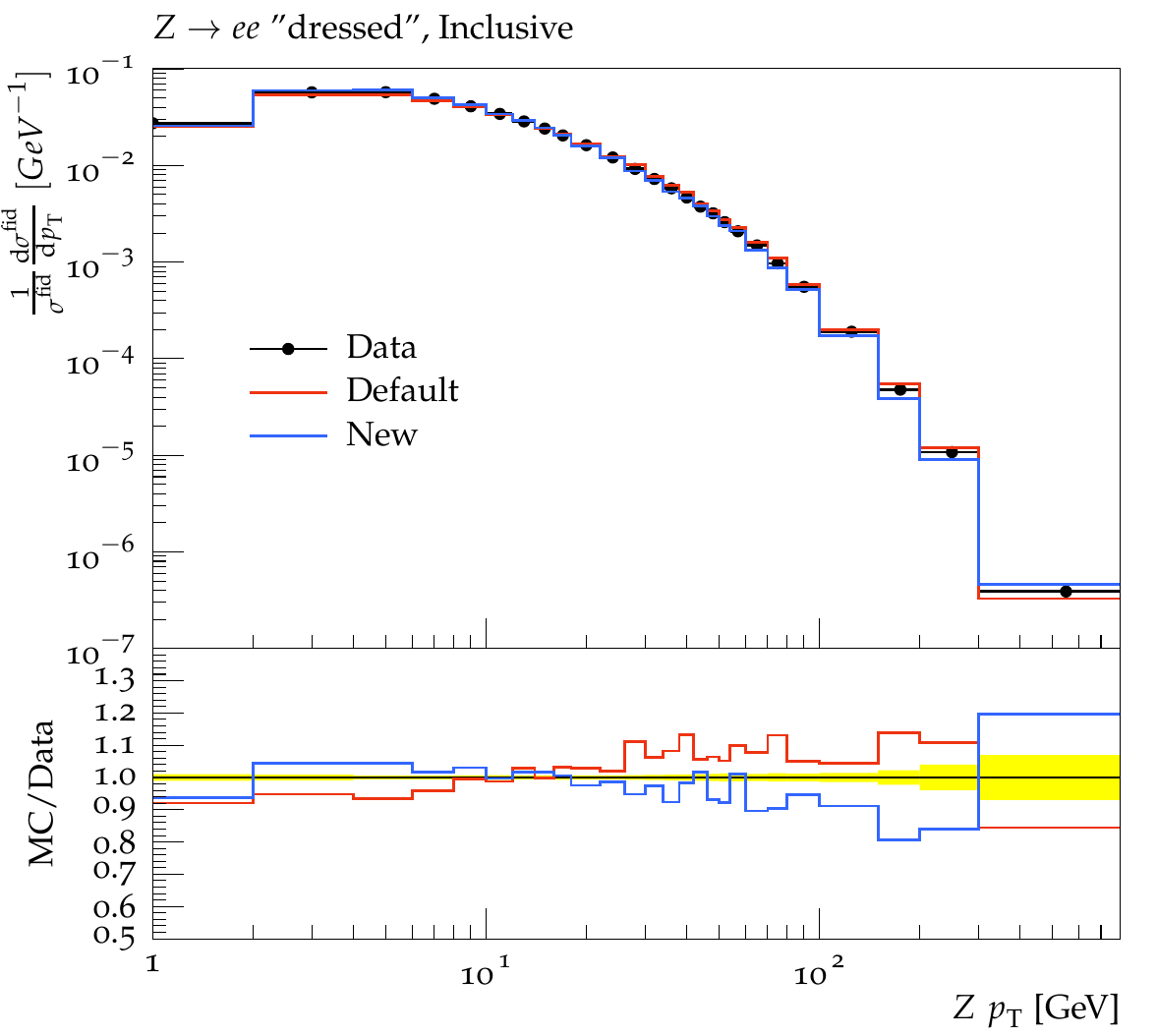}
\includegraphics[width=0.45\textwidth]{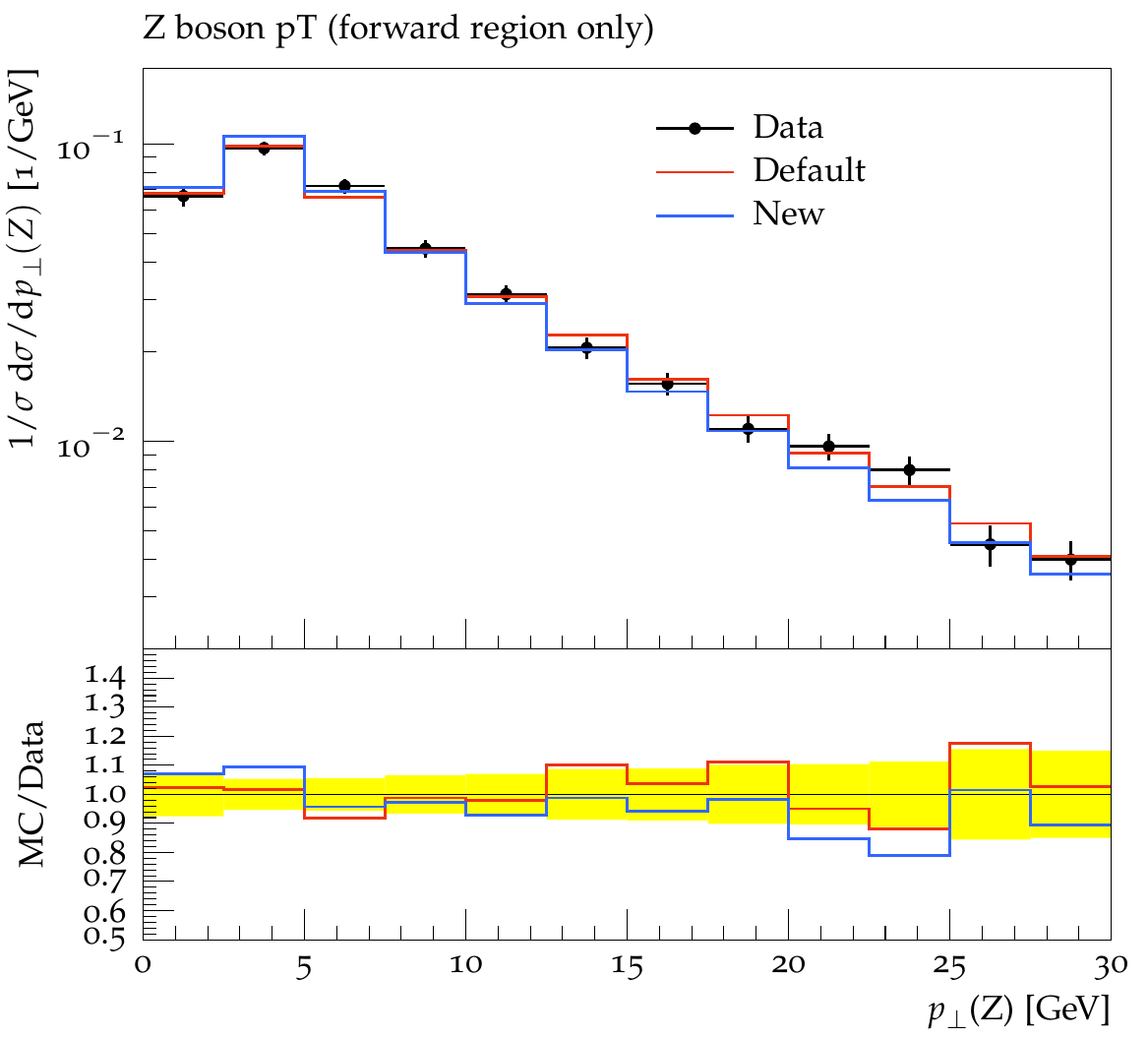} \\
\vspace{5pt}
(a) \hspace{200pt} (b)
\caption{Comparison between the new and old schemes for the $\gammaZ$ 
$\pT$ spectrum measured by (a) ATLAS for pp collisions at 7~TeV 
\cite{Buckley:2010ar, Aad:2014xaa}, (b) D0 for p$\overline{\mathrm{p}}$
collisions at 1.96~TeV \cite{Buckley:2010ar,Abazov:2007ac}.}
\label{Fig:pTZexp}
\end{figure}

The inclusive $2 \to 1$ $\gammaZ$ production process, followed by 
showers, can be compared with experimental data. Results are shown in 
Fig.~\ref{Fig:pTZexp} for the $\gammaZ$ $\pT$, compared with 
ATLAS \cite{Aad:2014xaa} and D0 \cite{Abazov:2007ac} data. 
The two shower procedures are here seen to lead to similar results,
but a tendency can be noted that the new scheme gives a spectrum
slightly shifted towards lower $\pT$ values, as could have been expected.

\begin{figure}[p] \centering
\includegraphics[width=0.44\textwidth]{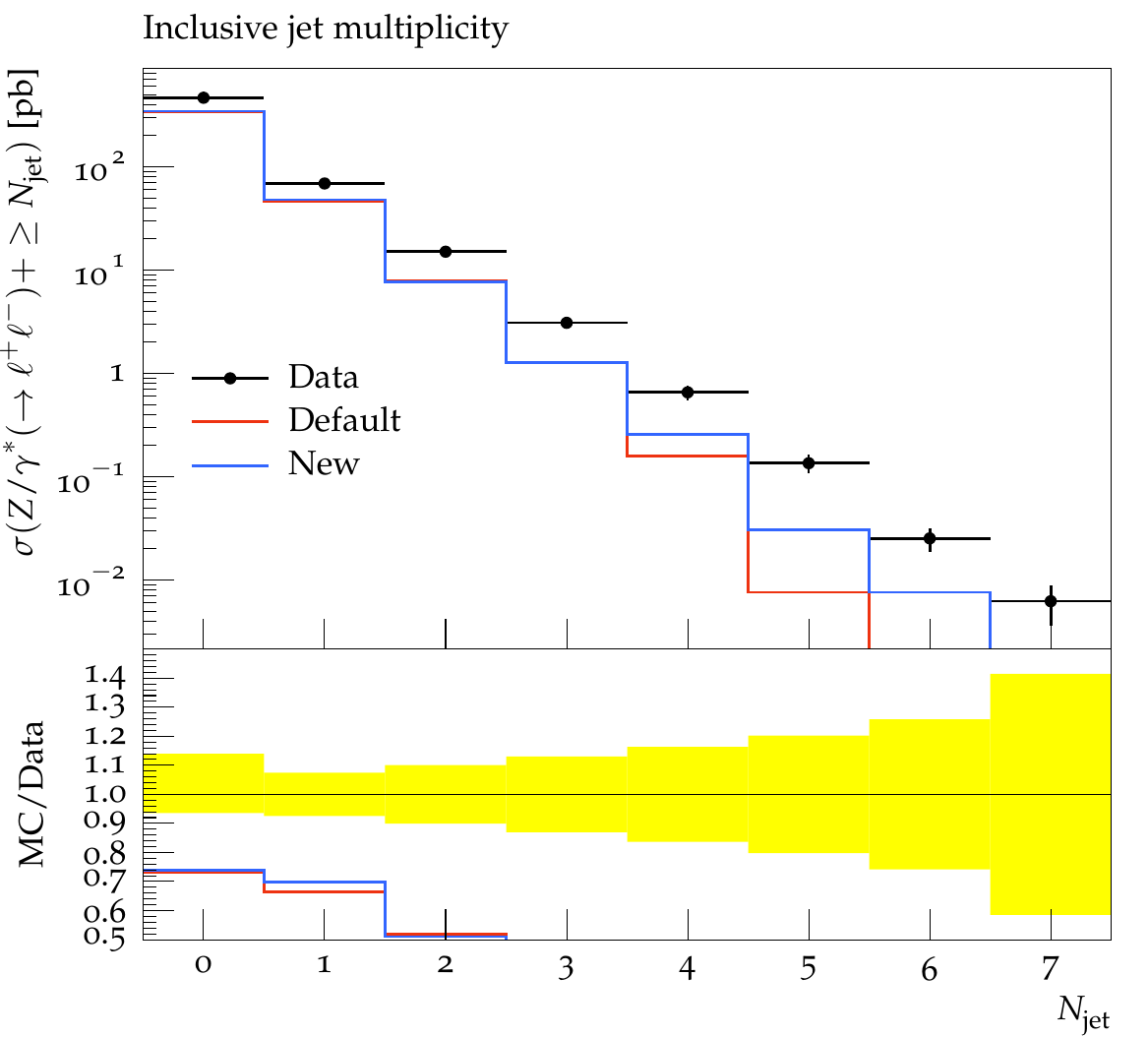}
\includegraphics[width=0.44\textwidth]{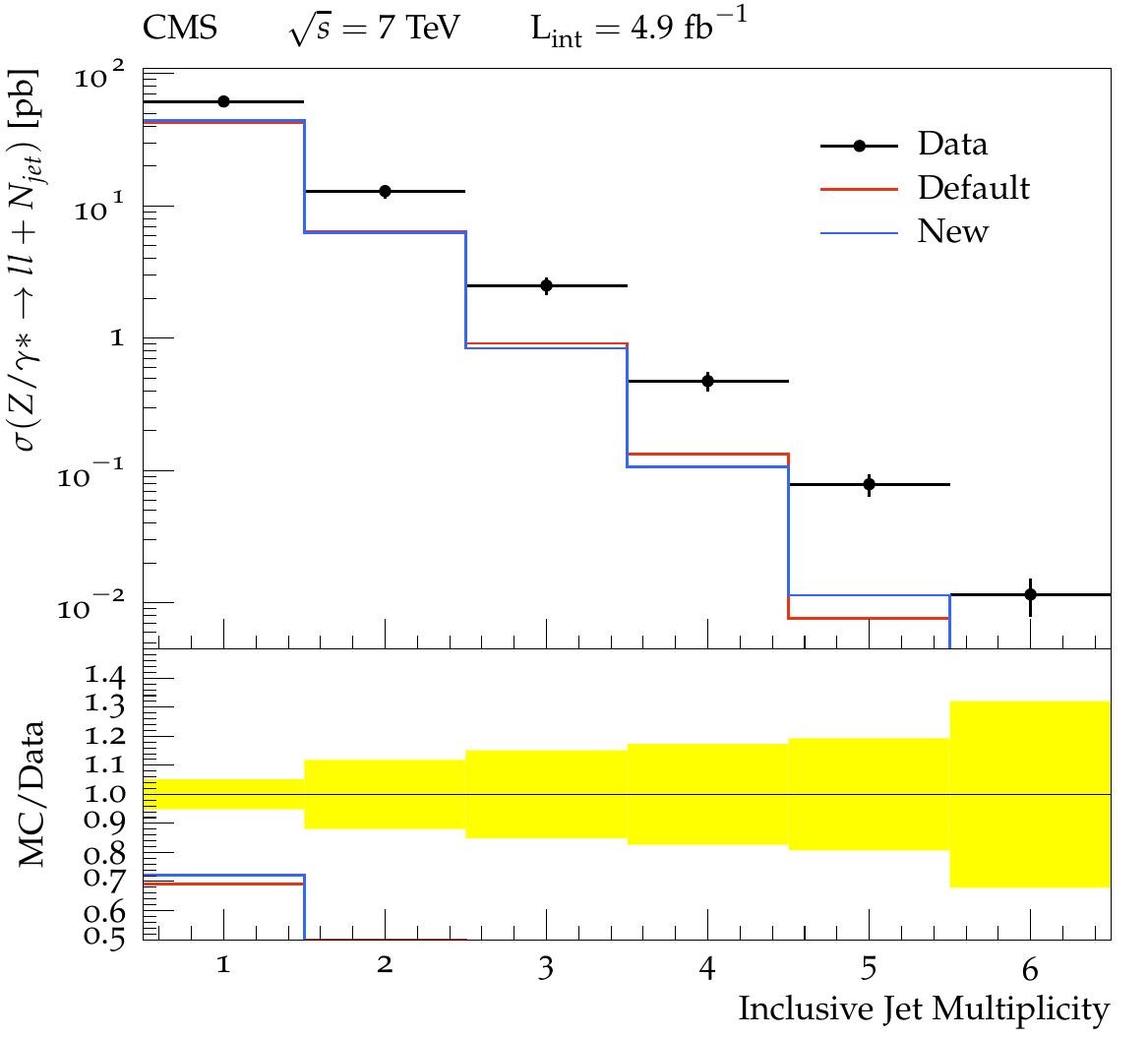}\\[2pt]
(a) \hspace{200pt} (b) \\
\includegraphics[width=0.44\textwidth]{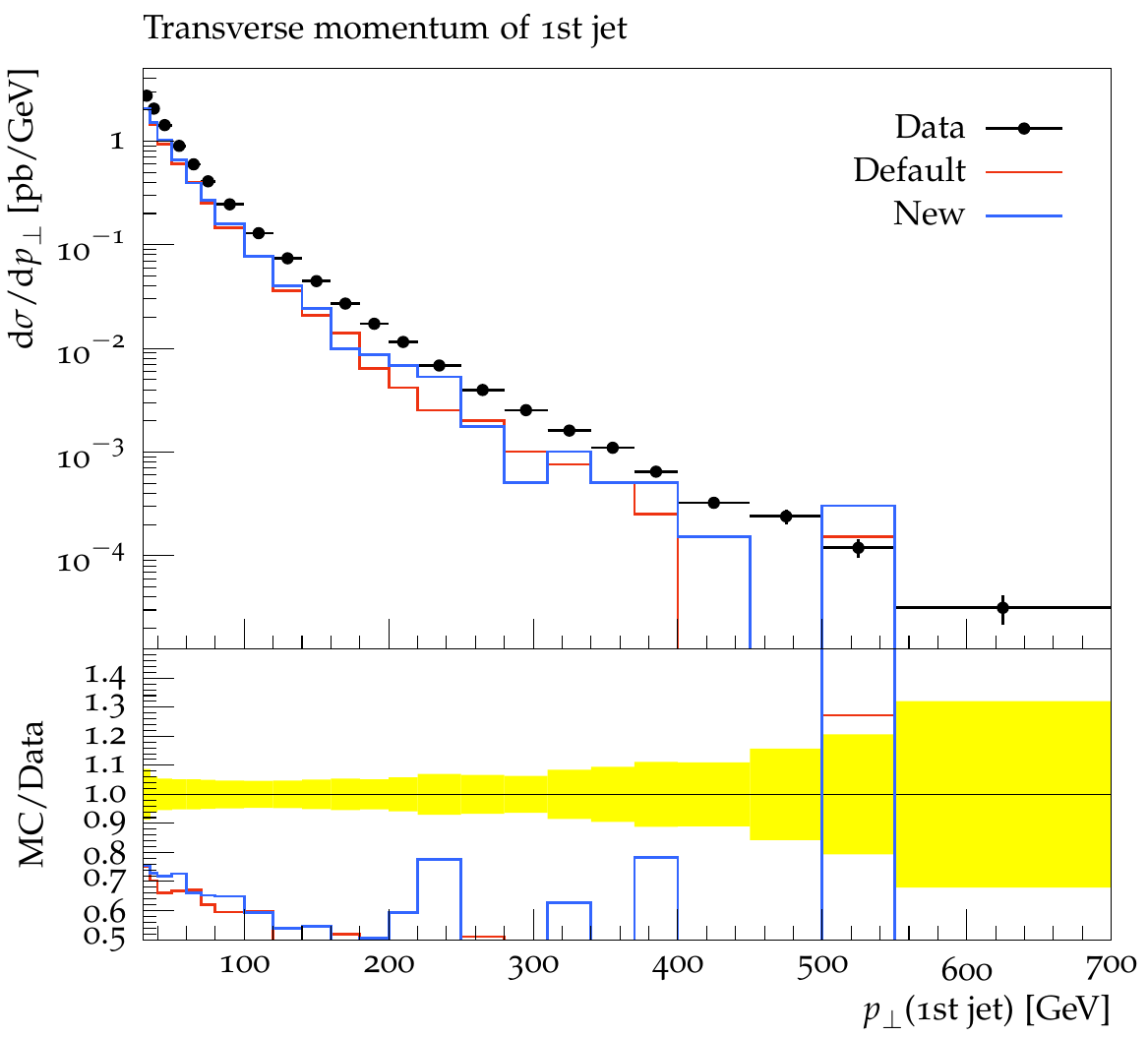}
\includegraphics[width=0.44\textwidth]{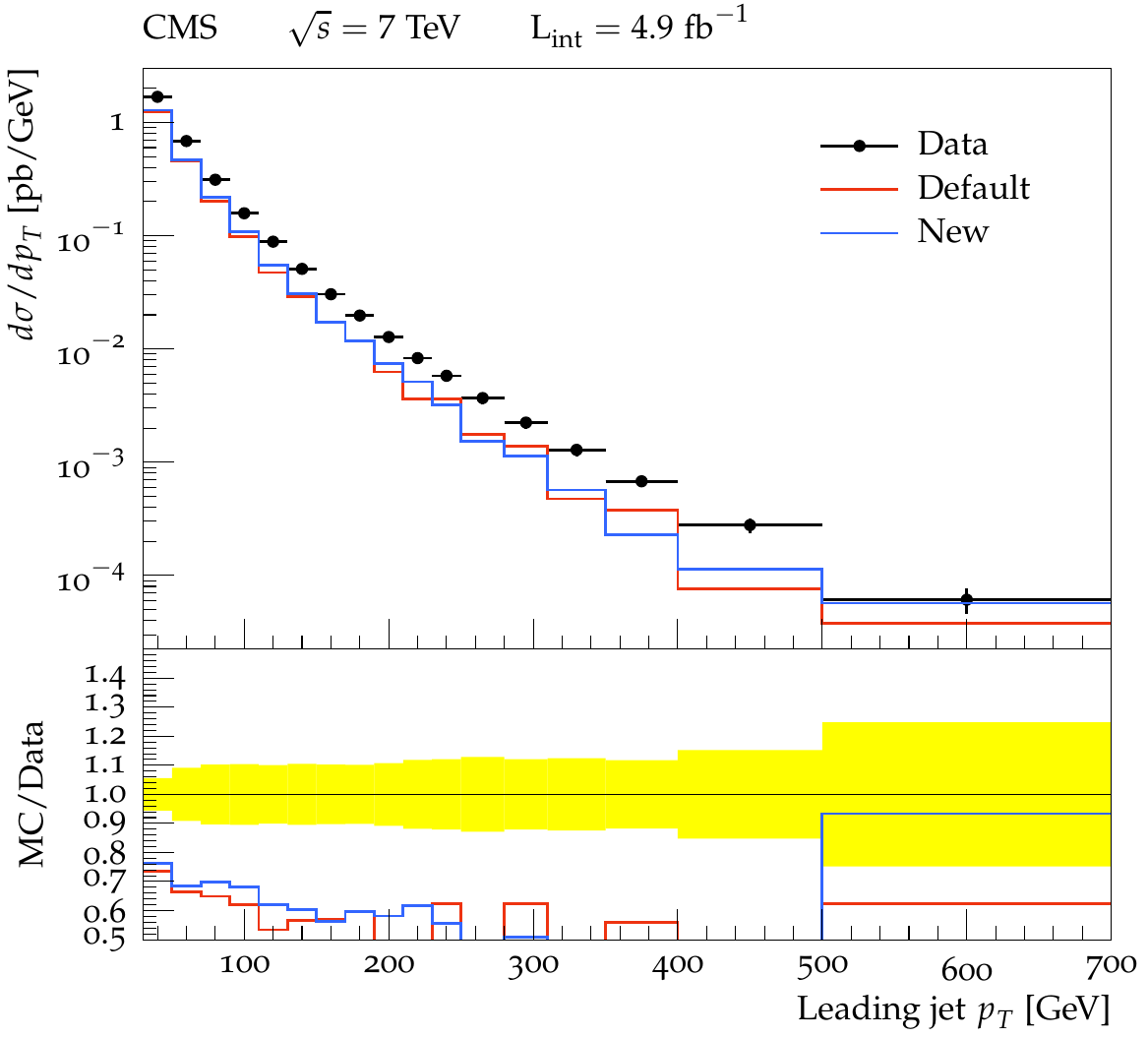}\\[2pt]
(c) \hspace{200pt} (d) \\
\includegraphics[width=0.44\textwidth]{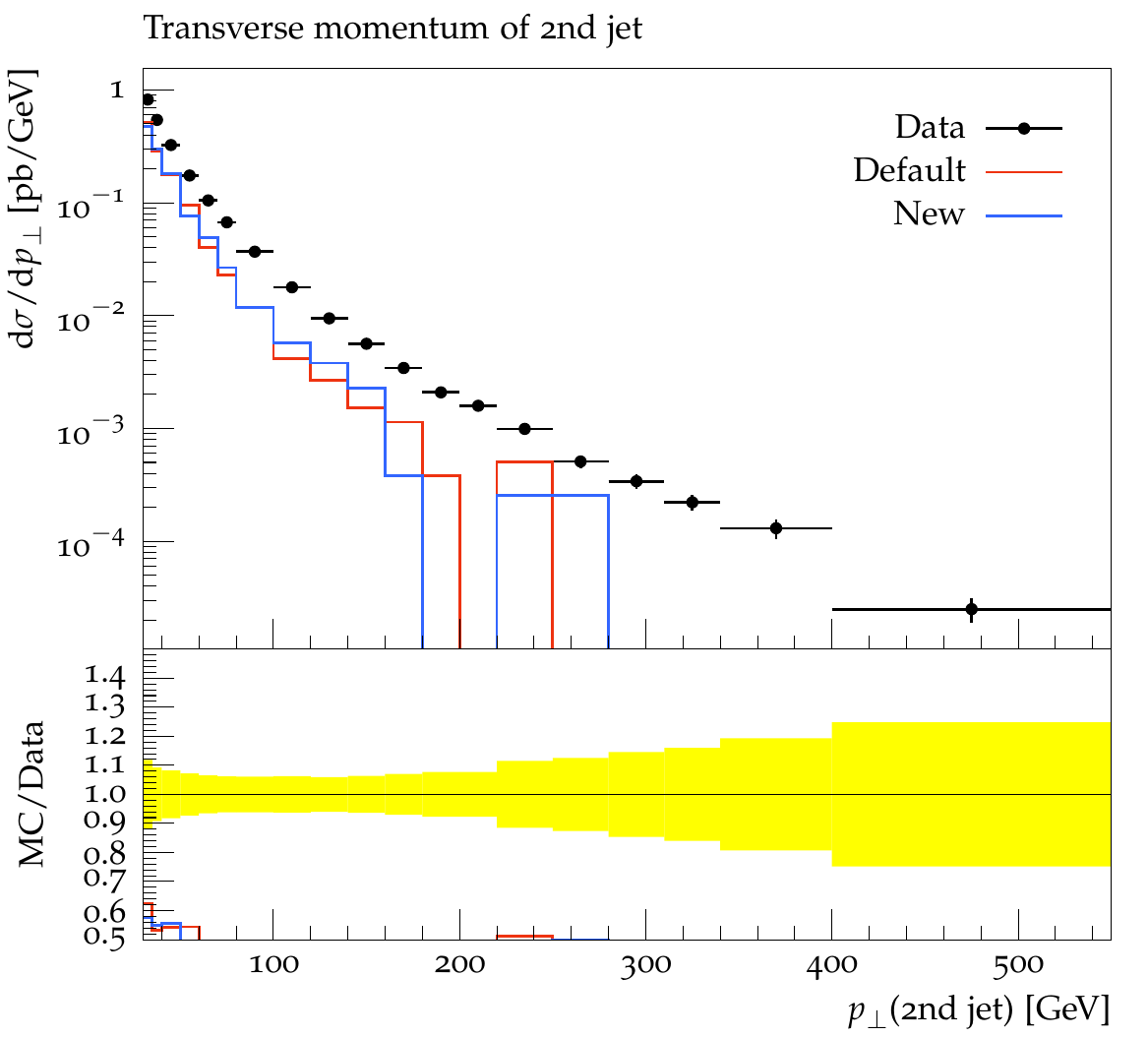}
\includegraphics[width=0.44\textwidth]{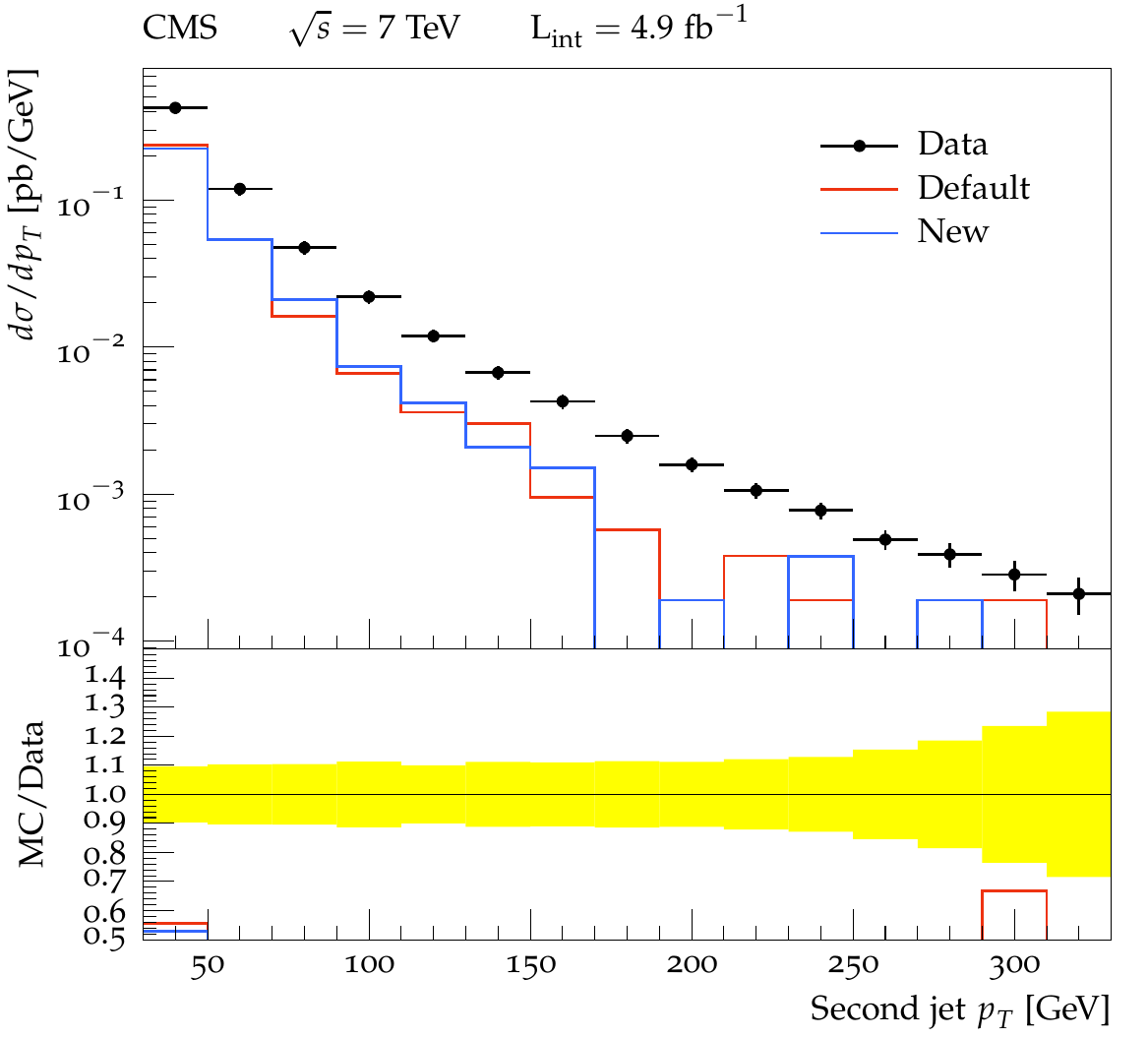}\\[2pt]
(e) \hspace{200pt} (f)
\caption{Jet production in inclusive $\gammaZ$ events, as measured by 
ATLAS (a,c,e) and CMS (b,d,f), respectively, for pp collisions at 7~TeV 
\cite{Buckley:2010ar, Aad:2013ysa,Khachatryan:2014zya}: 
(a,b) the number of jets, (c,d) $\pT$ of the first jet, 
(e,f) $\pT$ of the second jet.}
\label{Fig:ATLASCMSgmZ}
\end{figure}

The production of $\gammaZ$ is associated with jets. The multiplicity 
and $\pT$ spectra of such jets are given in Fig.~\ref{Fig:ATLASCMSgmZ},
compared with ATLAS \cite{Aad:2013ysa} and CMS \cite{Khachatryan:2014zya} 
data. It is interesting to note that the new procedure seems to lead 
to a slightly higher jet activity, at least for the ATLAS jet definition,
but overall differences are small. As before the high-multiplicity and
high-$\pT$ tails are underestimated in the purely shower-based approach, 
so for a better description the need to inject information from 
higher-order matrix-elements remains unchanged. 

\subsection{QCD jets}

\begin{figure}[t!] \centering
\includegraphics[width=0.3\textwidth,angle=-90]{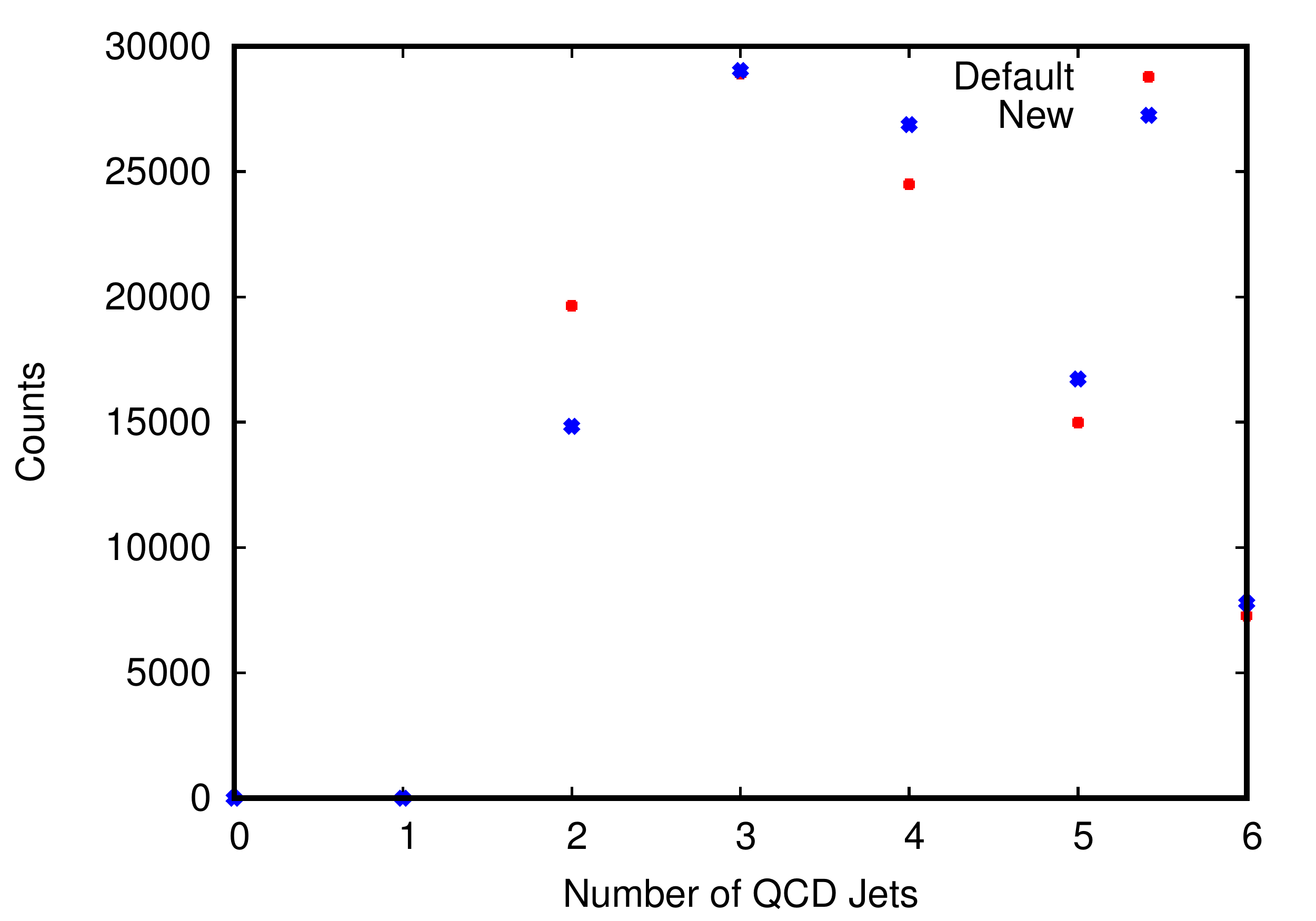}
\includegraphics[width=0.3\textwidth,angle=-90]{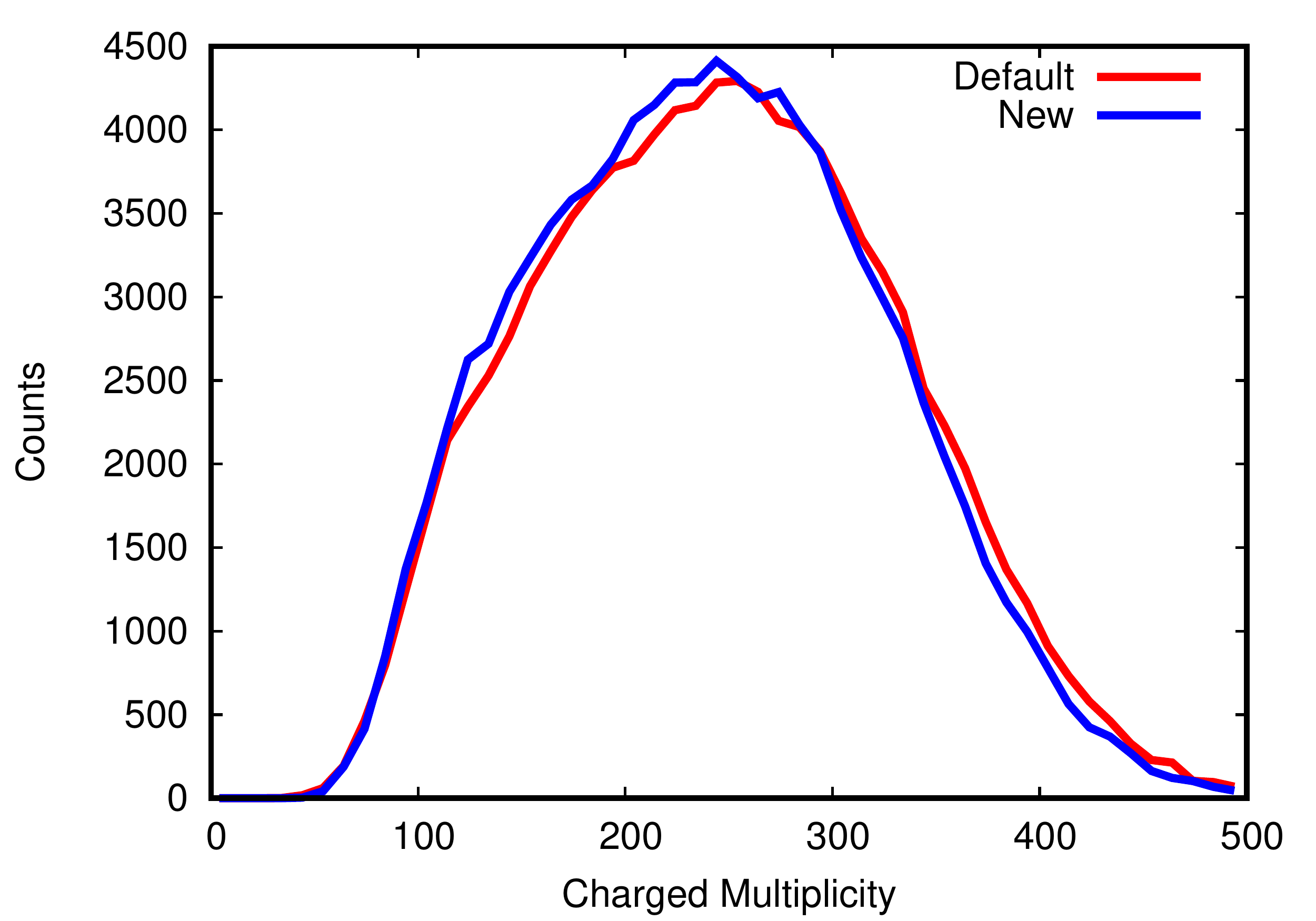}\\
\vspace{5pt}
(a) \hspace{200pt} (b) \\
\includegraphics[width=0.3\textwidth,angle=-90]{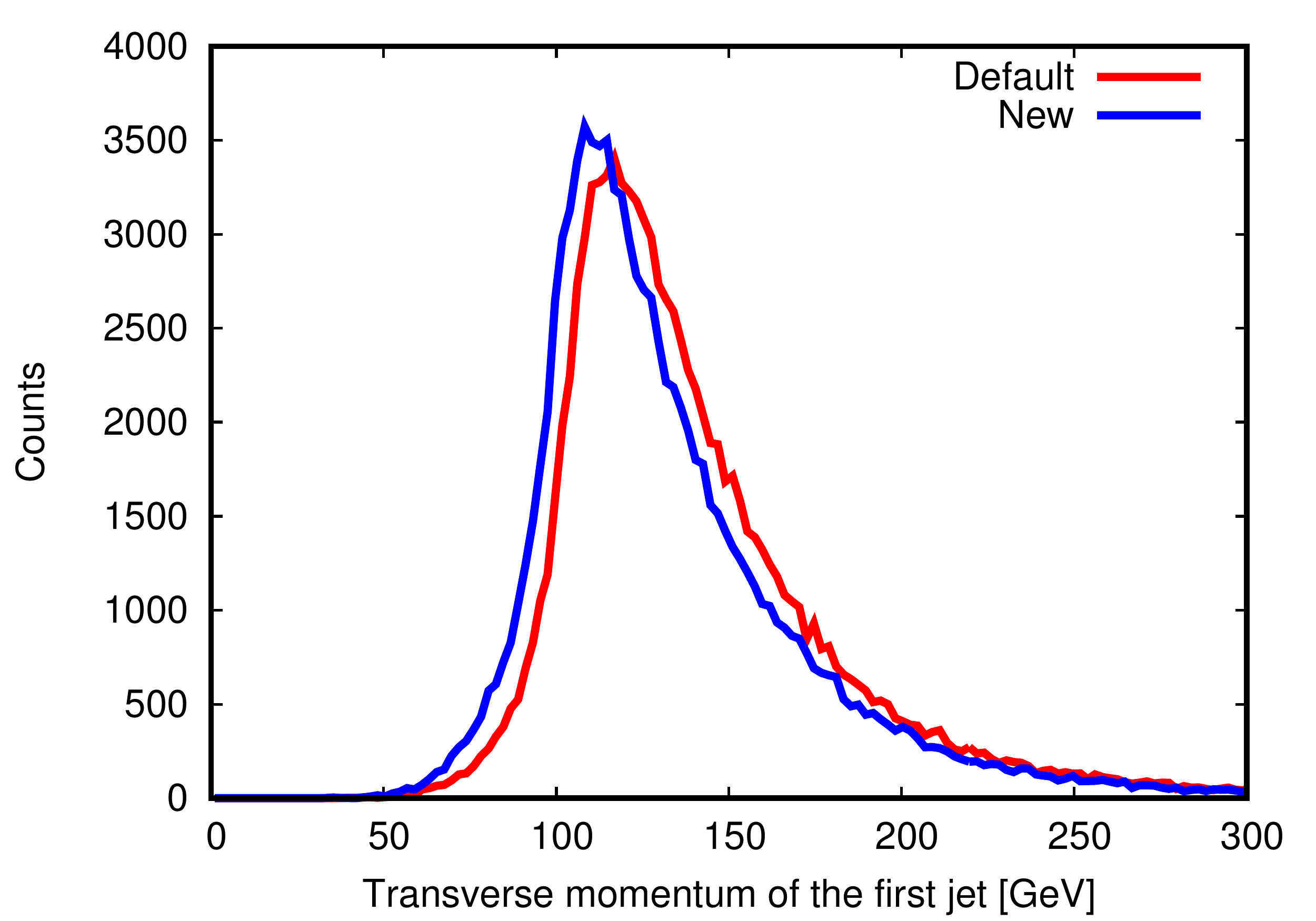}
\includegraphics[width=0.3\textwidth,angle=-90]{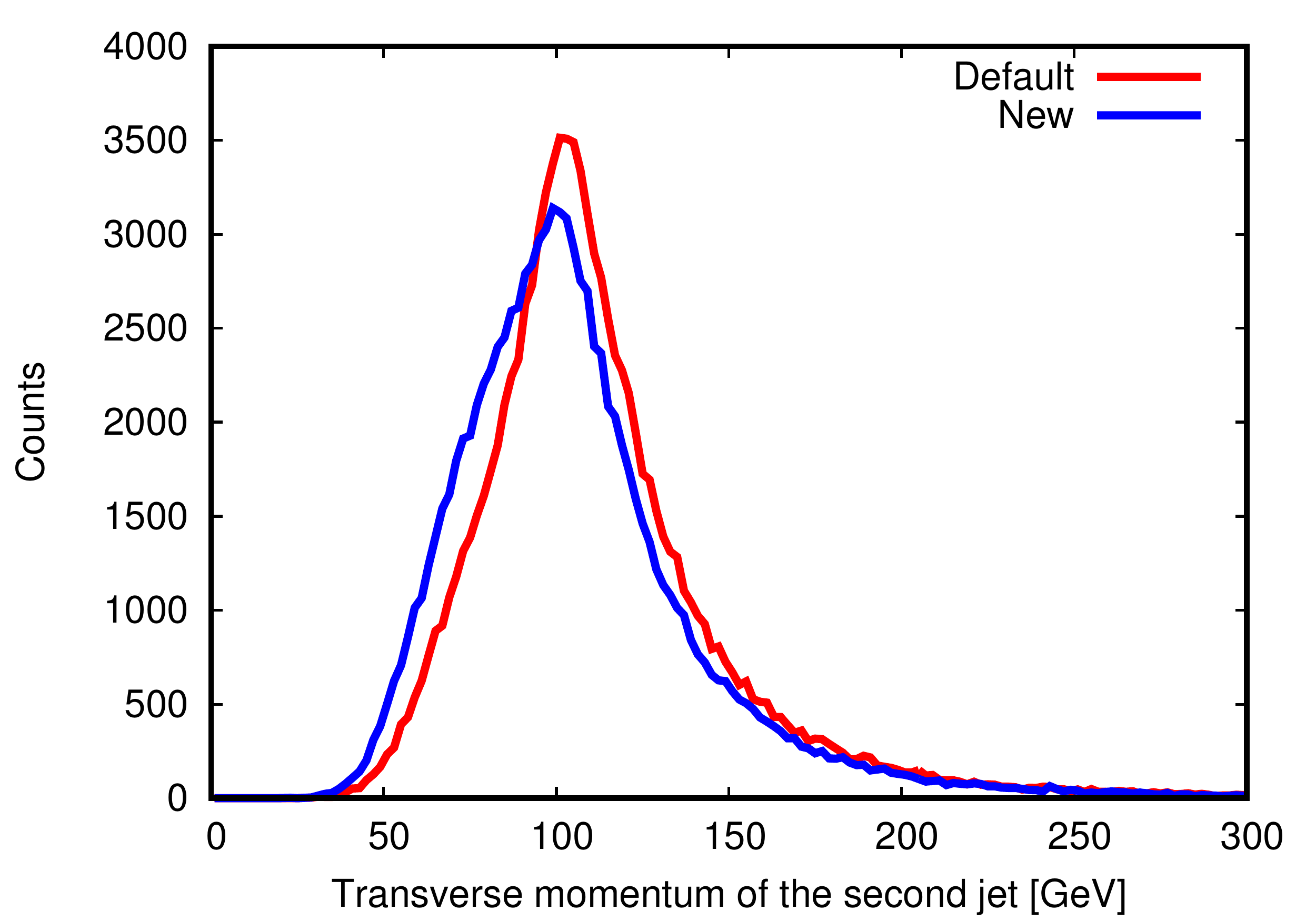}\\
\vspace{5pt}
(c) \hspace{200pt} (d) \\
\includegraphics[width=0.3\textwidth,angle=-90]{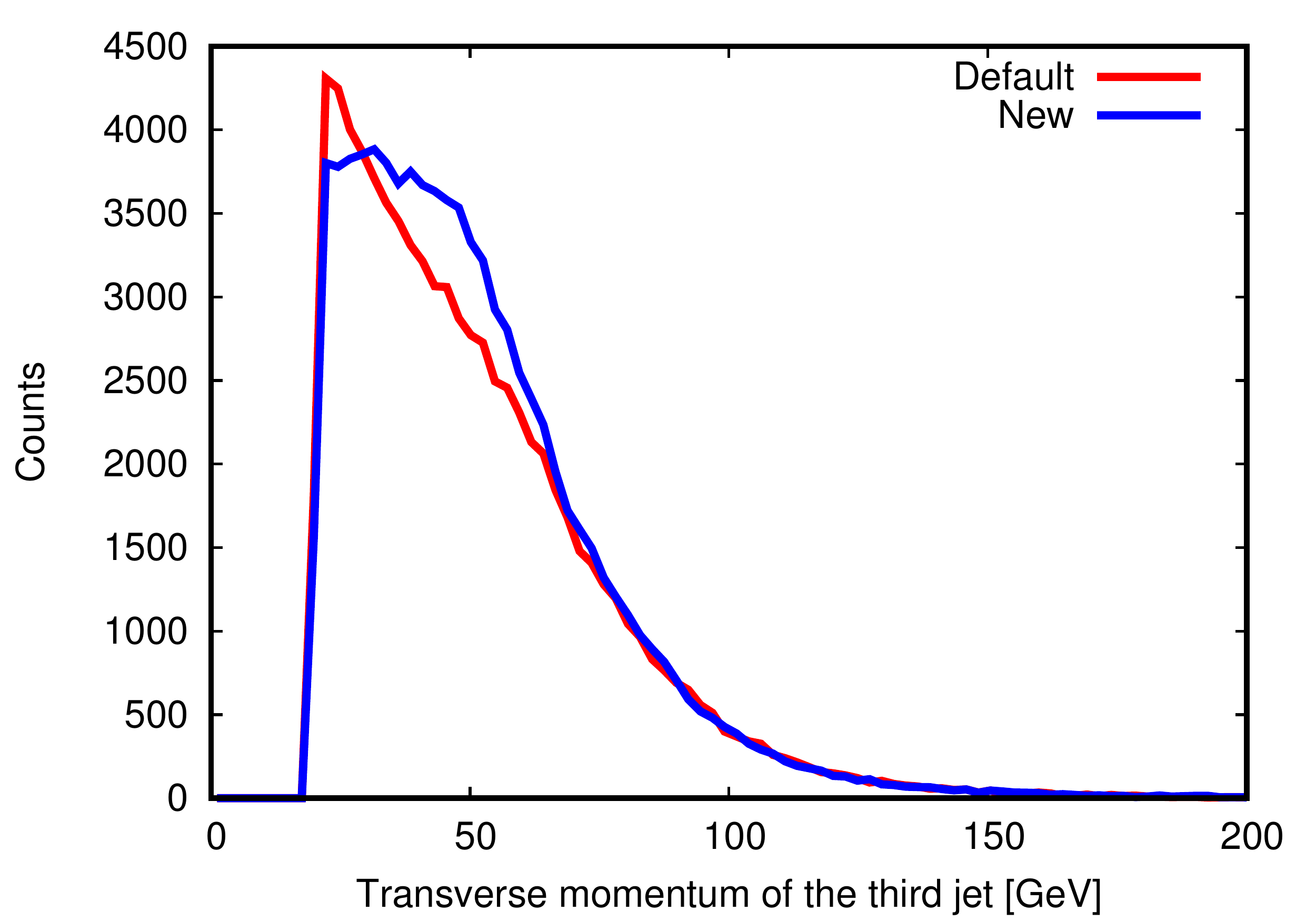}
\includegraphics[width=0.3\textwidth,angle=-90]{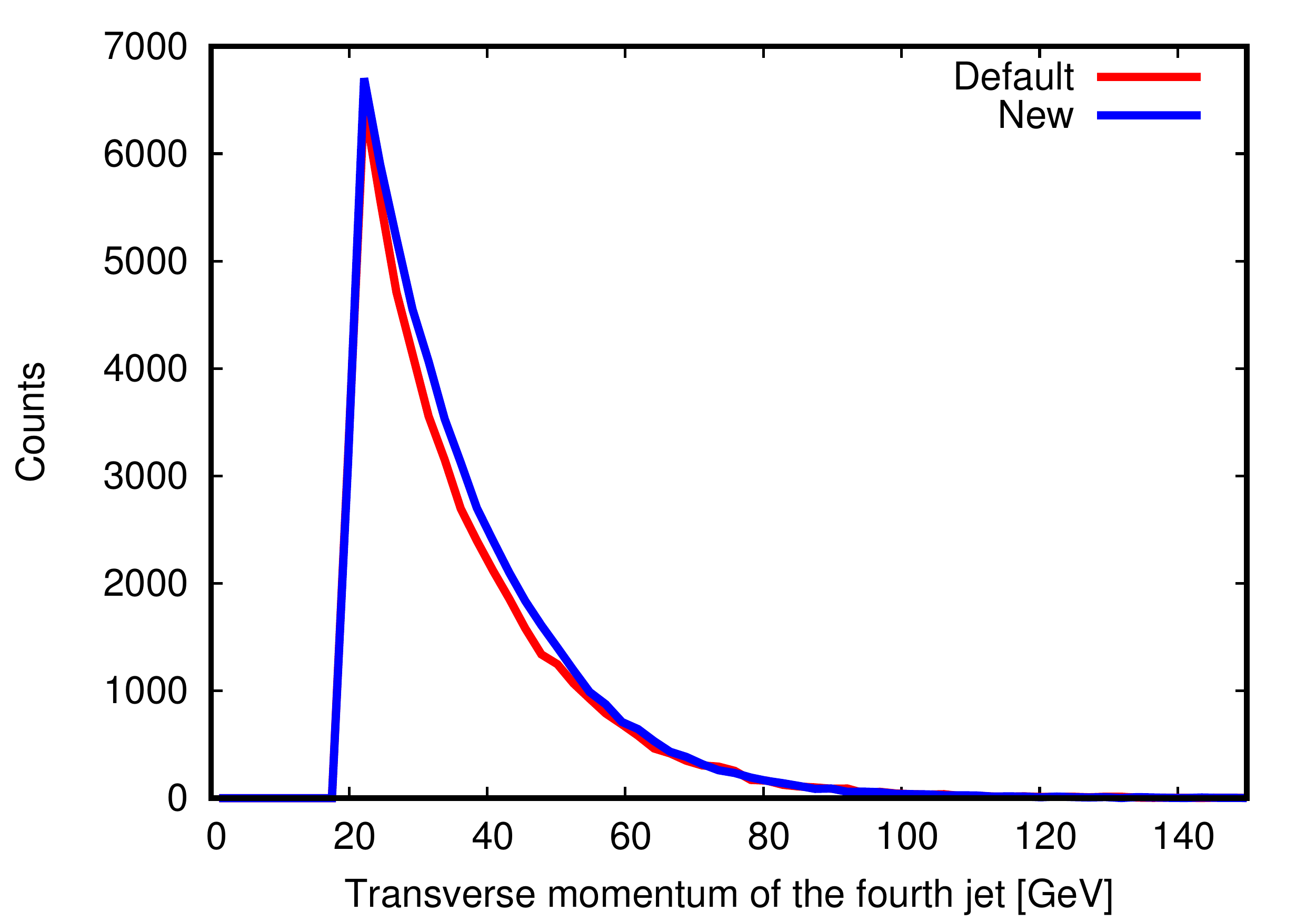}\\
\vspace{5pt}
(e) \hspace{200pt} (f)
\caption{Comparison of shower algorithm results for the toy study
described in the text: (a) number of QCD jets, (b) charged multiplicity,
transverse momentum of the (c) first jet, (d) second jet, (e) third jet, 
(f) fourth jet.}
\label{Fig:pTJets}
\end{figure}

Another relevant area for comparisons is QCD jet production by
$2 \to 2$ processes (\mbox{$\q\q \to \q\q$}, \mbox{$\g\g \to \g\g$}, 
\mbox{$\q\g \to \q\g$}, \ldots) followed by showers. These showers are 
evolved downwards from the $2 \to 2$ $\pThat$ scale, in order to avoid
doublecounting.  

Again we begin by a toy study, for LHC with $\sqrt{s}=14$~TeV and
$\pThat > 100$~GeV. Jets are defined by the anti-$k_{\bot}$ algorithm
\cite{Cacciari:2008gp}, with $R=0.7$ and $p_{\perp\mathrm{jet}} > 20$~GeV. 
Under these conditions the new procedure produces somewhat more jets 
than the default scheme, Fig.~\ref{Fig:pTJets}. Consistent with this 
the third and fourth jets (ordered by $\pT$) become harder, while the 
first two become softer. It is therefore slightly contradictory that 
the average charged multiplicity drops from 246 to 241 (with widths 
85 and 82, respectively). Further studies will be needed to sort out 
why some distributions suggest more activity and others less.

\begin{figure}[t!] \centering
\includegraphics[width=0.45\textwidth]{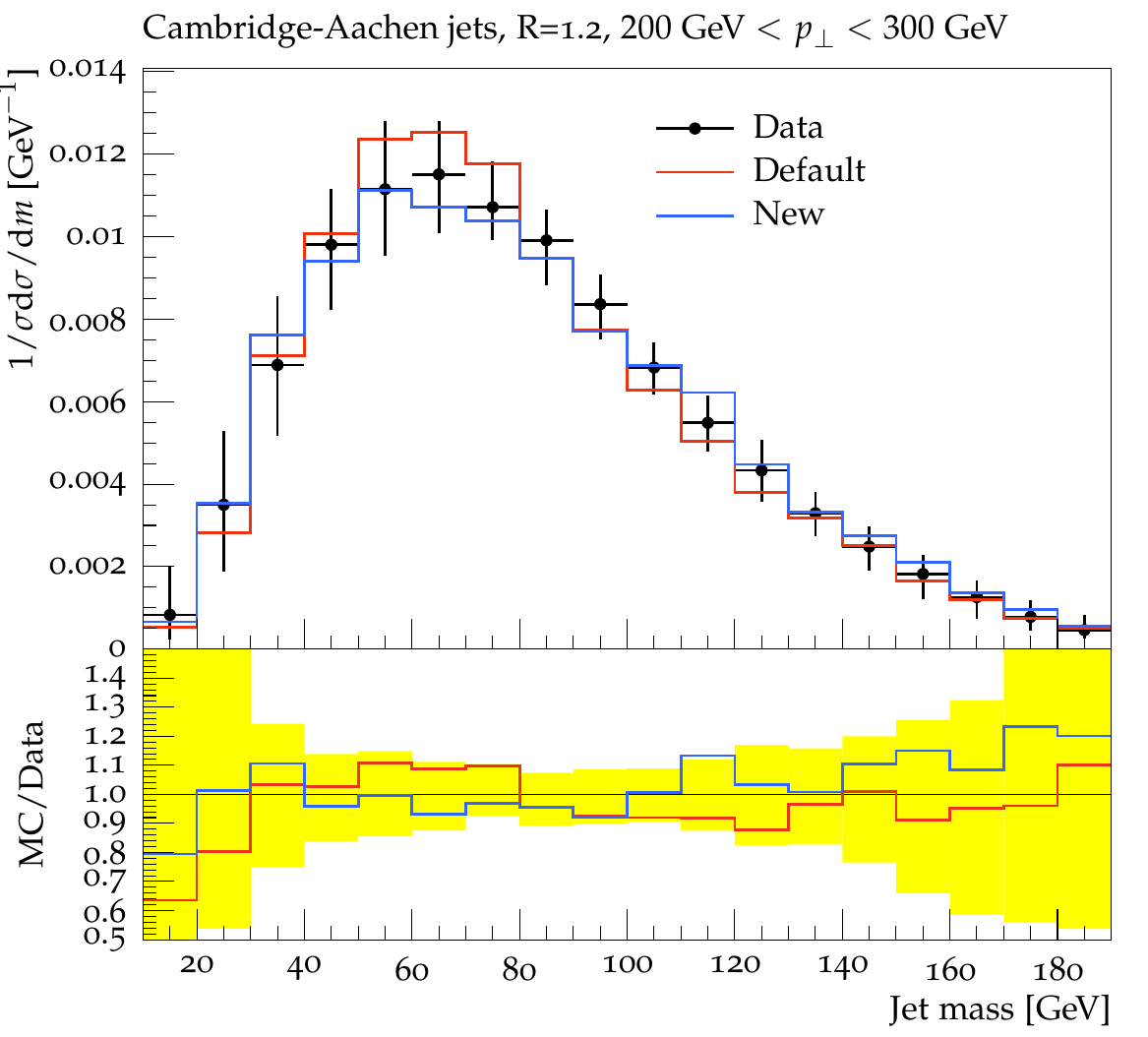}
\includegraphics[width=0.45\textwidth]{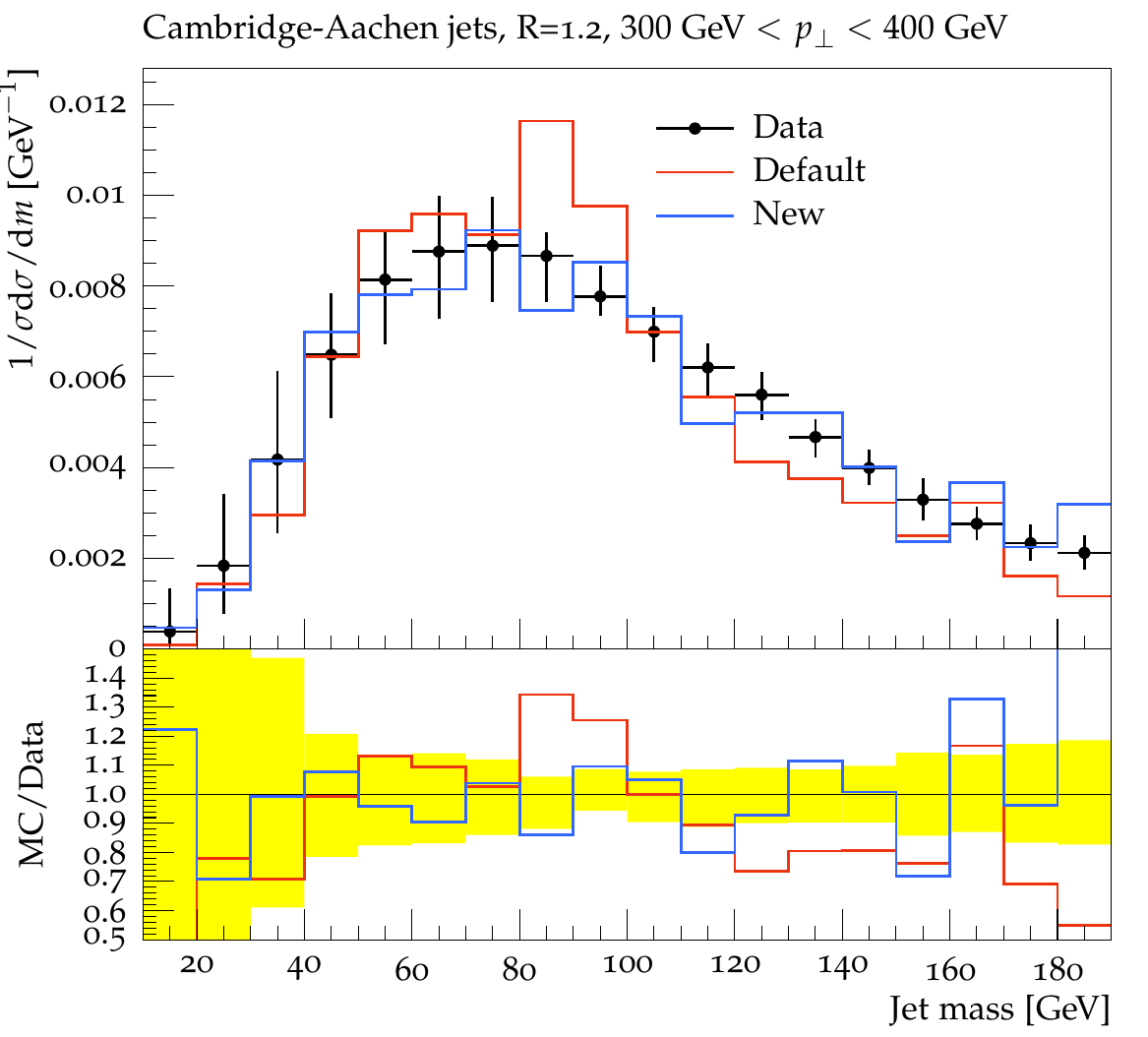}\\
\vspace{5pt}
(a) \hspace{200pt} (b) \\
\includegraphics[width=0.45\textwidth]{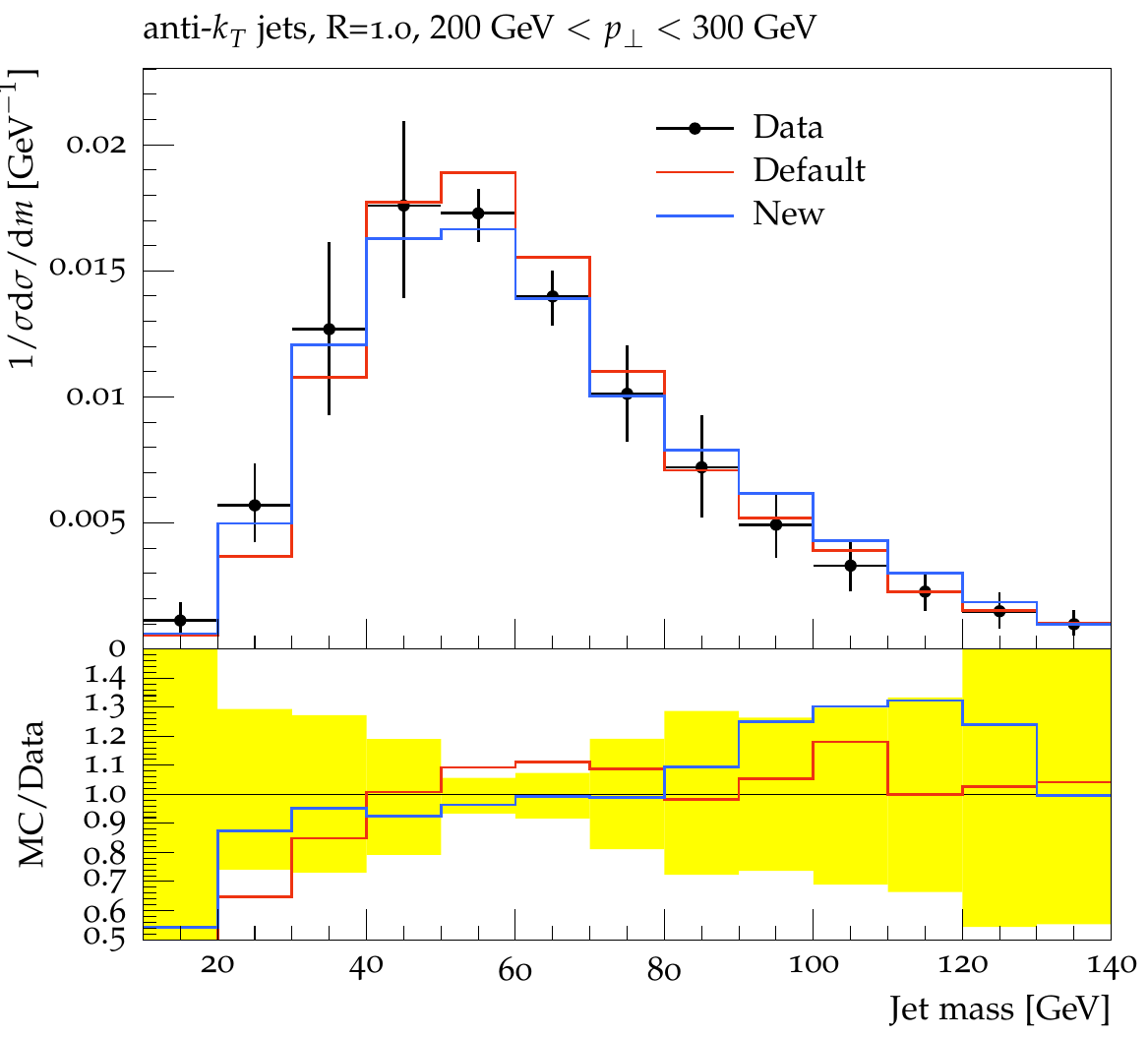}
\includegraphics[width=0.45\textwidth]{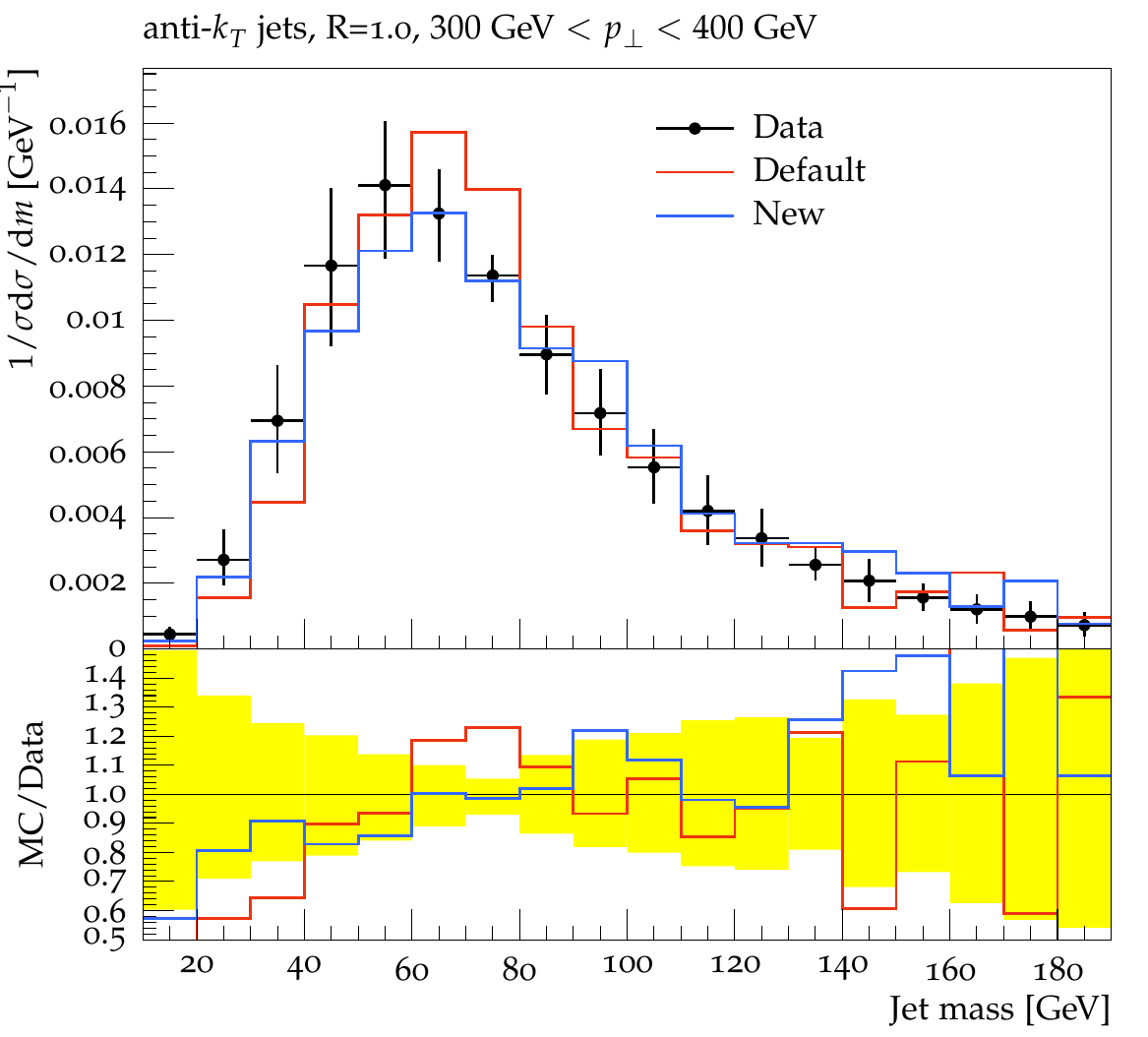}\\
\vspace{5pt}
(c) \hspace{200pt} (d)
\caption{Jet mass spectra measured in QCD events by ATLAS for pp 
collisions at 7~TeV \cite{Buckley:2010ar, ATLAS:2012am}. The jets 
have been reconstructed with: (a, b) Cambridge-Aachen with $R=1.2$, 
(c, d) anti-$k_{\perp}$ with $R=1.0$.}
\label{Fig:jetMass}
\end{figure}

\begin{figure}[t!] \centering
\includegraphics[width=0.45\textwidth]{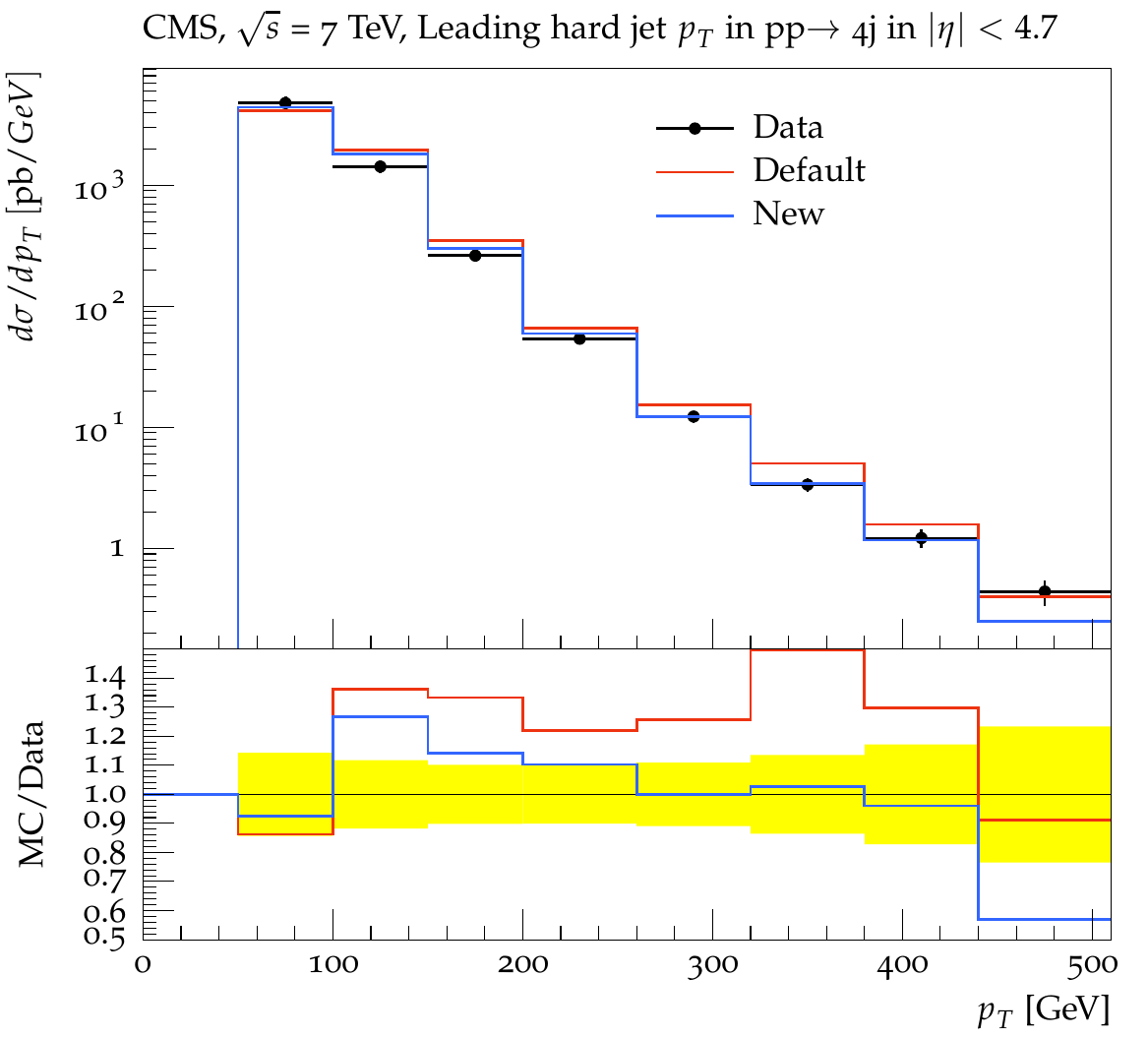}
\includegraphics[width=0.45\textwidth]{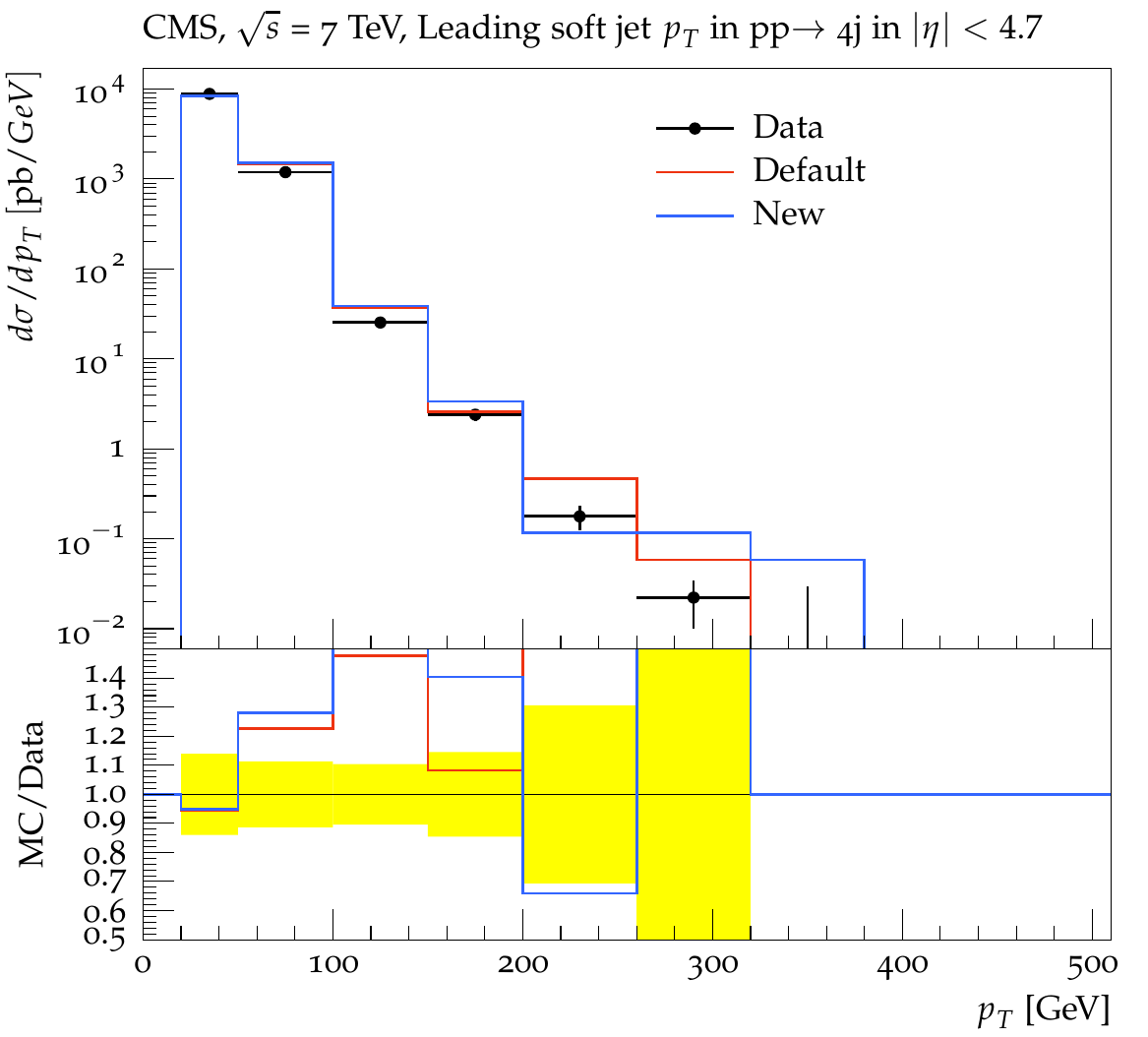}\\
\vspace{5pt}
(a) \hspace{200pt} (b) \\
\includegraphics[width=0.32\textwidth]{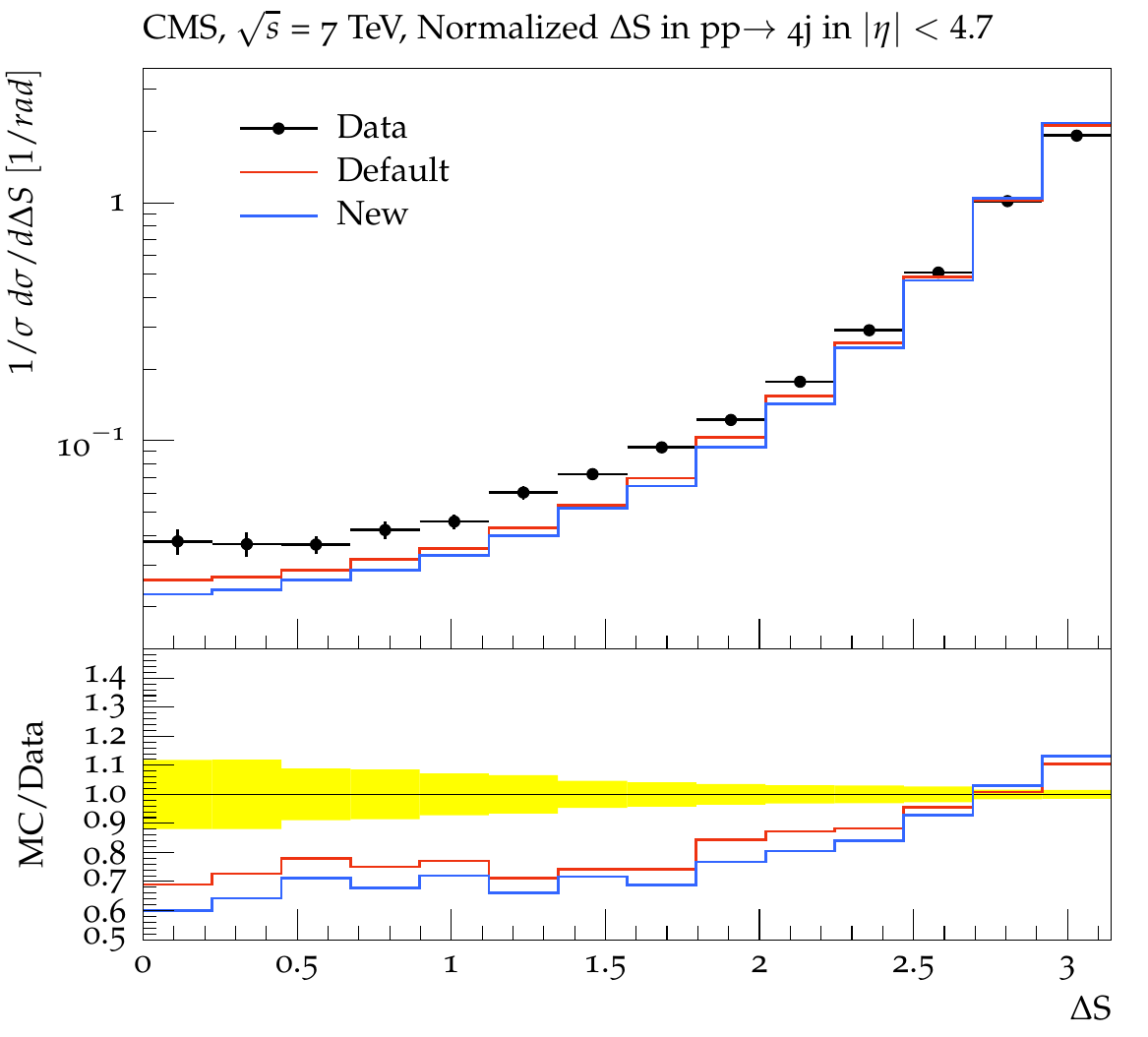}
\includegraphics[width=0.32\textwidth]{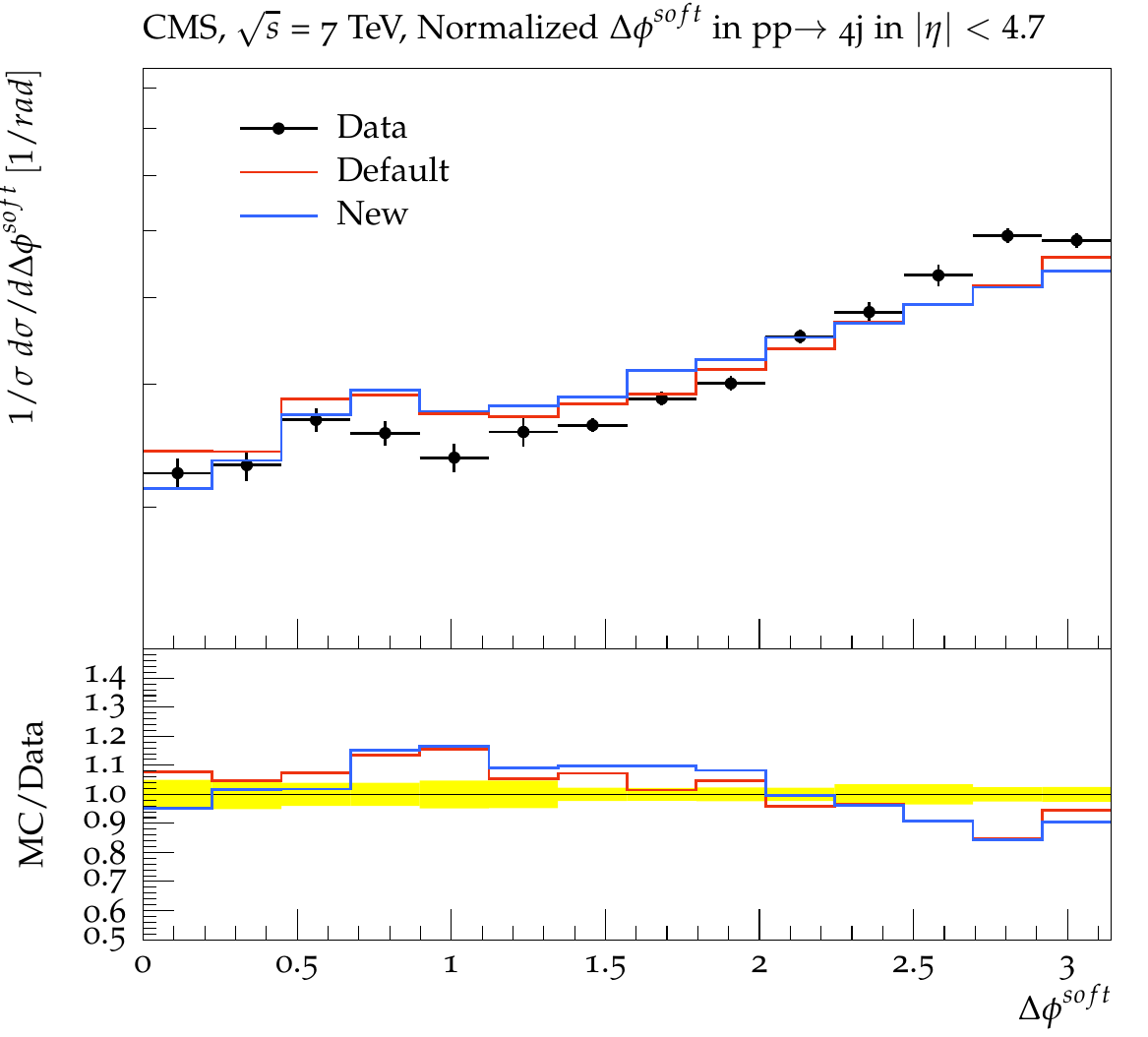}
\includegraphics[width=0.32\textwidth]{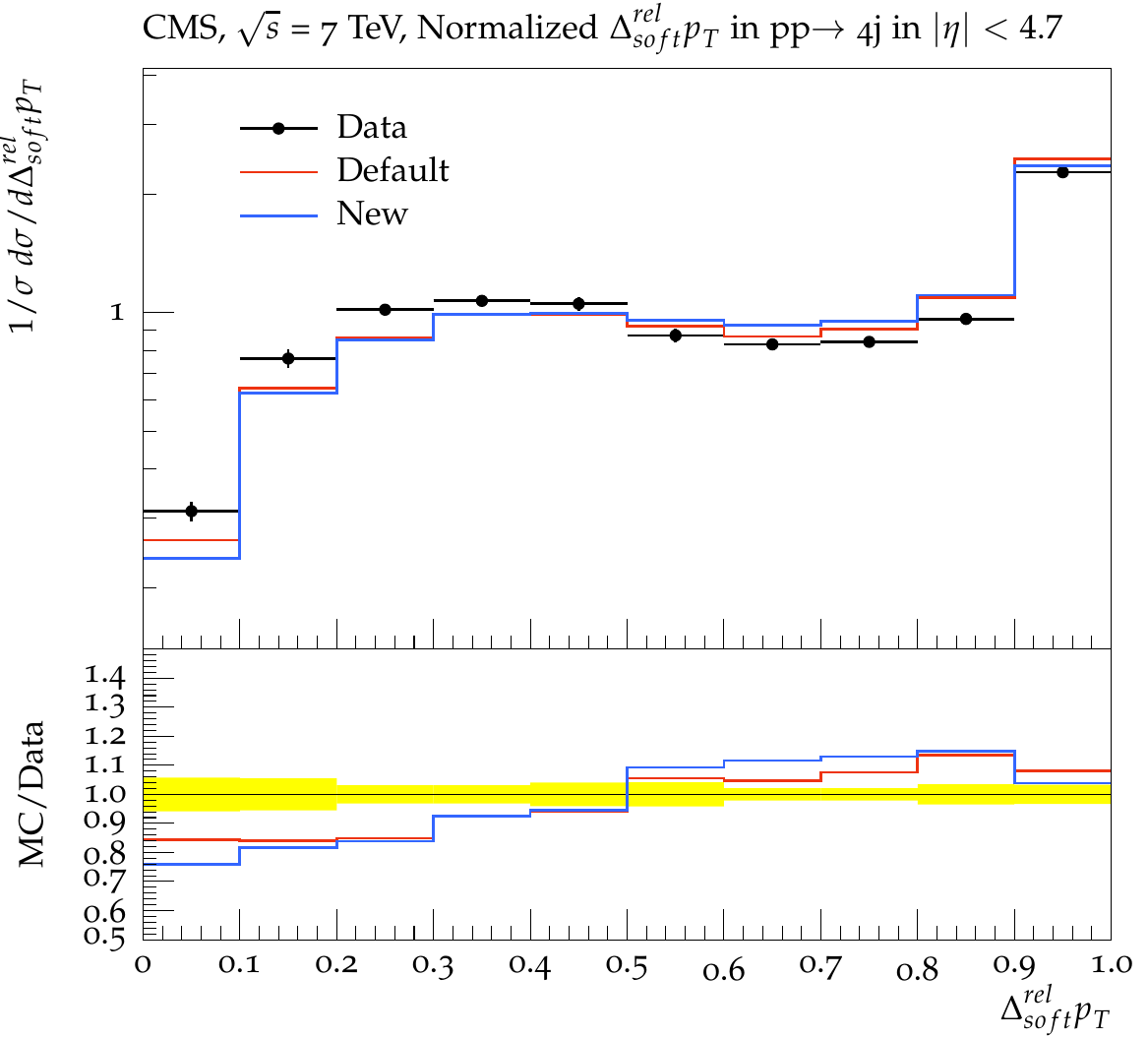}\\
\vspace{5pt}
(c) \hspace{150pt} (d) \hspace{150pt} (e) 
\caption{Four-jet cross-sections measured by CMS for pp collisions at 
7~TeV \cite{Buckley:2010ar, Chatrchyan:2013qza}, as a function of several 
observables defined in \cite{Chatrchyan:2013qza}.}
\label{Fig:CMSQCD}
\end{figure}

Turning to real data, in Fig.~\ref{Fig:jetMass} a few jet mass spectra 
measured by ATLAS \cite{ATLAS:2012am} are presented. It can be seen that 
both shower procedures describe the data well, with some hints of 
improvements in the new scheme. In Fig.~\ref{Fig:CMSQCD}, the
exclusive cross-section for the process pp $\to 4\, \mathrm{jets} + X$
is given as a function of several observables, as measured by CMS 
\cite{Chatrchyan:2013qza}. The $\pT$ spectra of the jets are well
reproduced. For the plots involving angular variables, where the 
agreement is somewhat worse, the errors may be related to the same
issues as already discussed for gluon polarization and colour coherence,
but again ultimately point to the need for four-jet matrix-element
input.

\subsection{DIS}

\begin{figure}[p] \centering
\includegraphics[width=0.45\textwidth]{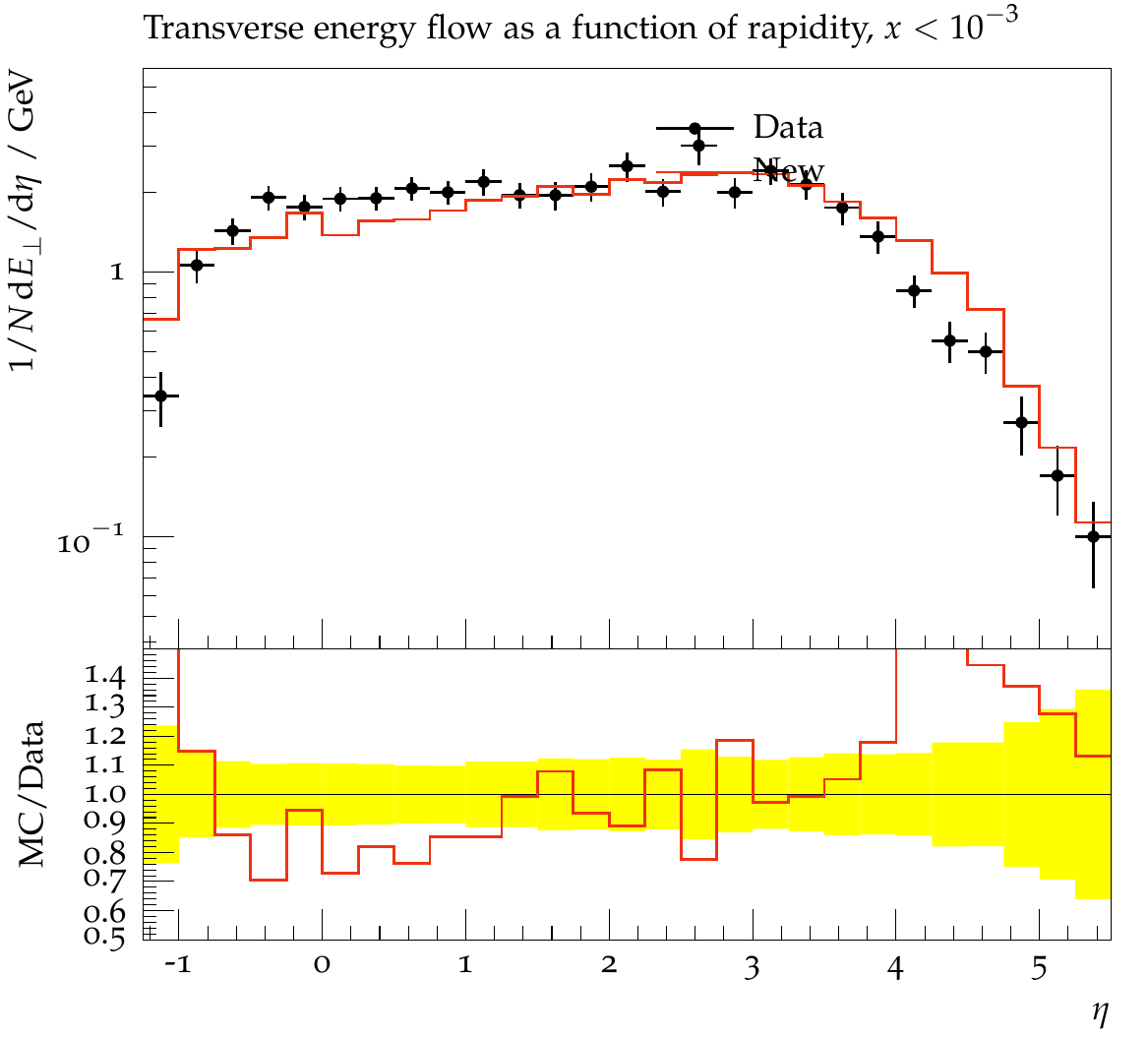}
\includegraphics[width=0.45\textwidth]{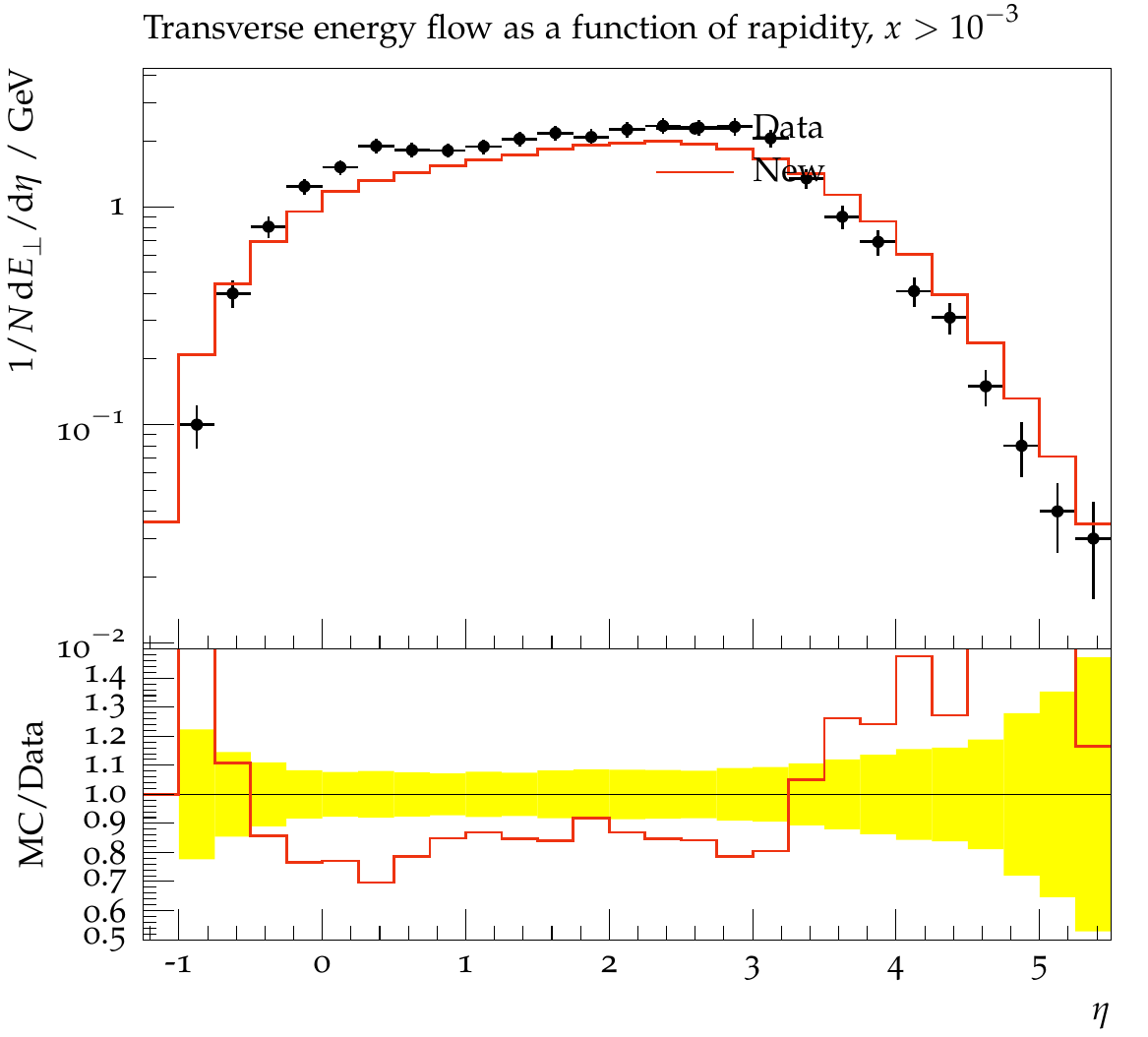}\\[3pt]
(a) \hspace{200pt} (b) \\
\includegraphics[width=0.45\textwidth]{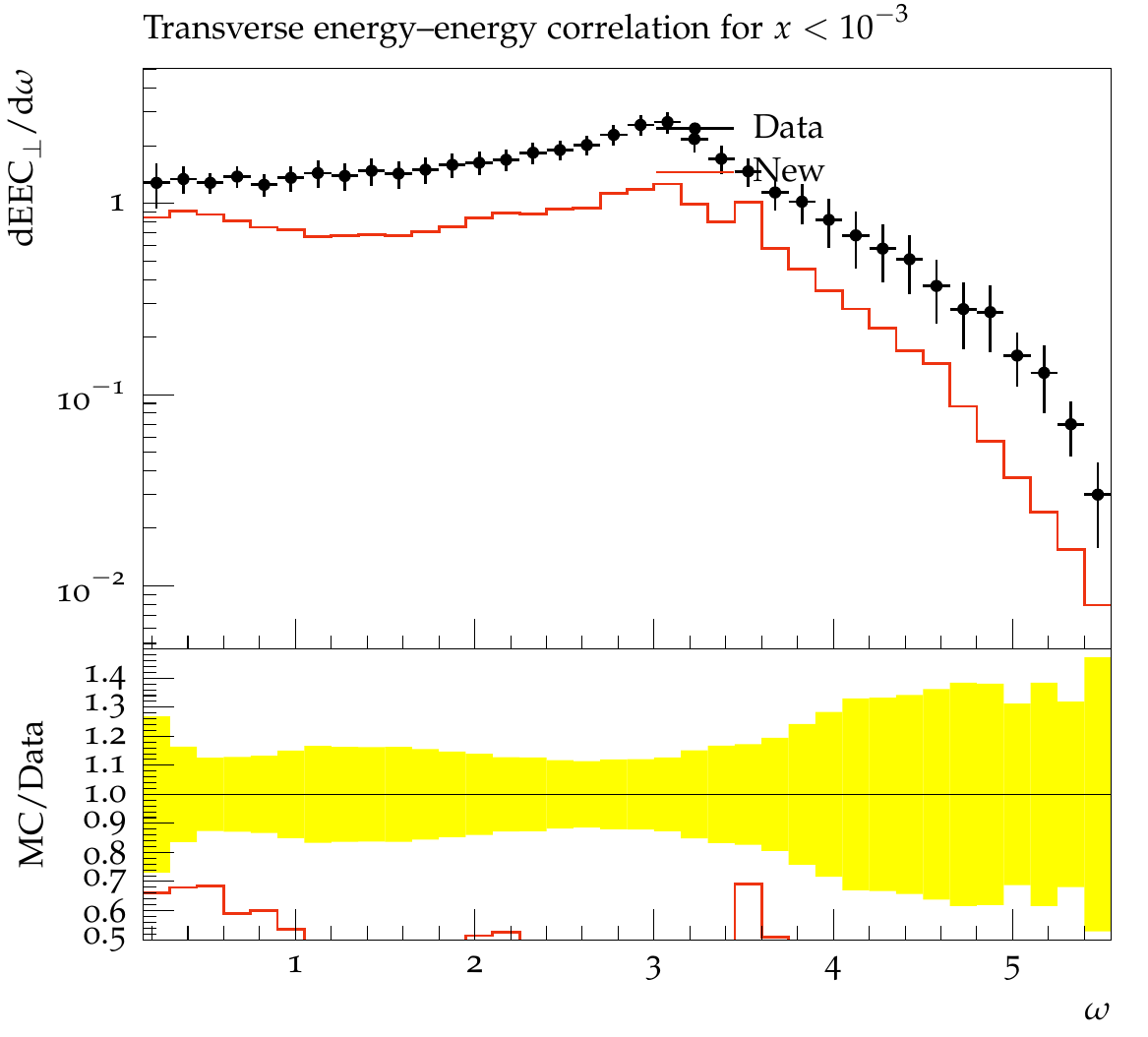}
\includegraphics[width=0.45\textwidth]{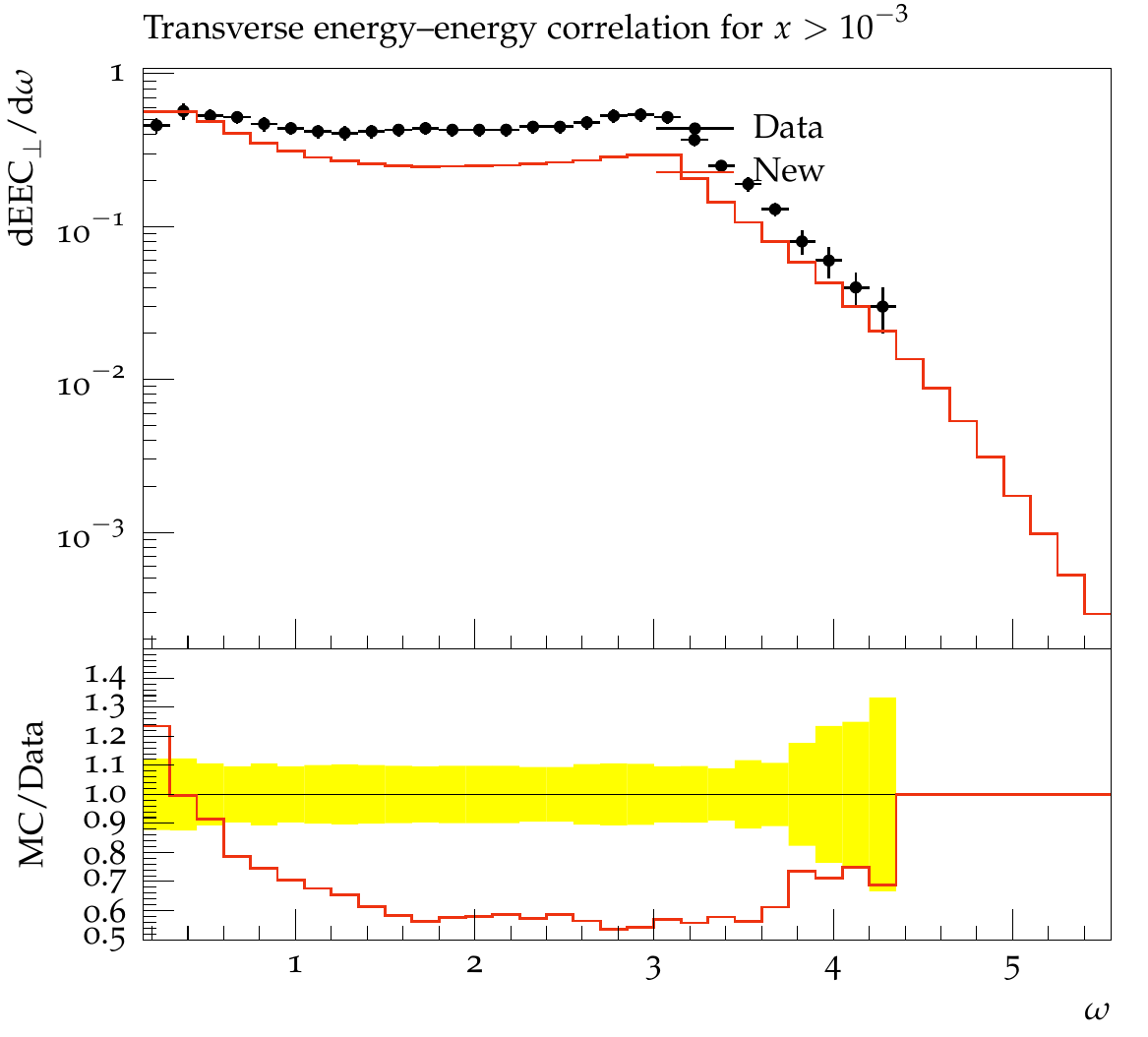}\\[3pt]
(c) \hspace{200pt} (d) \\
\includegraphics[width=0.45\textwidth]{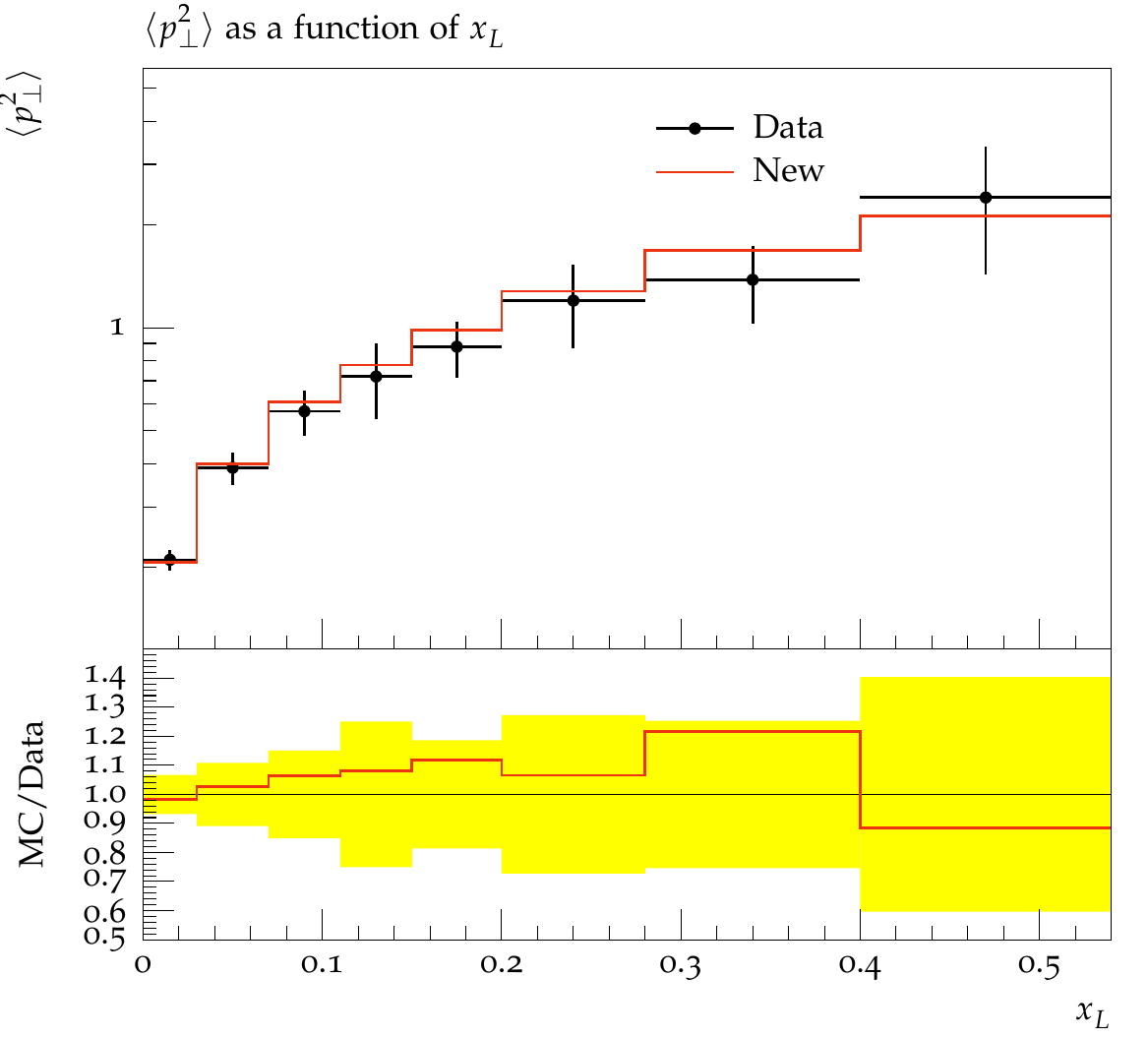}
\includegraphics[width=0.45\textwidth]{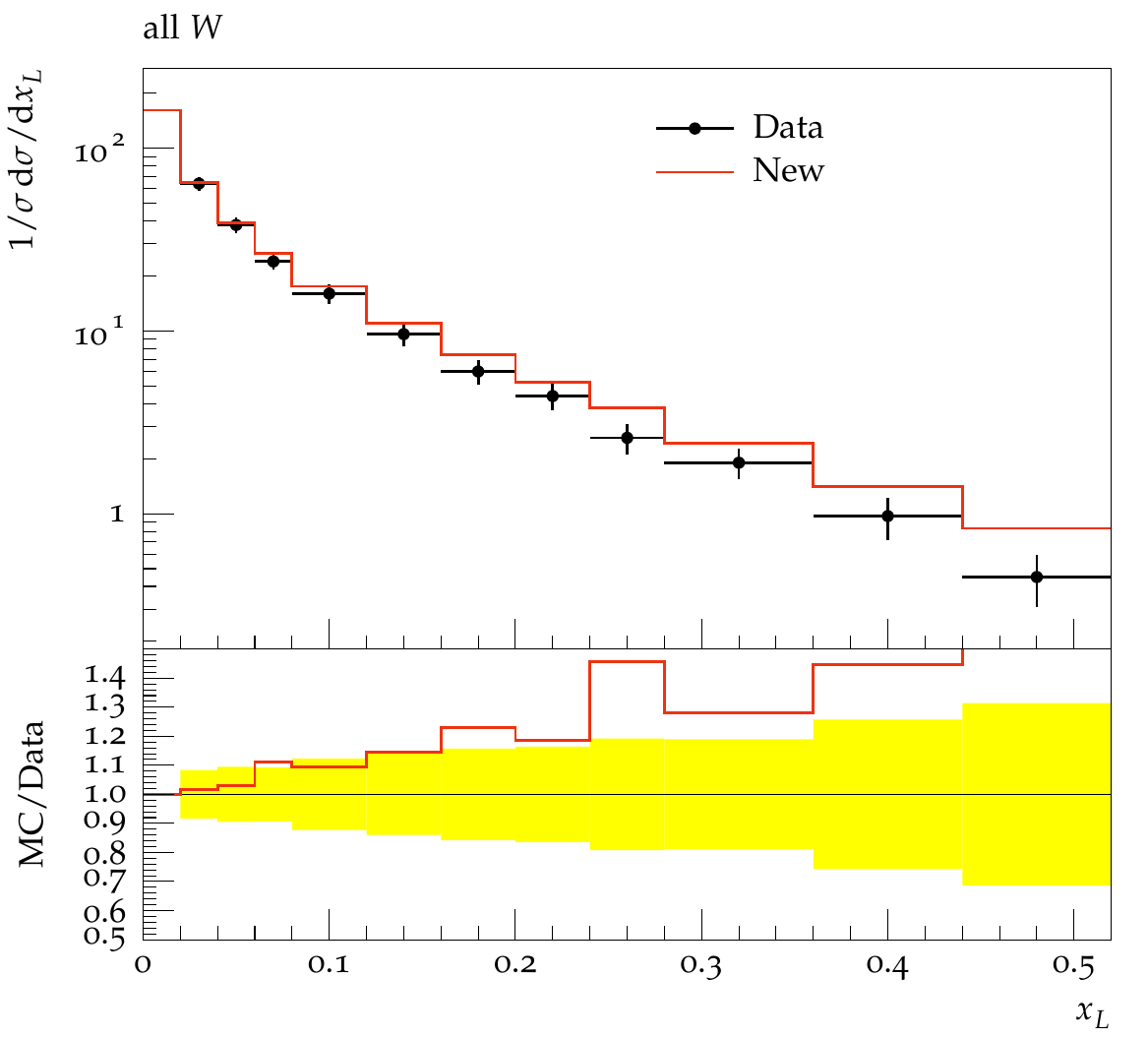}\\[3pt]
(e) \hspace{200pt} (f) 
\caption{DIS events at HERA \cite{Buckley:2010ar,Abt:1994ye}. 
The new scheme is compared with H1 data for $Q^2 > 40\,\mathrm{GeV}^2$. 
The definitions of the different observables 
can be found in \cite{Abt:1994ye}.}
\label{Fig:H1}
\end{figure}

The new scheme gives the opportunity to study DIS. As shown before, the
branching probability of an IF-type generates the full cross-section
in the case of gluon emission. The dipole approach, applied to DIS, is
then expected to reproduce the data decently. A first comparison has
been done for HERA with a 820 GeV proton beam colliding a 26.7 GeV
electron beam \cite{Abt:1994ye}, Fig.~\ref{Fig:H1}. As can be seen,
single-particle properties are reasonably well described, whereas
the energy-energy correlation undershoots data. This could be studied 
further, e.g. by comparing with the \textsc{Ariadne} dipole model
\cite{Lonnblad:1992tz} which was known to give a very accurate description
of these data. Unfortunately it cannot easily be combined with 
\textsc{Pythia~8}. 

\begin{figure}[t!] \centering
\includegraphics[width=0.34\textwidth,angle=-90]{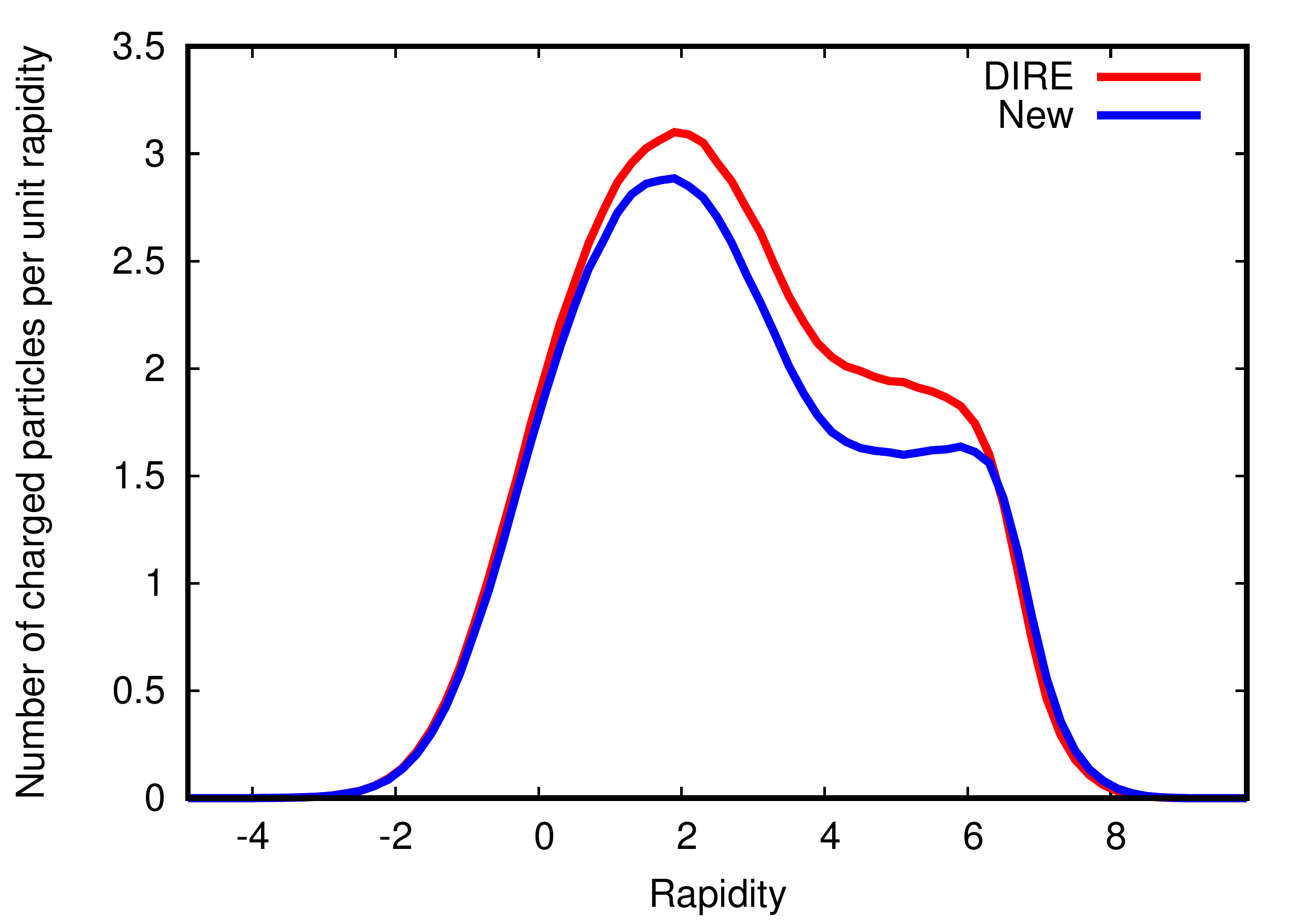}
\includegraphics[width=0.34\textwidth,angle=-90]{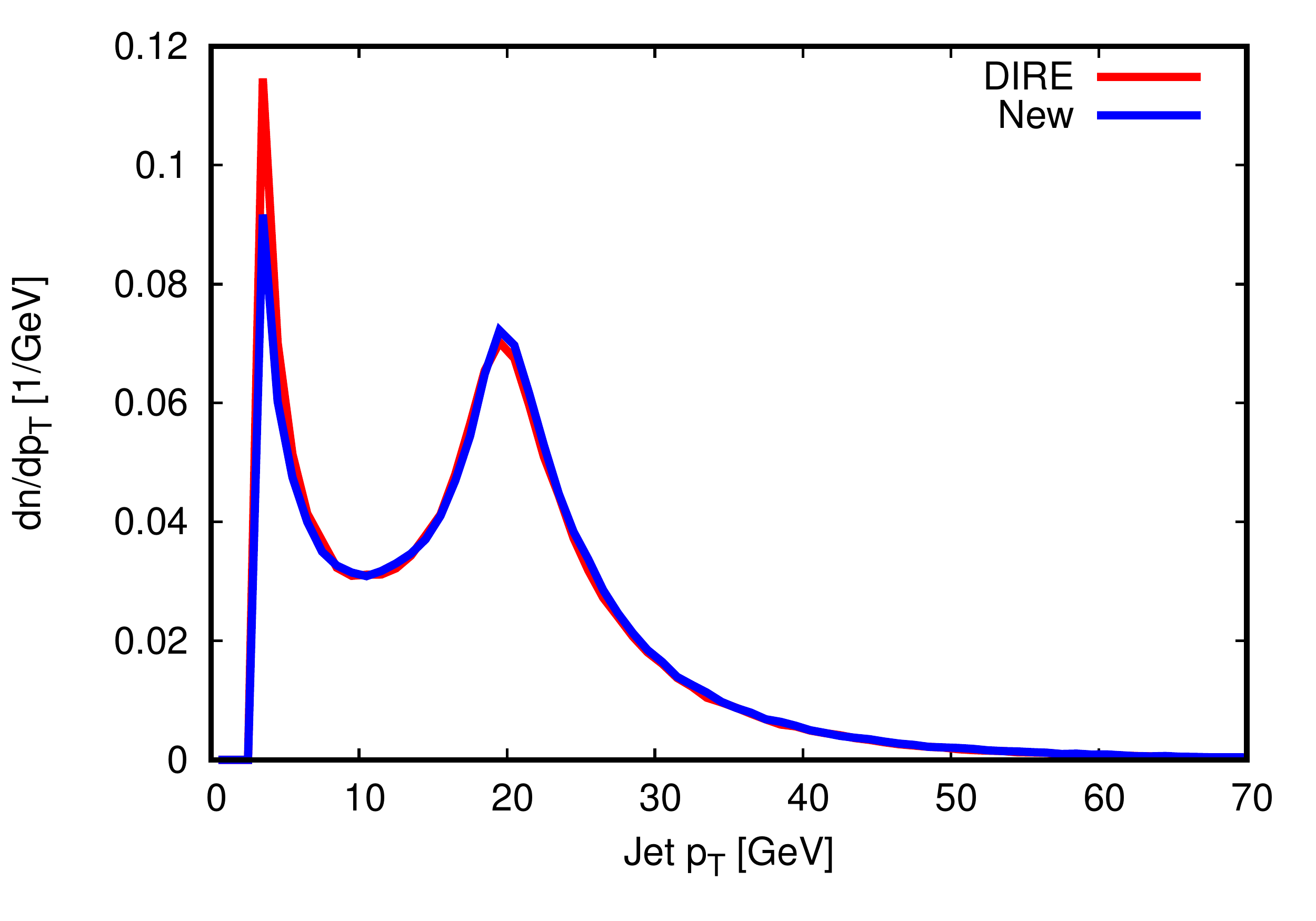}\\
\vspace{5pt}
(a) \hspace{200pt} (b) 
\caption{Comparison between the new approach and the \textsc{Dire} shower,
HERA ep collisions with beam energies 27.5 and 920~GeV and 
$Q^2 > 400 \, \mathrm{GeV}^2$: (a) charged rapidity spectrum and 
(b) jet $\pT$ spectrum using the anti-$k_{\perp}$ algorithm with 
$R = 0.7$ and $\pTmin = 3$~GeV.}
\label{Fig:Dire}
\end{figure}

An alternative for comparisons is instead offered by the \textsc{Dire}
dipole shower program \cite{Hoche:2015sya}, which can be used as a 
plugin to \textsc{Pythia~8}, such that the shower algorithms is the 
only difference. Results turn out to be closely similar to each other
in most variables, Fig.~\ref{Fig:Dire}. It may be noted, however, that 
the charged multiplicity is somewhat higher in \textsc{Dire}, even though
the jet rate is comparable. One reason is that \textsc{Pythia} by default
uses a smooth dampening of ISR at small scales, similar to the one for 
MPIs \cite{Sjostrand:1987su,Corke:2010yf}, while \textsc{Dire} has a lower 
sharp cutoff, giving it a larger partonic activity at small $\pT$ scales.  

\subsection{Particle production rates}

\begin{table}[t] 
\centering
\begin{tabular}{|c||c|c|c||c|c|c|} 
\hline
 & \multicolumn{3}{c||}{no MPI} & \multicolumn{3}{c|}{with MPI}
\\ \hline
showering & low $\theta_{\q\q}$ & high $\theta_{\q\q}$ & all $\theta_{\q\q}$ 
  & minbias & $\pThat > 25$~GeV & $\pThat > 200$~GeV \\
          & high $\theta_{\mathrm{col}}$ & low $\theta_{\mathrm{col}}$ 
          & all $\theta_{\mathrm{col}}$ & & & 
\\ \hline
no          &  $43 \pm 7$  &   $28 \pm 6$ &   $52 \pm 9$ 
            &  $76 \pm 40$ & $122 \pm 47$ & $126 \pm 46$ \\ 
old global  &  $75 \pm 17$ &  $56 \pm 16$ &  $83 \pm 18$ 
            & $113 \pm 74$ & $216 \pm 83$ & $253 \pm 83$ \\ 
new dipole  &  $82 \pm 17$ &  $40 \pm 10$ &  $82 \pm 18$ 
            & $110 \pm 72$ & $209 \pm 79$ & $248 \pm 80$ \\ 
\hline
\end{tabular}
\caption{Average charged event multiplicity and the width of the 
multiplicity distribution without showers, or with the old global
or new local showers. The first two columns are for 
$\q + \q' \to \q + \q'$ processes only, with cuts as described in 
the text, and the third for all $2 \to 2$ processes with 
$\pT > 25$~GeV. The last three columns are with MPIs also included,
for events of increasing (average) hardness.}
\label{table:nch}
\end{table}

To finish, it is useful to reflect on one of the 
key features that distinguish the dipole from the global-recoil approach,
that the amount of ISR depends on the invariant mass of the colour 
dipoles stretched out to the final state. To illustrate this, consider 
$\q + \q' \to \q + \q'$ with $\q \neq \q'$. Here only $t$-channel gluon
exchange contributes, so colour flows from the incoming $\q$ to the 
outgoing $\q'$. A small quark scattering angle $\theta_{\q\q}$ (in the 
rest frame of the collision) thus corresponds to a large colour-flow 
scattering angle $\theta_{\mathrm{col}} = \theta_{\q\q'} = \pi - \theta_{\q\q}$, 
and vice versa. 
With cuts $\hat{m} = \sqrt{\hat{s}} > 500$~GeV and $25 < \pThat < 50$~GeV 
for the hard $2 \to 2$ process, for LHC at 13 TeV, allowed scatterings 
split into one low-angle and one high-angle range. The total charged 
multiplicity for these cases is shown in Table~\ref{table:nch}. We see 
that, even without any showers or MPIs, the higher $\theta_{\mathrm{col}}$ 
range gives the larger multiplicity, because it implies higher-mass 
nonperturbative colour strings stretched between the scattered quarks 
and the beam remnants. The multiplicities come up when the old global
showers are added, slightly more for higher $\theta_{\mathrm{col}}$: while
the handling of the II dipole end is identical in the two cases, the
FI one does contain a dependence on the colour dipole masses. In the new
dipole shower the difference is much more pronounced, however. Even if
the $\pTe$ scale of the shower evolution is constrained from above by
the $\pT$ scale of the hard $2\to 2$ process in both cases, below that 
scale the phase space for emissions inside a dipole is (logarithmically) 
related to its mass, so a larger $\theta_{\mathrm{col}}$ opens up for more 
radiation.   

In real life it is not feasible to tag whether a quark scattering occured 
at a small or a large angle, and for the dominant $\g + \g \to \g + \g$ 
processes it is not even a meaningful question to ask. 
There is only a small net remaining multiplicity difference between 
the old and new shower approaches if all QCD $2 \to 2$ processes at all
angles are included, as we see in the third column of 
Table~\ref{table:nch}. A more differential picture can be obtained 
from the multiplicity dependence on the rapidity separation 
$|\Delta y| \approx -2 \ln \tan (\theta/2)$ between the two hard jets, 
while still not distinguishing $\theta$ from $\pi - \theta$. 
And, unfortunately, both shower options show almost identically the same 
rise of the multiplicity with increasing $|\Delta y|$, 
leaving no discriminating power.

When MPIs are included the differences are slightly larger in absolute 
numbers, since each MPI gives its contribution to the net difference;
see the last three columns of Table~\ref{table:nch} for inclusive
(nondiffractive) minimum-bias events, and jet events above two different
$\pT$ thresholds for the hard process. Relative to the no-shower baseline 
it is still notable that the old and new showers add almost the same amount 
of extra activity. It may suggest that many semi-inclusive observables
will also look rather similar, and that more specific observables will
be needed to distinguish the two. Furthermore, the charged-multiplicity 
discrepancies presumably could be resolved by some modest retuning, 
e.g.\ a slightly larger $\as$ for the new dipole showers. Such a retuning 
has not (yet) been done; at this stage of the studies it is useful 
to compare the two options under identical conditions. 

\section{Summary and Outlook}

The dipole approach to showers is not new, and in that sense the study 
in this article does not provide anything fundamentally new. It does 
offer a few new insights, however, and access to a new useful tool.

One of the interesting aspects is the constraints imposed on the 
recoil kinematics. For a final--final dipole the emission recoil 
can be shared between the two dipole ends in many ways.
But for an initial--final dipole a central constraint is that the 
initial incoming parton must be parallel with the beam axis. This 
enforces the same kinematics whether the process is viewed as that 
of final-state radiation with a recoil in the initial state or the 
other way around.

It could still be that the contributions from the initial- and final-state
emissions would need to be added to obtain the complete initial--final 
dipole emission pattern. It would then be important to combine the two
without gaps or doublecounting. The cleanest way is to compare with the
radiation pattern in Deeply Inelastic Scattering, notably for the 
gluon-emission process, $\gamma^* \q \to \q \g$. The pleasant surprise 
then is that initial-state emissions cover the full phase space on its 
own, with the correct denominator singularity structure and a finite 
numerator very close to the correct one. The final-state emissions 
do not give quite as simple an expression. A suitable reweighting could 
fix it, but the simple solution is to describe the full emission pattern
by ISR and omit FSR altogether. Unfortunately the results are not as
clean for gluon splittings, $\gamma^* \g \to \q \qbar$. This is no news;
gluon splittings have never fitted well inside the dipole framework. 

Some first comparisons with data have been presented in this article,
and look promising, but not so very different from the old non-dipole
approach. Partly this is because experimental procedures by necessity
average over different topologies, thereby largely cancelling effects
in the underlying dynamics, and partly because the old scheme approximated 
the boost effects by ISR azimuthal asymmetries. In fact, in some 
distributions the old approximate scheme gives larger effects 
than the new one does, and here data agrees better with the old one 
although the new one is theoretically better motivated.  

It should be remembered, however, that the new dipole framework has not 
yet been tuned, but is based on the existing default tune for the old
scheme, so disagreements were to be expected. Some difference thus may 
be tuned away, but others may remain.
Furthermore, no attempt has been made to include matching
and merging with higher-order matrix elements \cite{Buckley:2011ms}. 
In such a more complete framework the difference between alternative 
showers are partly masked, since the showers then are not providing 
the hard topologies. The ordering of emissions and the Sudakov factors 
that go with that do depend on the shower algorithm, however, so the 
possibility to compare different algorithms is useful to assess 
uncertainties. One may also want to combine global and local recoils 
by what technically is most convenient for the matching and merging 
schemes, similarly to what is already available for FSR. 

The new algorithm has been implemented in \textsc{Pythia}, and will
soon be publicly available. This will allow more detailed comparisons
to be made than the ones presented in this article. Comparisons with LHC 
data will here be the main application, needless to say. But it will
also open up for DIS studies, which could not be done with 
\textsc{Pythia}~8 previously, except by linking to the \textsc{Dire}
shower \cite{Hoche:2015sya}. Do note, however, that currently QED 
emission is not included. The $\mathrm{e} + \q \to \mathrm{e} + \q$ 
process implies quadrupole radiation, that could be approximated by 
a sum of dipoles. This is another example where further studies 
and extensions should follow.

In summary, our new dipole-based algorithm for ISR offers an interesting
alternative to the existing one. The new code can stand on its own right 
away for a number of interesting studies, but to realize the full potential 
it may require some further extensions. 

\section*{Acknowledgements}

This project has received funding in part by the Swedish Research 
Council, contracts number 621-2013-4287 and 2016-05996, 
in part from the European Research Council (ERC) under the European 
Union's Horizon 2020 research and innovation programme, grant agreement 
No 668679, and in part by the MCnetITN3 H2020 Marie Curie Initial 
Training Network, contract 722104.

\bibliographystyle{utphys}
\bibliography{lutp1728}

\end{document}